\documentclass[twoside,journey]{IEEEtran}
\usepackage{makecell}
\usepackage{array}
\usepackage{graphicx,amssymb,amsmath}
\usepackage{multicol}
\usepackage[noadjust]{cite}
\usepackage{setspace}
\usepackage{subfigure}
\usepackage{graphicx}
\usepackage{float}
\usepackage{url}
\usepackage{stfloats}
\usepackage{amsthm,pifont}
\usepackage{flushend}
\usepackage{cases,subeqnarray}
\usepackage{bm,multirow,bigstrut}
\usepackage{amsmath, amsthm, amssymb}
\usepackage{textcomp}
\usepackage{latexsym,bm}
\usepackage{booktabs}
\usepackage{xcolor}
\usepackage{mathtools}
\usepackage{dsfont}
\usepackage{extarrows}
\usepackage{epsfig}
\usepackage{epsfig}
\usepackage{epstopdf}
\usepackage[noend]{algpseudocode}
\usepackage{algorithmicx,algorithm}

\usepackage{cite}
\usepackage{bm}
\usepackage{multicol}       
\usepackage{multirow}       
\usepackage{array}          
\usepackage{colortbl}
\usepackage{makecell}
\definecolor{crimson}{RGB}{192,0,0}         
\definecolor{navy}{RGB}{47,85,151}         
\usepackage{bbding}
\usepackage{graphicx}
\usepackage{booktabs}
\usepackage{algorithm}
\usepackage{algpseudocode}
\usepackage{diagbox}
\usepackage{graphicx}
\usepackage{setspace}
\usepackage{cleveref}

\usepackage{url}
\usepackage{mathtools}
\def\BibTeX{{\rm B\kern-.05em{\sc i\kern-.025em b}\kern-.08em
 T\kern-.1667em\lower.7ex\hbox{E}\kern-.125emX}}

\usepackage{enumerate}

\newcommand{\tabitem}{~~\llap{\textbullet}~~}

\usepackage{multicol}
\usepackage[noend]{algpseudocode}
\usepackage{pgfplots}
\makeatletter
\def\BState{\State\hskip-\ALG@thistlm}
\makeatother
\usepackage{multirow}
\definecolor{Blues}{RGB}{0,0,0}
\definecolor{mygray}{gray}{.8}
\definecolor{mygray2}{gray}{.7}
\definecolor{mygray3}{gray}{.6}
\algdef{S}[FOR]{ForEach}[1]{\algorithmicforeach\ #1\ \algorithmicdo}

\definecolor{mygreen}{RGB}{143, 188, 187}
\definecolor{green1}{RGB}{147, 196, 125}
\definecolor{red1}{RGB}{213,88,88}

\theoremstyle{plain}

\theoremstyle{plain}

\usepackage{amsmath}

\IEEEoverridecommandlockouts

\begin{document}
\title{Multi-Agent Reinforcement Learning in Wireless Distributed Networks for 6G}
\author{{Jiayi~Zhang,~\IEEEmembership{Senior Member,~IEEE}, Ziheng~Liu, Yiyang~Zhu, Enyu~Shi, Bokai~Xu, Chau~Yuen,~\IEEEmembership{Fellow,~IEEE}, Dusit~Niyato,~\IEEEmembership{Fellow,~IEEE}, M\'{e}rouane Debbah,~\IEEEmembership{Fellow,~IEEE}, Shi~Jin,~\IEEEmembership{Fellow,~IEEE}, Bo~Ai,~\IEEEmembership{Fellow,~IEEE}, and Xuemin~(Sherman)~Shen,~\IEEEmembership{Fellow,~IEEE}}
\thanks{J. Zhang, Z. Liu, Y. Zhu, E. Shi, B. Xu, and B. Ai are with the School of Electronic and Information Engineering and the Frontiers Science Center for Smart High-Speed Railway System, Beijing jiaotong University, Beijing 100044, China (e-mail: {jiayizhang, zihengliu, yiyangzhu, enyushi, 20251197, boai}@bjtu.edu.cn).}
\thanks{C. Yuen is with the School of Electrical and Electronics Engineering, Nanyang Technological University, Singapore 639798 (e-mail: chau.yuen@ntu.edu.sg).}
\thanks{D. Niyato is with the College of Computing and Data Science, Nanyang Technological University, Singapore 639798 (e-mail: dniyato@ntu.edu.sg).}
\thanks{M. Debbah is with KU 6G Research Center, Department of Computer and Information Engineering, Khalifa University, Abu Dhabi 127788, UAE (email: merouane.debbah@ku.ac.ae) and also with CentraleSupelec, University Paris-Saclay, 91192 Gif-sur-Yvette, France.}
\thanks{S. Jin is with the National Mobile Communications Research Laboratory, Southeast University, Nanjing, 210096, P. R. China (email: jinshi@seu.edu.cn).}
\thanks{X. Shen is with the Department of Electrical and Computer Engineering, University of Waterloo, Waterloo, ON N2L 3G1, Canada (email: sshen@uwaterloo.ca).}}
\maketitle

\begin{abstract}
The introduction of intelligent interconnectivity between the physical and human worlds has attracted great attention for future sixth-generation (6G) networks, emphasizing massive capacity, ultra-low latency, and unparalleled reliability. Wireless distributed networks and multi-agent reinforcement learning
(MARL), both of which have evolved from centralized paradigms, are two promising solutions for the great attention. Given their distinct capabilities, such as decentralization and collaborative mechanisms, integrating these two paradigms holds great promise for unleashing the full power of 6G, attracting significant research and development attention. This paper provides a comprehensive study on MARL-assisted wireless distributed networks for 6G.
In particular, we introduce the basic mathematical background and evolution of wireless distributed networks and MARL, as well as demonstrate their interrelationships. Subsequently, we analyze different structures of wireless distributed networks from the perspectives of homogeneous and heterogeneous. Furthermore, we introduce the basic concepts of MARL and discuss two typical categories, including model-based and model-free.
We then present critical challenges faced by MARL-assisted wireless distributed networks, providing important guidance and insights for actual implementation. We also explore an interplay between MARL-assisted wireless distributed networks and emerging techniques, such as information bottleneck and mirror learning, delivering in-depth analyses and application scenarios. Finally, we outline several compelling research directions for future MARL-assisted wireless distributed networks.
\end{abstract}
\IEEEpeerreviewmaketitle
\section{Introduction}
\subsection{Motivation}
The fifth-generation (5G) networks have made significant strides since their global launch, delivering enormous potential in multiple application scenarios, including enhanced mobile broadband (eMBB), ultra-reliable low-latency communications (URLLC), and massive machine-type communications (mMTC) \cite{[7]}. These achievements have driven the transformation of applications from autonomous networks to immersive experiences \cite{[5],[6]}. Since the sixth-generation (6G) networks providing wide access to wireless connectivity, enhanced coverage, and sustainable communications, are expected to serve beyond 2030 \cite{[13],[14]}, customized key performance indicators (KPIs) for IMT-2030 (6G) have been outlined, comprising nine enhanced capabilities and six new capabilities. Compared to existing 5G networks, 6G networks are anticipated to exhibit dramatically enhanced communication capabilities, such as a 100-fold increase in peak data rate (reaching the Tb/s level), a tenfold reduction in latency, and an end-to-end reliability requirement of 99.99999\% \cite{[8],[9],[10],[11],[12],[715]}. In particular, these new requirements have driven the emergence of emerging application scenarios for 6G networks, including mobile broadband reliable low-latency communications (MBRLLC), i.e., eMBB+ and URLLC+, and massive URLLC (mURLLC), i.e., mMTC+ and URLLC+, which are extensions of application scenarios defined in traditional 5G networks \cite{[9],[13]}.

In response to meet these highly anticipated demands of 6G networks, several promising techniques are gaining significant attention, such as cell-free (CF) massive multiple-input multiple-output (MIMO) \cite{[226],[610],[611],[612],[16],[17]}, reconfigurable intelligent surface (RIS) or intelligent reflecting surface (IRS) \cite{[18],[19],[20],[632],[615]}, mobile edge computing (MEC) \cite{[21],[22]}, Terahertz (THz) communications \cite{[218],[25]}, and  UAV-assisted communications \cite{[207],[640],[645]}. All of these techniques demonstrate enormous potential and can be seamlessly integrated. However, with the proliferation of user equipments (UEs) and the increasing diversification of service demands, particularly driven by the advancement of 6G, traditional centralized paradigms are encountering significant challenges such as bandwidth bottlenecks, latency limitations, inefficient resource utilization, and privacy concerns. These issues highlight that relying solely on previous techniques is far from enough, and promoting the transformation of centralized paradigms is urgent. In this context, wireless distributed networks are emerging as a critical alternative due to their decentralized nature \cite{[29],[27],[31]}. They can offer significant advantages such as enhanced flexibility, lower latency, and improved fault tolerance, positioning them as essential techniques to meet the constantly changing demands of 6G \cite{[8],[9],[10],[11],[12]}.

As an emerging evolution of centralized paradigms, wireless distributed networks are a response to the limitations of traditional centralized computing models \cite{[28]}, no longer built solely around a central server that controls computing, communication, and data storage. Specifically, wireless distributed networks decompose traditional centralized computing models into multiple autonomous components, with each component responsible for a specific subset of tasks, while collaboratively coordinating to accomplish overall objectives \cite{[29],[28]}. This endows wireless distributed networks with unprecedented connectivity, intelligence, and autonomy, ensuring reliable and efficient communication across multiple interconnected components. Moreover, building on this evolution, wireless distributed networks can be further classified into homogeneous and heterogeneous types to meet the diverse demands of emerging applications \cite{[43]}. On the other hand, homogeneous networks are composed of components with consistent characteristics, most of which share the same hardware, software, and communication modes \cite{[37],[38]}. This consistency simplifies wireless distributed networks, making them suitable for application scenarios where components are evenly distributed or tasks are easy to handle. On the other hand, in contrast to homogeneous paradigms, heterogeneous networks typically consist of components with varying capabilities and functions \cite{[39]}, making them a better choice for complex scenarios that require different components to perform specific tasks. This indicates that heterogeneous networks can effectively meet the complex and constantly changing demands of actual application scenarios of 6G through flexible task allocation and dynamic resource management, ensuring greater network resilience while optimizing performance.

However, while wireless distributed networks reap the benefits of a decentralized nature, they also encounter substantial challenges in achieving effective coordination and communication as the network scale expands \cite{[43]}. In this context, multi-agent reinforcement learning (MARL) stands out from numerous artificial intelligence (AI) techniques due to its powerful collaborative mechanism, enabling autonomous components to seamlessly coordinate and adapt, thereby improving efficiency and resilience in complex dynamic environments \cite{[1],[44],[542],[543],[561]}. Meanwhile, the inherently decentralized nature of MARL aligns perfectly with wireless distributed networks, positioning it as a key enabler for promoting the development of 6G \cite{[2],[559],[560]}. Note that MARL evolved from centralized reinforcement learning (RL) \cite{[45],[46],[47]}, similar to the transition from wireless centralized networks to distributed networks, both referring to the transformation from a single control entity to multiple autonomous decision-makers \cite{[561]}. Thus, in this tutorial, we focus on MARL-assisted wireless distributed networks, exploring how this collaborative effect can effectively address various challenges of 6G \cite{[48],[609]}.

Naturally, two primary strategies are commonly employed for enabling effective decision-making, such as model-based MARL \cite{[801],[802],[808]} and model-free MARL \cite{[819],[821],[827]}. In model-based MARL, agents typically assume knowledge of the state transition function, which describes how the network transitions from one state to another and how rewards are generated \cite{[804],[805],[812]}. However, this assumption is not always feasible, especially in complex dynamic environments that are challenging to model or mathematically characterize accurately. This motivates the emergence of model-free MARL, in which agents can learn optimal policies without any prior knowledge about the network \cite{[820],[823],[826]}. Therefore, the selection of strategies needs to be comprehensively considered in conjunction with actual application scenarios.
Moreover, the basic idea of MARL is to enable multiple agents to make decisions autonomously in a shared environment. Three major implementation approaches to this idea have been investigated in existing works, such as centralized training with centralized execution (CTCE) \cite{[901],[902],[903]}, centralized training with decentralized execution (CTDE) \cite{[904],[905],[906]}, and decentralized training with decentralized execution (DTDE) \cite{[908],[909],[910],[911],[912],[913]}.
The first implementation approach \cite{[901],[902],[903]} belongs to a centralized mechanism, which mainly relies on a central server for unified control and optimization, resulting in an inability to balance performance and overhead. This prompts mainstream works to focus on the latter two approaches that belong to decentralized mechanisms. Specifically, the second implementation approach \cite{[904]} involves a centralized critic network that optimizes agent policies uniformly, while allowing them to make independent decisions during execution. This approach belongs to the most typical mechanism, which can achieve a balance between centralized learning efficiency and decentralized decision-making flexibility. On the other hand, excessive reliance on central servers may become bottlenecks, leading to high latency, network congestion, and scalability issues, especially in 6G networks. This drives the emergence of the third decentralized implementation approach \cite{[910]}, where each agent completes decision execution and policy optimization locally, highly matching the decentralized nature of 6G networks and providing potential for deployment in actual application scenarios.

\begin{table*}
\small
 \centering
 \caption{{Comparisons between relevant survey papers on MARL and our paper.}}
 \begin{tabular}{|c|m{0.08cm}|m{7.5cm}|m{7.5cm}|}
  \hline
  \multicolumn{2}{|c|}{} & \bfseries Key contributions  & \bfseries Main limitations \\
  \hline
\cite{[1]} & \raisebox{-7.1\normalbaselineskip}[0pt][0pt]{\!\rotatebox{90}{{{Comprehensive survey}}}}  & {{\quad}} \newline \vspace{ -0.6cm}
  \newline \tabitem \,Present basic mathematical background and preliminaries of RL, including single-agent and multi-agent
  \newline \tabitem Introduce the theoretical foundations and structural classifications for RL
  \newline \tabitem Summarize the potential challenges faced by MARL and introduce empowering collaborative techniques & { {\quad}} \newline \vspace{ -0.6cm}
  \newline \tabitem \!Focus mainly on wireless networks, lacking a definition of centralized and distributed paradigms
  \newline \tabitem Review the theoretical techniques of the improved MARL simply, without further exploring the combination of advanced techniques to address unique issues associated with actual implementation of MARL \\ \cline{1-1} \cline{3-4}
 \cite{[2]} & & {{\quad}} \newline \vspace{ -0.6cm}
 \newline \tabitem \!\!\!\!Present the mathematical background and theoretical foundations for single-agent and multi-agent
 \newline \tabitem \!\!\!Highlight challenges faced in complex multi-agent environments
 \newline \tabitem \!\!\!Explore and compare some implementations of MARL in solving emerging issues & {{\quad}} \newline \vspace{ -0.6cm}
 \newline \tabitem Omit the review and comparison for the improved MARL designs (so-called collaborative mechanisms in \cite{[2]})
 \newline \tabitem \!\!More emerging or advanced techniques should be investigated to overcome the limitations faced by traditional MARL, such as information bottleneck and mirror learning \\ \hline \hline

 \multicolumn{2}{|c|}{} &  \multicolumn{2}{m{15cm}|}{\bfseries Key contributions} \\
  \hline
 \cite{[561]}  & \raisebox{-5.1\normalbaselineskip}[0pt][0pt]{\!\rotatebox{90}{{{Short brief}}}}  & \multicolumn{2}{m{15cm}|}{{\quad} \newline \vspace{ -0.6cm}
 \newline \tabitem Review networked MARL for learning communication and collaboration in large-scale application scenarios
 \newline \tabitem Introduce some key future directions on networked MARL, such as development and theoretical understanding} \\ \cline{1-1} \cline{3-4}
 \cite{[559]}  &  &  \multicolumn{2}{m{15cm}|} {{\quad} \newline \vspace{ -0.6cm}
 \newline \tabitem Highlight the importance of collaboration in solving high-dimensional continuous control problems
 \newline \tabitem Explain how to collaborate effectively from five perspectives: \emph{whom}, \emph{when}, \emph{what}, \emph{how}, and \emph{where} to aggregate}\\ \cline{1-1} \cline{3-4}
 \cite{[560]}  &  & \multicolumn{2}{m{15cm}|} {{\quad} \newline \vspace{ -0.6cm}
 \newline \tabitem Address the technical problem ``\emph{how to achieve effective collaboration among agents?}", with the difference being the introduction of GNNs to enhance collaboration, focusing on selectively aggregating received messages} \\ \cline{1-1} \cline{3-4}
 \cite{[609]}  &  &  \multicolumn{2}{m{15cm}|} {{\quad} \newline \vspace{ -0.6cm}
 \newline \tabitem Shift the research direction of MARL from far-field to near-field and analyze their relationships and characteristics
 \newline \tabitem Introduce the application of MARL in the near-field region from power control and antenna selection}\\ \hline \hline
\raisebox{-8\normalbaselineskip}[0pt][0pt]{\rotatebox{90}{{{Our paper}}}} & \raisebox{-10.3\normalbaselineskip}[0pt][0pt]{\!\rotatebox{90}{{{Comprehensive survey}}}}& \bfseries Overlapping Contributions & \bfseries Distinct Contributions\\ \cline{3-4}
& &{\quad} \newline \vspace{ -0.6cm}
 \newline \tabitem Summarize the mathematical background and evolution of MARL: from single agent to multi-agent
 \newline \tabitem Review the basic concepts and structural classifications of MARL from the perspective of model-based MARL and model-free MARL
 \newline \tabitem \!\!Discuss collaborative techniques for MARL-assisted wireless distributed networks to guide collaboration
 \newline \tabitem \!Outline a series of application scenarios and potential future directions for MARL
& {\quad} \newline \vspace{ -0.6cm}
 \newline \tabitem \!Comprehensively summarize the preliminaries and basic components for wireless distributed networks and MARL
 \newline \tabitem \!\!Introduce different structures of wireless distributed networks from homogeneous and heterogeneous
 \newline \tabitem Demonstrate the relationship between wireless distributed networks and MARL
 \newline \tabitem \!\!\!Review emerging techniques to enhance MARL and facilitate implementation in wireless disributed networks
 \newline \tabitem \!Provide many tutorials for enhanced-MARL modeling
 \newline \tabitem \!Motivate future directions on MARL-assisted wireless disributed networks from three perspectives  \\
 \hline
 \end{tabular}
 \label{Comparison}
\end{table*}
\begin{figure*}[t]
\centering
    \includegraphics[scale=0.34]{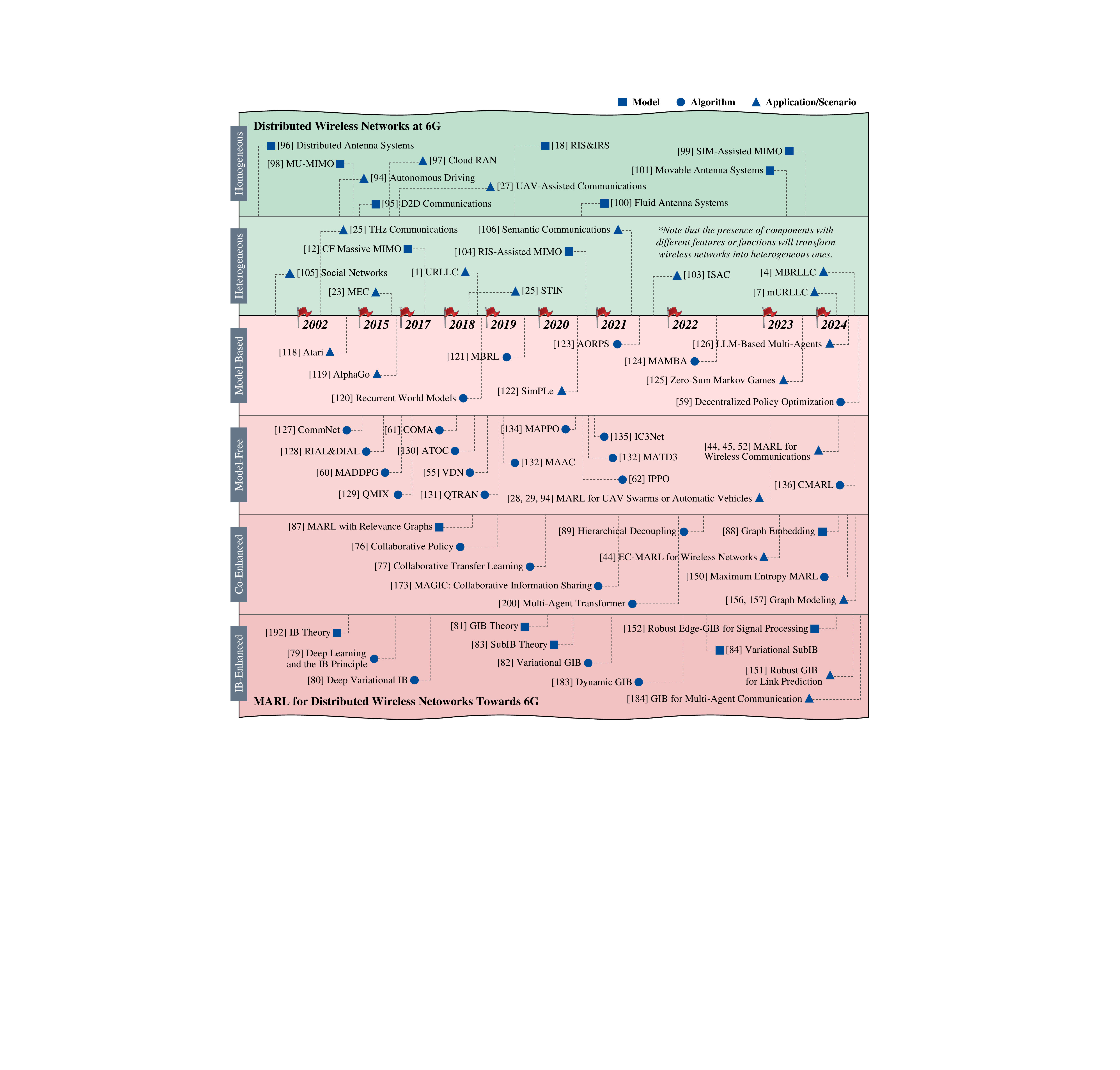}
    \caption{The development roadmap of wireless distributed networks and MARL from 2002 to Dec 2024. From the perspective of wireless distributed network development, networks have evolved from a single homogeneous type to a complex heterogeneous type of interconnected everything. From the perspective of MARL development, MARL has gradually shifted from a single network architecture to integration with various emerging techniques. Note that ``IB" is defined as an ``information bottleneck".
    \label{fig1}}
\end{figure*}
\begin{figure*}[t]
\centering
    \includegraphics[scale=0.29]{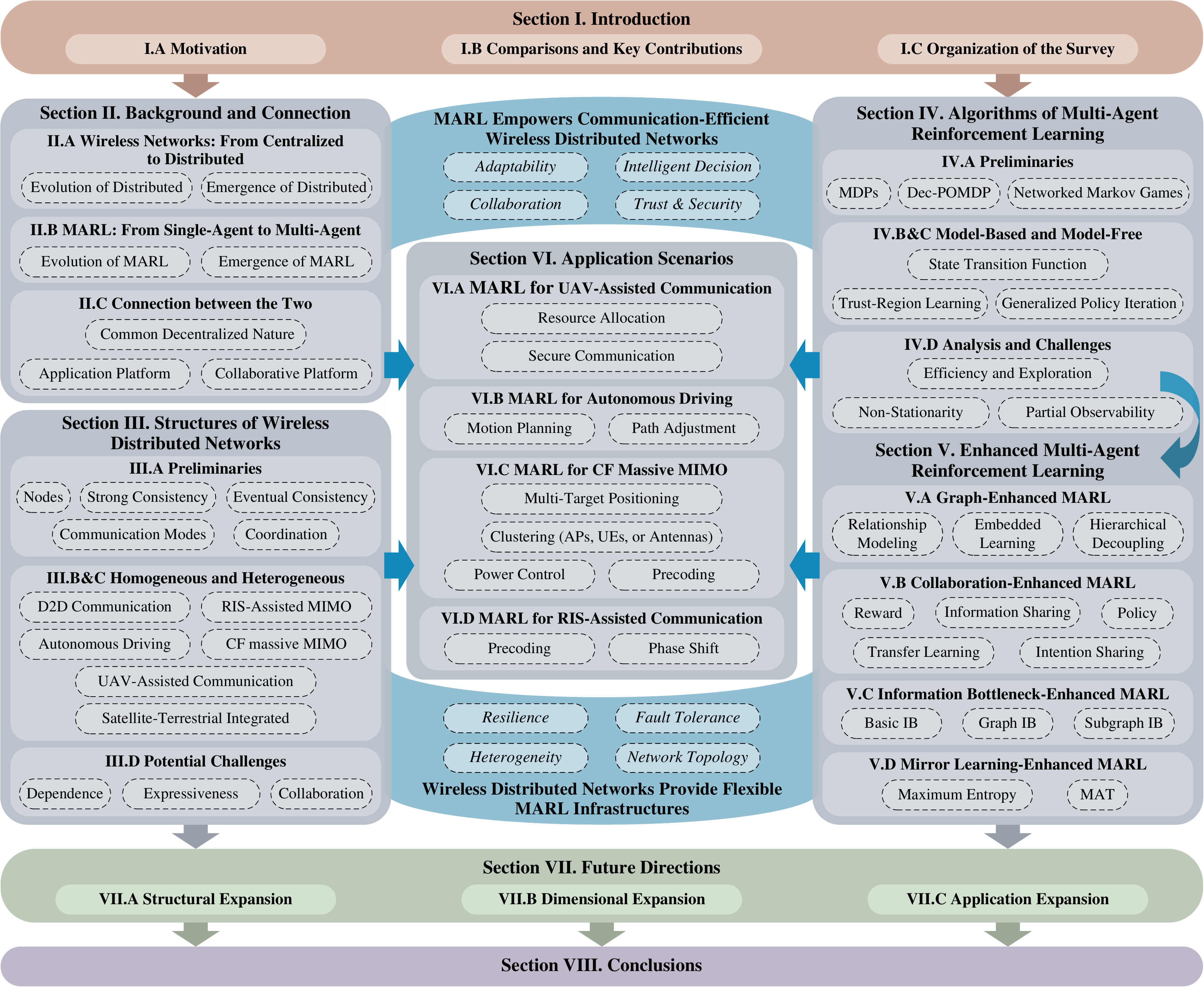}
    \caption{The outline of this tutorial, where we introduce the provisioning of different MARL algorithms on wireless distributed networks for 6G, and highlight some essential implementation challenges about MARL-assisted wireless distributed networks and emerging techniques that can be adopted.
    \label{fig1}}
\end{figure*}
Despite the appealing advantages, MARL-assisted wireless distributed networks still face significant challenges, such as poor communication efficiency and sparse information expressiveness \cite{[914],[904],[910]}. These call for further research into emerging techniques to effectively unleash the full power of MARL-assisted wireless distributed networks. One widely integrated emerging technique is to design effective collaborative protocols \cite{[546],[559],[560],[566],[545]}, mainly focusing on information sharing to guide agents to collaborate effectively from multiple perspectives, including \emph{whom}, \emph{when}, \emph{what}, and \emph{how} \cite{[559],[560]}. In addition to collaborative techniques, recent works also explore advanced techniques that enhance robustness and scalability, such as information bottleneck (IB) \cite{[505],[506],[509],[510],[520],[521]}, and mirror learning \cite{[406],[411]}, as well as graph-assisted techniques \cite{[572],[576],[583]}. These techniques collectively hold the potential to enable MARL-assisted wireless distributed networks to more effectively address the challenges posed by complex dynamic environments, thereby laying the foundation for meeting the diverse demands of 6G.
\subsection{Comparisons and Key Contributions}
Recently, MARL has attracted considerable research in 6G networks, and some review papers focus on this topic. In this tutorial, we provide an overview of research activities related to wireless distributed networks and MARL, as illustrated in Fig. 1. Meanwhile, Table \uppercase\expandafter{\romannumeral1} presents a detailed comparison between our paper and the existing survey papers.

Specifically, the authors in \cite{[1]} focused on applications of RL algorithms in wireless networks. Firstly, the mathematical background and preliminaries of RL were presented from the perspectives of single-agent and multi-agent. Then, the theoretical foundations for RL frameworks were comprehensively introduced from model-free and model-based, where the former focuses on learning (e.g., policy-based and value-based), while the latter focuses on planning (e.g., background and decision-time). Moreover, the potential challenges of MARL are summarized, indicating the necessity of adopting collaborative MARL to address them. To promote the actual implementation of MARL, some enabling techniques for 6G networks were reviewed such as CF massive MIMO and RIS\&IRS. However, the authors mainly focused on wireless networks, lacking a definition of centralized and distributed paradigms. Thus, a survey/tutorial focusing on a more general wireless distributed networks for 6G is needed. Additionally, the theoretical techniques of the improved MARL were reviewed in \cite{[1]} briefly, focusing only on collaborative MARL. Some critical issues for the implementation of MARL also need to be addressed in conjunction with emerging techniques.

In particular, the authors in \cite{[2]} expanded the application of MARL to the future Internet-of-Thing (IoT) involving several emerging techniques, presenting the mathematical background and theoretical foundations for single-agent and multi-agent, including Markov decision process (MDP) and partially observable Markov decision process (POMDP) for the former, and decentralized POMDP (Dec-POMDP) for the latter. It emphasizes challenges faced in multi-agent environments, such as partial observability, non-stationarity, and scalability, and details different MARL algorithms to overcome these challenges, which typically belong to the classical CTDE mechanism.
Additionally, the authors explored and compared some implementations of MARL in solving emerging issues in the future Internet, such as dynamic network access and motion planning. However, this survey \cite{[2]} focused on the application aspect and lacked a comparative analysis of the improved MARL designs, which calls for further research on emerging techniques to address various limitations.

In addition to \cite{[1]} and \cite{[2]}, there are several short overview papers \cite{[561],[559],[560],[609]} giving details of MARL from different perspectives.
Specifically, the authors in \cite{[561]} reviewed networked MARL for learning communication and collaboration in large-scale control networks and communication networks. The optimization within decentralized settings was particularly emphasized, including decentralized rewards and communication protocols. Then, some important future directions on networked MARL were introduced, such as robustness, resilience, development, and theoretical understanding. The authors in \cite{[559]} highlighted the importance of collaboration in solving high-dimensional continuous control problems with partially observable states, and explained how to effectively collaborate, including \emph{whom}, \emph{when}, \emph{what}, \emph{how}, and \emph{where} to aggregate received messages. Moreover, the technical problem ``\emph{how to achieve effective collaboration among agents?}" was further discussed in \cite{[560]}, with the difference being the introduction of graph neural networks (GNNs) that seamlessly switch with the communication topology to enhance collaboration, focusing on selectively aggregating received messages. Alternatively, the authors in \cite{[609]} shifted the research direction of MARL from far-field to near-field and analyzed their relationships and characteristics. Furthermore, the applications of MARL in the near-field region were explored with respect to power control and antenna selection. All these studies offer valuable insights into MARL and wireless distributed networks. However, they primarily focus on the evolution, collaboration, and application design of MARL, or present reviews from specific perspectives, lacking a comprehensive introduction of the technical aspects and enhancement tutorials of MARL-assisted wireless distributed networks.

To this end, in this paper, we comprehensively review MARL-assisted wireless distributed networks towards 6G. The major contributions of our tutorial are summarized as follows:
\begin{itemize}
\item We introduce the background of wireless distributed networks and MARL respectively. More importantly, we demonstrate the relationship between wireless distributed networks and MARL, such as their common decentralized nature, offering important insights for the implementation of MARL-assisted wireless distributed networks.

\item We present the preliminaries and basic components for wireless distributed networks and introduce different structures from the perspective of homogeneous and heterogeneous. Moreover, we also discuss the basic concepts of MARL and introduce two typical categories, i.e., model-based and model-free, which systematically summarize the key features of MARL and provide guidance for integrating into of wireless distributed networks.

\item We review the emerging techniques to enhance MARL from the perspective of graph-enhanced, collaboration-enhanced, IB-enhanced, and mirror learning-enhanced. Notably, the IB-enhanced MARL designs are presented and motivated to gain insights into the actual implementation of MARL-assisted wireless distributed networks.

\item Finally, we discuss a series of MARL-assisted application scenarios for wireless distributed networks, such as unmanned aerial vehicle (UAV)-assisted communications, autonomous driving, CF massive MIMO, and RIS-assisted MIMO, and provide key future research directions on MARL-assisted wireless distributed networks, which are divided into three categories: structural, dimensional, and application extension.
\end{itemize}
\subsection{Organization of the Tutorial}
The organization of this tutorial is illustrated in Fig. 2. Section \uppercase\expandafter{\romannumeral2} introduces the background of wireless distributed networks and MARL, and analyzes the connection between the two. Then, structures of wireless distributed networks are reviewed in Section \uppercase\expandafter{\romannumeral3}. The preliminaries and basic components for wireless distributed networks are first presented in Section \uppercase\expandafter{\romannumeral3}-A, and different structures are introduced from the perspective of homogeneous in Section \uppercase\expandafter{\romannumeral3}-B and heterogeneous in Section \uppercase\expandafter{\romannumeral3}-C, respectively. More significantly, the potential challenges of wireless distributed networks are presented in Section \uppercase\expandafter{\romannumeral3}-D. Section \uppercase\expandafter{\romannumeral4} focuses on innovative MARL algorithms. In Section \uppercase\expandafter{\romannumeral4}-A, we review the basic concepts of MARL. Section \uppercase\expandafter{\romannumeral4}-B and Section \uppercase\expandafter{\romannumeral4}-C discuss two categories of MARL, i.e., model-based and model-free. Besides, the theoretical analysis of MARL is presented in Section \uppercase\expandafter{\romannumeral4}-D. Subsequently, the integration of enhanced-MARL with various emerging techniques to further unleash the potential of wireless distributed networks is presented in Section \uppercase\expandafter{\romannumeral5}, including graph-enhanced, collaboration-enhanced, IB-enhanced, and mirror learning-enhanced, while Section \uppercase\expandafter{\romannumeral6} discusses four typical application scenarios, such as autonomous driving and wireless communications. Finally, several future directions on MARL-assisted wireless distributed networks are discussed in Section \uppercase\expandafter{\romannumeral7}.
\begin{figure}[t]
\centering
    \includegraphics[scale=0.35]{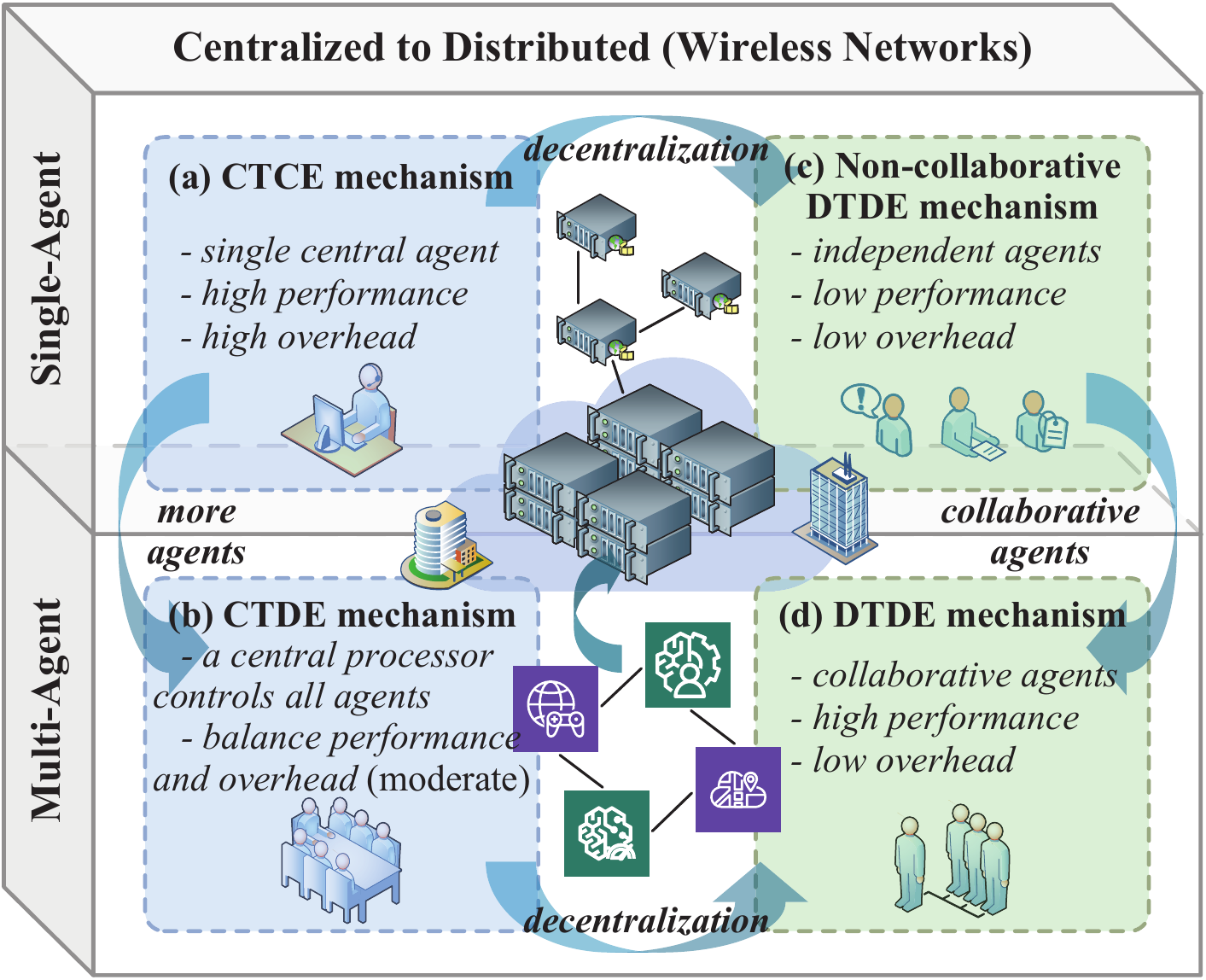}
    \caption{The connection between wireless distributed networks and MARL reflects the four evolutionary mechanisms of MARL, including high-overhead CTCE, classical CTDE, non-collaborative DTDE, and collaborative DTDE.
    \label{fig1}}
\end{figure}
\section{Background and Connection}
The rise of wireless distributed networks \cite{[29],[27],[31],[28],[43]} and MARL \cite{[801],[802],[808],[819],[821],[827]} reflects the growing demand for intelligent, scalable, and autonomous frameworks in various applications. Specifically, both fields have developed independently, with wireless distributed networks being utilized to handle the increasingly complex and scaled problems, while MARL is utilized to achieve communication and coordination among multiple autonomous agents to collectively accomplish tasks. Despite having different origins, the decentralized nature of MARL perfectly aligns with wireless distributed networks, driving these two fields to become increasingly interconnected.

In this section, we first explore the evolution of wireless distributed networks, emphasizing their role in addressing challenges such as scalability, fault tolerance, and robustness. Then, we turn to MARL and track its development from single-agent to multi-agent, as well as the unique challenges faced by these techniques. Moreover, we also analyze their connection and how to effectively integrate them to meet the strict demands of future 6G networks.
\subsection{Wireless Networks: Centralized to Distributed}
\subsubsection{Evolution of Wireless Distributed Networks}
The rapid development of communication techniques has paved the way for the future development of 6G networks, which aim to support diverse emerging application scenarios such as intelligent transportation, immersive virtual reality, and autonomous networks \cite{[301],[302],[440]}. The core of these advancements lies in wireless distributed networks rather than traditional centralized paradigms \cite{[300]}, characterized by unprecedented connectivity, intelligence, and autonomy, ensuring reliable and efficient communication across multiple components.

Specifically, the concept of wireless distributed networks emerged as a response to the limitations of traditional centralized computing models \cite{[29],[27],[31],[28],[43]}. In the early stages of computing, most networks were typically built around a central server that controlled computing, communication, and data storage. However, with the emergence of various applications and the increase in network scale, centralized paradigms face significant challenges related to flexibility and scalability.
\subsubsection{Emergence of Wireless Distributed Networks}
Clearly, wireless distributed networks can effectively address the above challenges by decoupling the original centralized computing model or server into multiple independent components, where each component is only responsible for a portion and collaborates with each other to complete tasks \cite{[29],[27],[31],[28],[43]}.
Moreover, unlike centralized paradigms, wireless distributed networks focus on addressing interaction problems involving a set of interdependent components, where the network environment is dynamically influenced by the joint actions or interactions of all components. In other words, the performance of a given component no longer depends solely on its own observed states or actions, but on the joint information and interactions of all other components within the network. This requires each component to consider the behavior and policies of other components to optimize overall performance or achieve long-term goals. By contrast, the main advantages of wireless distributed networks \cite{[29],[27],[31],[28],[43]} are as follows:
\begin{itemize}
\item \emph{Flexibility:} Distributed paradigms deploy multiple independent components to work together, allowing them to flexibly add, remove, or adjust components and functions without changing the overall architecture, making them more flexible than centralized paradigms.

\item \emph{Scalability:} Distributed paradigms can cope with scale growth by horizontally scaling (e.g. adding more independent components), allowing their network capacity and processing power to flexibly expand according to demand without relying on a single entity, thus having better scalability while balancing performance.

\item \emph{Fault Tolerance:} Distributed paradigms ensure that the network can still operate normally when some components fail through component redundancy and data backup. This decentralized design significantly outperforms centralized paradigms in terms of fault tolerance.
\end{itemize}

These fundamental advantages play a crucial role in designing and optimizing wireless distributed networks, ensuring that they meet the strict demands of future 6G.
\subsection{Reinforcement Learning: Single-Agent to Multi-Agent}
\subsubsection{Evolution of MARL}
Correspondingly, the rapid development of communication techniques has also led to a demand for intelligent networks that can make decisions in dynamic and interconnected environments. RL networks have emerged, which train a single agent to autonomously optimize behavior based on its interaction with the environment, especially performing better in isolated settings. However, with an expansion of application scenarios, especially in heterogeneous environments, RL have gradually exposed the following limitations: poor scalability, poor reasoning ability, and difficulty in handling cooperation and competition among agents. Moreover, this also brings about a sharp increase in computational burden and network overhead, limiting their application in actual complex environments. Specifically, in emerging applications such as UAV swarms \cite{[640],[645]} and autonomous vehicles \cite{[201]}, each agent needs to coordinate with each other instead of making isolated decisions. This has led to the emergence of MARL \cite{[820]}, which can effectively respond to situations where agents cooperate or compete with each other in a shared environment. By contrast, MARL can better adapt to the increase in network scale compared to RL.
\begin{table*}[t!]
  \centering
  \fontsize{8.5}{12}\selectfont
  \caption{Characteristics for different wireless distributed networks, including homogeneous and heterogeneous.}
  \label{CE}
   \begin{tabular}{ !{\vrule width1.2pt}  m{2.075 cm}<{\centering}
   !{\vrule width1.2pt}  m{0.65 cm}<{\centering}
   !{\vrule width1.2pt}  m{4.55 cm}<{\centering}
   !{\vrule width1.2pt}  m{3.75 cm}<{\centering}
   !{\vrule width1.2pt}  m{4.7 cm}<{\centering}   !{\vrule width1.2pt} }

    \Xhline{1.2pt}
        \rowcolor{gray!30} \bf Frameworks  &  \bf Ref. &  \bf Descriptions &  \bf Composed Nodes    & \bf Communication Modes \cr
    \Xhline{1.2pt}
        \multirow{10}{*}{\bf \shortstack{Homogeneous\\ Distributed\\ Networks}}
                    & \cite{[201]}  & *Autonomous Driving & Autonomous Vehicles  & Synchronous, Multicast, and Broadcast \\
        \cline{2-5} & \cite{[202]}  & D2D Communications & Consistent Devices, e.g., UEs  & Asynchronous, Unicast, and Multicast \\
        \cline{2-5} & \cite{[207]}  & *UAV-Assisted Communications & UAV Swarms   & Synchronous, Unicast, and Multicast \\
        \cline{2-5} & \cite{[213]}  & *Distributed Antenna Systems & BSs or APs  & Synchronous and Broadcast \\
        \cline{2-5} & \cite{[210]}  & *Cloud RAN & Remote Units   & Synchronous and Multicast  \\
        \cline{2-5} & \cite{[211]}  & Multi-User (MU)-MIMO  & UEs (Single BS)  & Synchronous and Broadcast  \\
        \cline{2-5} & \cite{[18]}  & RIS \& IRS (MIMO) & RISs \& IRSs  & Synchronous and Broadcast \\
        \cline{2-5} & \cite{[212]}  & *SIM-Assisted MIMO & UEs (SIM-Assisted BS)  & Synchronous and Broadcast \\
        \cline{2-5} & \cite{[214],[215]}  & *Fluid Antenna Systems \& *Movable Antenna Systems & Antennas (belonging to BS)    & Synchronous and Broadcast \cr\Xhline{1.2pt}

        \multirow{10}{*}{\bf \shortstack{Heterogeneous\\ Distributed\\ Networks}}
                    & \cite{[7]}  & Ultra-Reliable Low-Latency Communications & Sensors, BSs, and UEs & Asynchronous and Multicast \\
        \cline{2-5} & \cite{[21]}  & Mobile Edge Computing & Servers, Gateways, and UEs  & Asynchronous and Multicast \\
        \cline{2-5} & \cite{[221]}  & Space-Terrestrial Integrated Networks  &  Satellite \& Terrestrial Nodes & Asynchronous and Multicast  \\
        \cline{2-5} & \cite{[216]}  & Integrated Sensing and Communications & Sensors, Radars, BSs, and UEs & Asynchronous and Multicast \\
        \cline{2-5} & \cite{[218]}  & THz Communications (CF)  & CPUs, BSs, and UEs   & Synchronous and Broadcast \\
        \cline{2-5} & \cite{[226]}  & CF Massive MIMO & CPUs, APs, and UEs   & Synchronous and Broadcast  \\
        \cline{2-5} & \cite{[227]}  & RIS-Assisted MIMO (CF) & CPUs, APs, UEs, and RISs   & Synchronous, Unicast, and Broadcast \\
        \cline{2-5} & \cite{[220]}  & Social Networks & Organizations and UEs  & Asynchronous, Unicast, and Multicast \\
        \cline{2-5} & \cite{[217]}  & Semantic Communications & Encoding/Decoding/Semantic   & Synchronous, Unicast, and Multicast  \cr\Xhline{1.2pt}
        \multicolumn{5}{l}{\footnotesize * denotes that nodes with different features or characteristics will transform homogeneous distributed networks into heterogeneous ones.}\\
    \end{tabular}
  \vspace{0cm}
\end{table*}
\subsubsection{Emergence of MARL}
Clearly, MARL belongs to an extended form of RL \cite{[801],[802],[808],[819],[821],[827]}, which allows agents to learn in a collaborative manner rather than independently. Moreover, unlike traditional RL, each agent in MARL not only relies on its own behavior but also needs to consider the behavior of other agents. This collaborative mechanism effectively helps agents adjust policies and manage resources. Correspondingly, the main advantages of MARL \cite{[801],[802],[808],[819],[821],[827]} are as follows:
\begin{itemize}
\item \emph{Collaboration:} RL typically deploys a single agent or multiple independent agents, which results in an inability to balance performance and overhead due to a lack of collaboration with other agents. In contrast, MARL allows multiple agents to jointly optimize global objectives by designing collaborative mechanisms, which can effectively improve collaboration efficiency and significantly surpass the single-agent limitations of RL networks.

\item \emph{Dynamism and Adaptability:} Agent in RL typically rely on local observed information for optimization, resulting in poor adaptability to environmental changes, especially when multiple agents participate simultaneously. In contrast, MARL can quickly adapt to dynamic environmental changes and the policies of other agents through interaction and information sharing \cite{[559],[560]}. This is beneficial for enhancing the global adaptability and real-time decision-making ability of all agents, ensuring that the network can operate efficiently in competitive or cooperative dynamic environments.
\end{itemize}

However, MARL also brings challenging problems that do not exist in single-agent RL, such as non-stationarity and partial observability, where the behavior of each agent changes over time, leading to dynamic and unpredictable environments. This prompts us to seek emerging techniques to improve existing MARL, with specific details in Section \uppercase\expandafter{\romannumeral5}.
\subsection{Connection between Wireless Distributed Networks and Multi-Agent Reinforcement Learning}
The connection between wireless distributed networks and MARL fundamentally stems from their common decentralization and collaborative mechanisms, as shown in Fig. 3. Together, they provide an effective framework for future 6G networks, enabling dispersed components or agents to autonomously learn how to cooperate or compete to optimize their collective behavior in actual application scenarios. Their connection can be summarized as follows:
\subsubsection{Wireless Distributed Networks as a Platform for MARL}
Wireless distributed networks provide the computing and communication infrastructure required to support large-scale MARL applications due to their decentralized nature. By decentralizing computing and storage across multiple independent components, an ideal environment has been created for multi-agent deployment. These components can act as individual agents in MARL, learning their optimal policies by interacting with other components.
\subsubsection{MARL as a Collaborative Framework for Wireless Distributed Networks}
MARL serves as a powerful collaborative framework in wireless distributed networks. By allowing all agents to share observed information rather than being completely independent, they can dynamically adjust policies and collaborate to optimize global goals, becoming an effective tool for addressing challenges in complex environments.

Furthermore, MARL-assisted wireless distributed networks can be divided into four progressive mechanisms to meet different application demands, including CTCE \cite{[901],[902],[903]}, classical CTDE \cite{[904],[905],[906]}, non-collaborative DTDE \cite{[908],[909],[911]}, and collaborative DTDE \cite{[910],[912],[913]}, gradually transitioning from centralized to decentralized, better balancing performance and overhead. Therefore, wireless distributed networks and MARL are two complementary and compatible decentralized techniques that are collaborating to reshape the future of autonomous and intelligent networks.
\section{Structures of Wireless Distributed Networks}
In this section, we delve into the basic components of wireless distributed networks, categorizing them into two typical types: homogeneous and heterogeneous, as shown in Table \uppercase\expandafter{\romannumeral2}. Each type presents unique network paradigms, which play a crucial role in determining their applicability. Additionally, the successful deployment of wireless distributed networks largely depends on identifying and addressing potential challenges, such as complex interactive relationships, as shown in Table \uppercase\expandafter{\romannumeral3}. Effectively overcoming these challenges is crucial for fully unleashing the power of wireless distributed networks.
\subsection{Preliminaries}
We first discuss and analyze several basic components and key characteristics of wireless distributed networks, which provides readers with a clearer understanding.
\subsubsection{Nodes}
Refers to the basic component of wireless distributed networks, belonging to independent computing units, including network hardware (e.g., central processing units (CPUs)), IoT devices (e.g., sensors, radars, and gateways), communication devices (e.g., base stations (BSs), access points (APs), UEs, and RISs) \cite{[31],[559],[560]}. Nodes are mainly responsible for performing computing tasks, storing data, providing services, or acting as communication intermediaries among nodes \cite{[559],[560]}. Note that each node can operate autonomously or collaborate, contributing to the network's collective functionality.
\subsubsection{Communication Modes}
Refers to how messages are transmitted and shared among nodes in wireless distributed networks, which determines the flow of transmission, synchronous behavior, and dependency relationships. Typical communication modes \cite{[560]} can be divided into:
\begin{itemize}
\item \emph{Synchronous:} The transmitter waits for the receiver's response before continuing, suitable for scenarios that require immediate feedback.
\item \emph{Asynchronous:} The transmitter does not wait for a response and continues to execute other tasks, suitable for high concurrency or long-term processing.
\item \emph{Unicast:} One-to-one communication, suitable for point-to-point request response.
\item \emph{Broadcast:} One to many communication, suitable for transmitting messages to all nodes.
\item \emph{Multicast:} One to some specific nodes, suitable for transmitting messages only to specific groups.
\end{itemize}

\begin{figure*}[t]
\centering
    \includegraphics[scale=0.4]{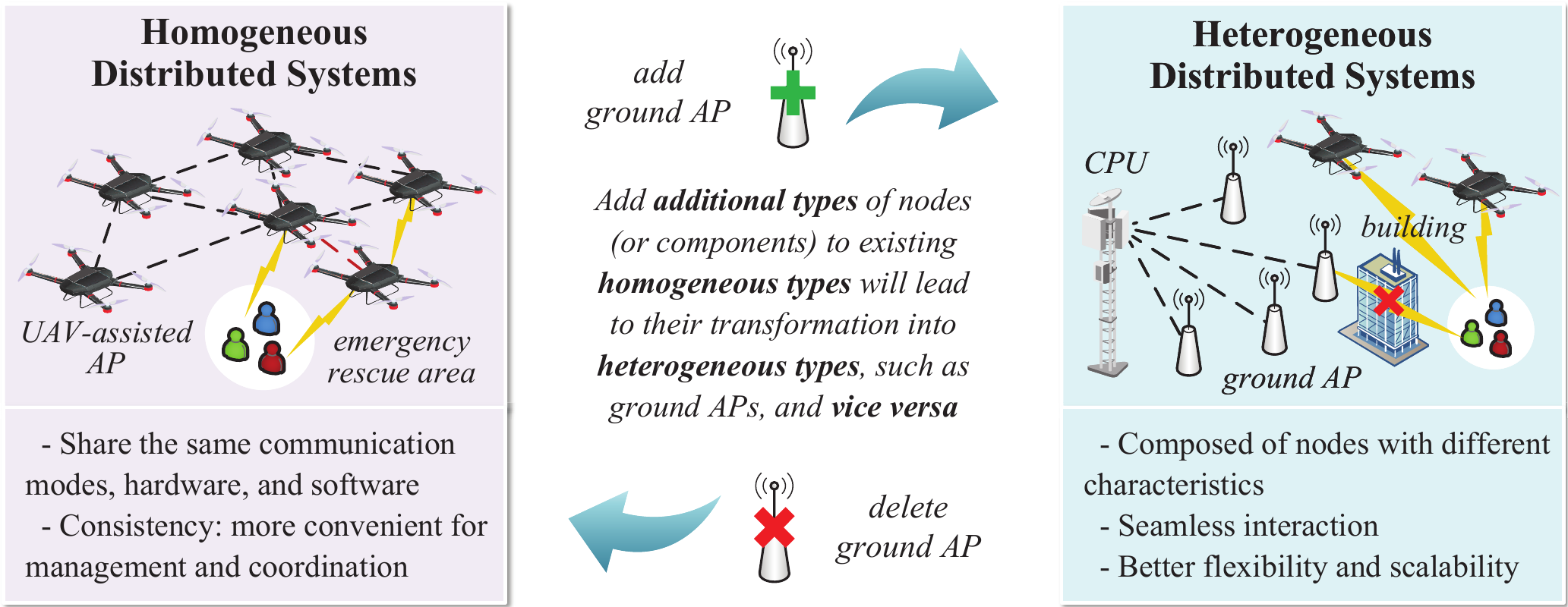}
    \caption{The network architecture and connection of different wireless distributed networks, including homogeneous types (e.g., sharing the same communication modes, hardware, and software) and heterogeneous types (e.g., composed of nodes with different characteristics).
    \label{fig1}}
\end{figure*}
Note that the choice of communication mode depends on specific network demands, such as performance, latency, and reliability, and the appropriate communication mode plays a crucial role in optimizing wireless distributed networks.
\subsubsection{Consistency and Coordination}
Consistency refers to the state in which data across multiple nodes in wireless distributed networks remains consistent, including strong consistency and eventual consistency \cite{[27]}, as follows:
\begin{itemize}
\item \emph{Strong Consistency:} The network requires all nodes to observe consistent data at any given time, which ensures accuracy but often results in performance loss.
\item \emph{Eventual Consistency:} The network allows data to be inconsistent at certain points in time but requires all replicas to converge to the same value, which is beneficial for improving the network's availability and performance.
\end{itemize}

\begin{figure}[t]
\centering
    \includegraphics[scale=0.55]{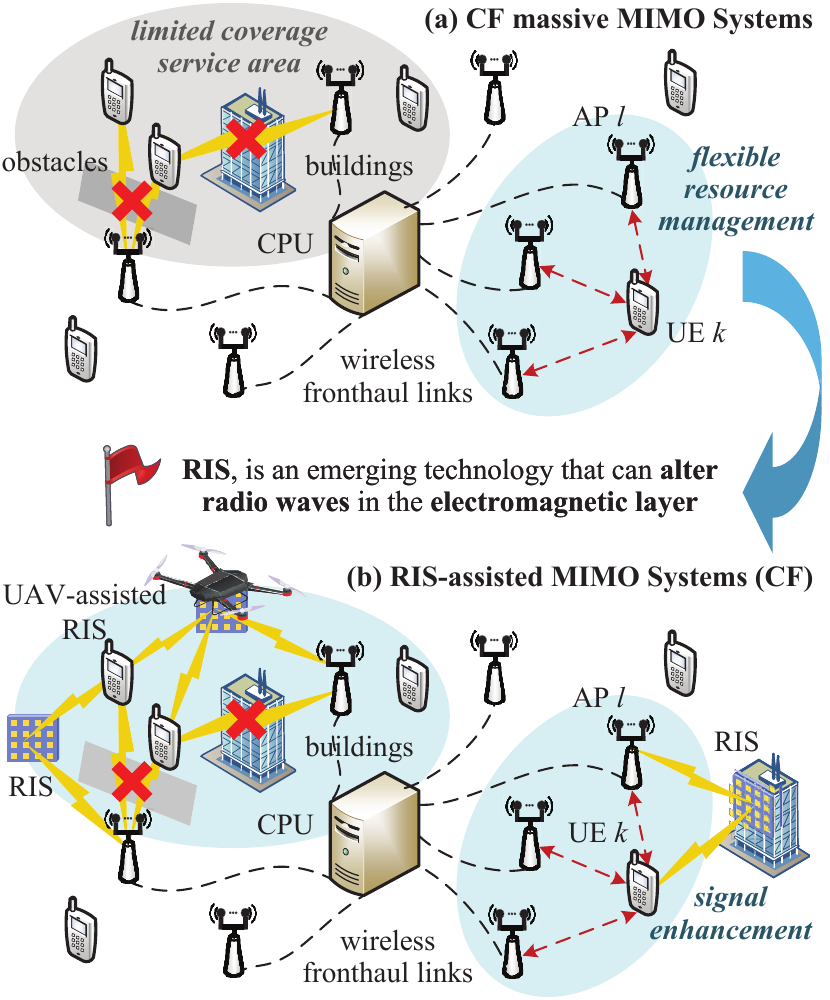}
    \caption{Two typical heterogeneous distributed networks for 6G, including CF massive MIMO networks and RIS-assisted MIMO networks.
    \label{fig1}}
\end{figure}
Coordination \cite{[559],[560]} refers to the collaboration among multiple nodes, helping wireless distributed networks address challenges related to message sharing (similar to \emph{communication modes}), state synchronization, and conflict avoidance.
\subsection{Homogeneous Distributed Networks}
Homogeneous distributed networks are characterized by consistent nodes that share the same hardware, software, and communication modes. This consistency simplifies wireless distributed networks, making them easier to manage, coordinate, and optimize. Meanwhile, standardized resource allocation, task scheduling, and data communication ensure consistent performance while reducing the risk of compatibility issues and synchronization overhead. This effectively enhances network reliability and communication efficiency (\emph{consistency}), allowing new nodes to integrate seamlessly.
Therefore, this indicates that homogeneous distributed networks are suitable for scenarios that require high reliability and efficiency, such as autonomous driving \cite{[201]}, device-to-device (D2D) communications \cite{[202]}, UAV-assisted communications \cite{[207]}, cloud radio access network (RAN) \cite{[210]}, and stacked intelligent metasurface (SIM)-assisted MIMO \cite{[212],[441]}, as shown in Table \uppercase\expandafter{\romannumeral2}. Note that integrating different types of nodes into these networks will prompt them to transform into heterogeneous distributed networks, such as roadside units (RSUs) in autonomous driving and ground-AP in UAV-assisted communications, as shown in Fig. 4. Typical homogeneous types are described as follows:
\subsubsection{UAV-Assisted Communication Networks}
They are composed of numerous UAVs with the same characteristics, as shown in Fig. 4 (a). With the continuous miniaturization of communication equipment and the increased endurance of UAVs \cite{[207]}, multiple UAVs can cooperate with each other to form UAV-assisted (i.e., aerial AP or BS) mobile communication networks \cite{[208]}, which is conducive to enhancing coverage and increasing capacity. Compared with ground APs with fixed geographical locations, aerial UAV-assisted APs can adaptively coordinate flight paths and resource allocation to better respond on demand \cite{[639]}, especially suitable for emergency response or disaster recovery scenarios \cite{[637]}.

Moreover, when UAV-assisted APs coexist with ground APs (or others) to enhance network coverage, this promotes the transformation of existing homogeneous distributed networks into heterogeneous ones, as shown in Fig. 4 (b).
\subsubsection{Autonomous Driving Networks}
They are composed of uniform-type vehicles, forming homogeneous networks that collaborate to ensure seamless communication and enhance coverage. Specifically, by coordinating movement and sharing information, all vehicles can be allowed to dynamically adjust network parameters, improve connectivity, and support critical functions like vehicle-to-vehicle (V2V) communications \cite{[201],[651]}. This collaboration also further reduces reliance on static infrastructure, ensuring the reliability of autonomous driving networks \cite{[652]}. Moreover, typical computing networks for autonomous driving can be divided into three categories, including ego-only, modular-based, and end-to-end \cite{[229],[230],[231]}, among which modular-based computing networks stand out due to their flexibility and scalability \cite{[231]}, and have been widely applied in most actual prototype designs.

Correspondingly, when vehicles coexist with RSUs (or others) to enhance network connectivity, this also promotes the transformation of existing networks into heterogeneous ones, similar to UAV-assisted communication networks.
\begin{table*}[tp]
  \centering
  \fontsize{8.5}{12}\selectfont
  \caption{Potential challenges and feasible techniques for wireless distributed networks.}
  \label{System_ULA}
    \begin{tabular}{
    !{\vrule width1.2pt} m{4.9 cm}<{\centering}
    !{\vrule width1.2pt} m{12.2 cm}<{\centering}
    !{\vrule width1.2pt} }

    \Xhline{1.2pt}
        \rowcolor{gray!30} \bf Potential Challenges  &  \bf Feasible techniques\cr
    \Xhline{1.2pt}
        Excessive Complexity & \makecell[l]{$\bullet$ Section \uppercase\expandafter{\romannumeral4} of ``{algorithms of multi-agent reinforcement learning}" \\$\bullet$ Section \uppercase\expandafter{\romannumeral5}-C of ``{information bottleneck-enhanced MARL}", i.e., SubIB} \cr\hline

        Complex Interactive Relationships
        & \makecell[l]{$\bullet$ Section \uppercase\expandafter{\romannumeral4}-C of ``{model-free MARL algorithms}" \\$\bullet$ Section \uppercase\expandafter{\romannumeral5}-A of ``{graph-enhanced MARL}", i.e., relationship modeling} \cr\hline

        Cooperation and Competition Trade-off
        & \makecell[l]{$\bullet$ Section \uppercase\expandafter{\romannumeral4} of ``{algorithms of multi-agent reinforcement learning}" \\$\bullet$ Section \uppercase\expandafter{\romannumeral5}-B of ``{collaboration-enhanced MARL}"} \cr\hline

        Complex Dynamism
        & \makecell[l]{$\bullet$ Section \uppercase\expandafter{\romannumeral4} of ``{algorithms of multi-agent reinforcement learning}"} \cr\hline

        Complex Dependence
        & \makecell[l]{$\bullet$ Section \uppercase\expandafter{\romannumeral4} of ``{algorithms of multi-agent reinforcement learning}" \\$\bullet$ Section \uppercase\expandafter{\romannumeral5}-A of ``{graph-enhanced MARL}", i.e., relationship modeling and hierarchical decoupling} \cr\hline

        Poor Stability
        & \makecell[l]{$\bullet$ Section \uppercase\expandafter{\romannumeral4} of ``{algorithms of multi-agent reinforcement learning}" \\$\bullet$ Section \uppercase\expandafter{\romannumeral5}-A of ``{graph-enhanced MARL}", i.e., hierarchical decoupling} \cr\hline

        Severe Noise Interference
        & \makecell[l]{$\bullet$ Section \uppercase\expandafter{\romannumeral5}-B of ``{collaboration-enhanced MARL}" \\$\bullet$ Section \uppercase\expandafter{\romannumeral5}-C of ``{information bottleneck-enhanced MARL}"} \cr\hline

        Partial Observability
        & \makecell[l]{$\bullet$ Section \uppercase\expandafter{\romannumeral5}-B of ``{collaboration-enhanced MARL}", i.e., collaborative policy and information sharing \\$\bullet$ Section \uppercase\expandafter{\romannumeral5}-D of ``{mirror learning-enhanced MARL}"} \cr\hline

        Poor Robustness and Generalization
        & \makecell[l]{$\bullet$ Section \uppercase\expandafter{\romannumeral5}-B of ``{collaboration-enhanced MARL}" \\$\bullet$ Section \uppercase\expandafter{\romannumeral5}-C of ``{information bottleneck-enhanced MARL}"} \cr\hline

        Poor Communication Efficiency
        & \makecell[l]{$\bullet$ Section \uppercase\expandafter{\romannumeral5}-B of ``{collaboration-enhanced MARL}", i.e., collaborative information sharing \\$\bullet$ Section \uppercase\expandafter{\romannumeral5}-D of ``{mirror learning-enhanced MARL}"} \cr\hline

        Sparse Information Expressiveness
        & \makecell[l]{$\bullet$ Section \uppercase\expandafter{\romannumeral5}-A of ``{graph-enhanced MARL}", i.e., embedded learning \\$\bullet$ Section \uppercase\expandafter{\romannumeral5}-C of ``{information bottleneck-enhanced MARL}"} \cr\hline
    \Xhline{1.2pt}
    \end{tabular}
  \vspace{0cm}
\end{table*}
\subsection{Heterogeneous Distributed Networks}
In contrast to homogeneous paradigms, heterogeneous distributed networks are composed of nodes with different characteristics, indicating that the features may differ in hardware, software, or communication modes. This diversity of nodes provides greater flexibility for wireless distributed networks, enabling them to manage, coordinate, and optimize a wider range of downstream tasks. However, the presence of diverse nodes adds complexity to network design, necessitating advanced algorithms or frameworks for efficient resource management and load balancing. Additionally, integrating diverse node types often requires appropriate communication modes to ensure seamless interaction among them.
Therefore, this indicates that heterogeneous distributed networks are well-suited for large-scale dynamic environments, such as MEC \cite{[22]}, space-terrestrial integrated networks (STIN) \cite{[221]}, integrated sensing and communications (ISAC) \cite{[216]}, CF massive MIMO \cite{[226]}, and RIS-assisted MIMO \cite{[227]}, where different node types and communication modes must operate together to meet diverse application demands, as shown in Table \uppercase\expandafter{\romannumeral2}. Typical heterogeneous types are described as follows:
\subsubsection{CF massive MIMO Networks}
They are composed of heterogeneous nodes, including a large number of APs and UEs \cite{[226]}, where all APs coherently provide services to UEs on the same time-frequency resources \cite{[610],[611],[612]}, as shown in Fig. 5 (a). Specifically, this network operates in time-division duplex (TDD) mode and exploits uplink-downlink channel reciprocity to help all APs obtain their own channel state information (CSI) with all UEs from the uplink pilot, which is sufficient for coherent transmission and reception \cite{[16],[17]}. Moreover, all APs are assumed to be connected via fronthaul links to one or more CPUs, which are responsible for data encoding and decoding \cite{[616]}. Note that it is usually assumed that the CPUs only know the long-term channel quality, i.e., statistical CSI, while only the APs have instantaneous CSI.

Additionally, the system model of CF massive MIMO mainly consists of uplink channel estimation, uplink data transmission, and downlink data transmission \cite{[16]}. Specifically, channel estimation provides estimated channel information for network operation, uplink data transmission is responsible for transmitting data from UEs to CPUs or APs, while downlink transmits data from APs or CPUs to UEs. These three complement each other and jointly support the efficient operation.
\subsubsection{RIS-Assisted MIMO Networks}
They consist of one CPU, one AP, one UE, and several auxiliary RISs, belonging to heterogeneous nodes, where the AP and several RISs are connected to the CPU via fronthaul links, which facilitates the exchange of payload data between the CPU and deployed AP/RISs \cite{[227],[230]}. Note that RIS is an emerging technique that can alter radio waves at the electromagnetic level without the need for active power amplifiers and complex digital signal processing \cite{[634]}. This enables the AP to transmit data to the UE through two downlink paths, including a direct link and an aggregated link through RISs, where each RIS plays an auxiliary enhancement role \cite{[615]}. Similarly, the system model of RIS-assisted MIMO mainly consists of uplink channel estimation and downlink data transmission.

Moreover, they can also be extended to CF RIS-assisted massive MIMO networks \cite{[614],[634]}, where the number of APs and UEs is no longer single, similar to previous CF massive MIMO neworks, as shown in Fig. 5 (b). Based on this paradigm, spatial multiplexing is leveraged, facilitated by a significantly larger number of APs than UEs, allowing the CPU to enable seamless communication among all nodes.
\subsection{Potential Challenges}
Wireless distributed networks, owing to their decentralized nature, offer significant advantages in flexibility, scalability, and fault tolerance. However, despite these benefits, they still face numerous challenges that hinder their potential in actual applications \cite{[29],[27],[31],[28],[43]}, particularly within the context of 6G, as shown in Table \uppercase\expandafter{\romannumeral3}. Typical challenges are as follows:
\begin{itemize}
\item \emph{Excessive Complexity:} The complexity of wireless distributed networks arises from the large number of interconnected nodes, each with different optimization objectives and constraints. This leads to the ``curse of dimensionality", where the joint state-action space grows exponentially with the number of nodes, rendering traditional optimization algorithms inefficient and difficult to scale, especially in large-scale scenarios.

\item \emph{Complex Interactive Relationships:} Refers to the rapidly changing environment in which multiple autonomous nodes collaborate or compete with each other. These complex interactions make the overall behavior of wireless distributed networks difficult to predict and optimize, especially in large-scale heterogeneous scenarios.

\item \emph{Cooperation and Competition Trade-off:} Multiple autonomous nodes typically need to collaborate to achieve common goals, while competing for limited resources. This makes it a challenge for nodes to effectively share limited resources while ensuring superior performance, even in highly competitive scenarios.

\item \emph{Complex Dynamism:} The rapidly changing environment also brings about dynamic changes in communication topology and task execution for wireless distributed networks, where the former is often unstable due to signal fluctuations or network congestion. This requires nodes to be able to perceive these dynamics and adjust strategies promptly to maintain reliable services.

\item \emph{Complex Dependence:} The performance of each autonomous node is closely related to the observed states and actions of other nodes. However, traditional algorithms often struggle to explain these interactions, resulting in poor performance. This interdependence poses challenges in modeling and optimizing node behavior.

\item \emph{Poor Stability:} The dynamic characteristics of wireless distributed networks often lead to instability in node behavior and environment prediction, resulting in oscillatory or divergent when applying traditional algorithms.

\item \emph{Severe Noise Interference:} Wireless distributed networks face significant interference, including environmental interference and random interference among nodes. The challenge lies in the difficulty of distinguishing between signals and interference, which often leads to inaccurate information reception. Therefore, it is essential to effectively suppress interference and extract useful signals.
\end{itemize}

\begin{table*}[t!]
  \centering
  \fontsize{8.5}{12}\selectfont
  \caption{Characteristics for different MARL algorithms, including model-based MARL and model-free MARL.}
  \label{CE}
   \begin{tabular}{ !{\vrule width1.2pt}  m{1.9 cm}<{\centering}
   !{\vrule width1.2pt}  m{1.45 cm}<{\centering}
   !{\vrule width1.2pt}  m{3.85 cm}<{\centering}
   !{\vrule width1.2pt}  m{3.85 cm}<{\centering}
   !{\vrule width1.2pt}  m{4.4 cm}<{\centering}   !{\vrule width1.2pt} }

    \Xhline{1.2pt}
        \rowcolor{gray!30} \bf Frameworks  &  \bf Ref. &  \bf Classifications &  \bf Communication Modes  & \bf Application Scenarios \cr
    \Xhline{1.2pt}
        \multirow{9}{*}{\bf \shortstack{Model-Based\\ MARL}}
                    & \cite{[813],[814]}  & Atari \& AlphaGo & Independent  & Dynamic Programming \\
        \cline{2-5} & \cite{[810]}  & Recirrent World Models & Cooperative   & Model Predictive Control \\
        \cline{2-5} & \cite{[809]}  & MBRL & Hybrid  & Model Predictive Control \\
        \cline{2-5} & \cite{[811],[851]}  & SimPLe \& AORPO & Cooperative  & Model Predictive Control  \\
        \cline{2-5} & \cite{[815]}  & MAMBA & Hybrid  & Dynamic Programming \\
        \cline{2-5} & \cite{[807]}  & Zero-Sum Markov Games & Independent \& Adversarial  & Dynamic Programming \\
        \cline{2-5} & \cite{[816]}  & LLM-Based Multi-Agents & Hybrid  & Dynamic Programming \\
        \cline{2-5} & \cite{[812]}  & Decentralized Policy Optimization & Cooperative \& Adversarial   & Model Predictive Control \& Dynamic Programming \cr\Xhline{1.2pt}

        \multirow{11}{*}{\bf \shortstack{Model-Free\\ MARL}}
                    & \cite{[828]}    & CommNet & Cooperative & Communication Methods \\
        \cline{2-5} & \cite{[829]}   & RIAL \& DIAL & Independent \& Cooperative  & Communication Methods \\
        \cline{2-5} & \cite{[820],[822]}  & MADDPG \& QMIX  & Cooperative \& Adversarial & Direct Methods  \\
        \cline{2-5} & \cite{[823]}  & COMA  & Hybrid & Direct Methods \\
        \cline{2-5} & \cite{[830]}  & ATOC & Cooperative  & Communication Methods \\
        \cline{2-5} & \cite{[821],[824]}  & VDN \& QTRAN & Cooperative   & Communication Methods \\
        \cline{2-5} & \cite{[832]}  & MAAC \& MATD3 &  Cooperative \& Adversarial  & Communication Methods \\
        \cline{2-5} & \cite{[825]}  & MATD3 &  Cooperative \& Adversarial  & Direct Methods \\
        \cline{2-5} & \cite{[831]}  & MAPPO & Cooperative  & Communication Methods \\
        \cline{2-5} & \cite{[826]}  & IPPO & Independent  & Direct Methods \\
        \cline{2-5} & \cite{[833],[850]}  & IC3Net \& CMARL & Hybrid & Direct \& Communication Methods \cr\Xhline{1.2pt}
        \multicolumn{5}{l}{\footnotesize * Note that the specific algorithm descriptions are shown in Section \uppercase\expandafter{\romannumeral4}-B and Section \uppercase\expandafter{\romannumeral4}-C.}\\
    \end{tabular}
  \vspace{0cm}
\end{table*}
Moreover, because individual nodes in wireless distributed networks can usually only perceive local information and cannot effectively infer global information, they may also face challenges such as \emph{partial observability}, \emph{poor communication efficiency}, and \emph{sparse information expressiveness}, which are detailed in Section \uppercase\expandafter{\romannumeral4}-D.
In contrast, MARL with the ability to facilitate decentralized decision-making and continuous adaptation, emerges as a natural approach to address these challenges. By harnessing the power of MARL, wireless distributed networks can achieve enhanced performance, greater scalability, and improved reliability, making them more suitable for the complex demands of 6G.
\section{Algorithms of Multi-Agent Reinforcement Learning}
In this section, we provide an overview of the fundamental MARL algorithms, categorized into two typical types: model-based and model-free, as summarized in Table \uppercase\expandafter{\romannumeral4}. Model-based MARL relies on explicit environmental models to support planning and decision-making, offering high efficiency in structured scenarios. In contrast, model-free MARL learns policies directly through interactions with the environment, excelling in complex dynamic scenarios, as shown in Fig 6.
Combining the strengths of both paradigms has the potential to fully unleash the power of wireless distributed networks. However, this also presents several challenges, such as non-stationarity, partial observability, and balancing exploration with exploitation in multi-agent environments.
\subsection{Preliminaries}
As an advanced approach within distributed AI, MARL algorithms offer promising solutions \cite{[801]} and have driven significant progress in various applications \cite{[802]}, such as wireless communications \cite{[559],[560]}, UAV-assisted communications \cite{[640],[645]}, and autonomous driving \cite{[201]}. They specifically addresses sequential decision-making problems involving multiple agents, where the network dynamics emerge from their interactions. Note that in MARL-empowered applications, the rewards assigned to each agent depend not only on their individual behavior but also on the collective behavior of all agents. This requires each agent to account for the policies adopted by others to effectively optimize long-term returns. To provide a clear framework for understanding the MARL algorithms discussed in the following part, we first define the key concepts and principles underlying this approach.
\subsubsection{Markov Decision Process}
Within traditional RL algorithms,
agents explore their environment to address sequential decision-making tasks \cite{[561]}. Typically, environments \emph{fully observable} are formulated as Markov decision processes (MDPs) described by the tuple $<\mathcal{S}, \mathcal{A}, \mathcal{P}, \mathcal{R}, \gamma>$ \cite{[820],[625]}. Herein, $\mathcal{S}$ represents the set of states and $\mathcal{A}$ the set of actions. The transition dynamics $\mathcal{P}: \mathcal{S} \times \mathcal{A}$ specifies the probability of transitioning from state $s_{t}$ to state $s_{t+1}$ after performing the action $a_{t}$ at time $t$. The reward function $\mathcal{R}: \mathcal{S} \times \mathcal{A} \times \mathcal{S}$ assigns rewards based on actions $a_{t}$ in state $s_{t}$ leading to $s_{t+1}$, while the discount factor $\gamma \in [0, 1]$ modulates the present value of future rewards. Assuming global observability \cite{[902]}, agents have precise access to the current state $s_{t}$, enabling the selection of action $a_{t}$ and resulting in a state transition to $s_{t+1}$, sampled via probability $\mathcal{P}(\cdot | s_t, a_t)$. Consequently, agents receive an immediate reward $\mathcal{R}(s_t,a_t,s_{t+1})$ and the expected return can be represented as
\begin{align}
\mathbb{E} \bigg[\sum_{t=0}^{\infty} \gamma_t \mathcal{R}(s_t, a_t, s_{t+1}) \mid a_{t} \sim \pi(\cdot|s_{t}), s_0 \bigg],
\end{align}
where the policy $\pi(\cdot|s_t)$ represents the action distribution under a given current state $s_t$, which is referred to as the \emph{infinite-horizon discounted return} \cite{[803]}. Correspondingly, another widely utilized formulation is the \emph{undiscounted finite-horizon return} \cite{[540]}, given by
\begin{align}
\mathbb{E} \bigg[\sum_{t=0}^{T} \mathcal{R}(s_t, a_t, s_{t+1}) \mid a_t \sim \pi(\cdot|s_t), s_0 \bigg],
\end{align}
in which the return is calculated over a fixed time horizon $ T $, a formulation that is frequently applied in \emph{episodic tasks} \cite{[580]}, where the task concludes at a predetermined endpoint.

\begin{figure*}[t]
\centering
    \includegraphics[scale=0.4]{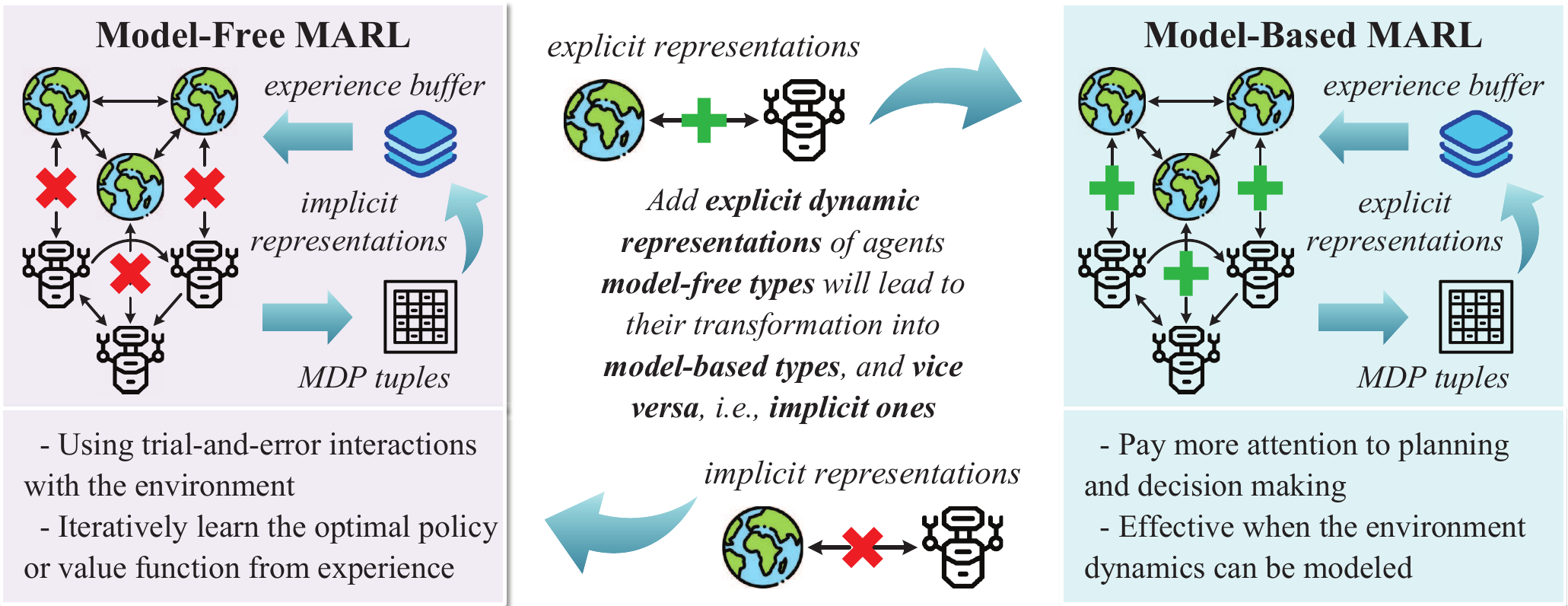}
    \caption{The network architecture and connection of different MARL algorithms, including model-free types (e.g., any prior knowledge about wireless distributed networks is unknown) and model-based types (e.g., assuming that agents know the state transition function in advance).
    \label{fig1}}
\end{figure*}
To better determine the optimal policy $\pi^*$, we can define a corresponding \emph{$Q$-function} (or \emph{value function}) to measure the expected accumulated rewards starting from a specific state-action pair $(s_t, a_t)$-\emph{$Q$-function} (or a state $s_t$-\emph{value function}), both of which can be modeled as
\begin{subequations}
\begin{align}
\!\!\!\!Q^{\pi}(s,a)\! &=\! \mathbb{E} \bigg[\sum_{t=0}^{\infty} \gamma_t \mathcal{R}(s_t, a_t, s_{t+1})|a_t \sim \pi(\cdot|s_t) \bigg]_{s_0, a_0},\\
V^{\pi}(s)\! &=\! \mathbb{E} \bigg[\sum_{t=0}^{\infty} \gamma_t \mathcal{R}(s_t, a_t, s_{t+1})|a_t \sim \pi(\cdot|s_t)\bigg]_{s_0}.
\end{align}
\end{subequations}

Note that in MARL settings, the two key functions are commonly used to evaluate and optimize agent behaviors. The $Q$-function evaluates the expected return of a state-action pair, while the state-value function, assesses the overall value of a state. These functions serve distinct yet complementary roles in MARL, with their applicability and utility varying based on the environment's structure and the specific approach being utilized, which are compared as follows:
\begin{itemize}
\item $Q$-functions $Q^{\pi}(s,a)$ excel in evaluating state-action pairs, ideal for discrete action spaces, like in model-based MARL. They refine policies iteratively and aid decision-making when action evaluation is needed, crucial for agent coordination in structured contexts \cite{[401]}.
\item Value functions $V^{\pi}(s)$ excel at estimating a state's overall value, which is especially useful in expansive or continuous action spaces where evaluating all state-action pairs is unfeasible \cite{[801]}. Therefore, they are used primarily in model-free MARL to guide strategic decisions.
\end{itemize}

To sum up, integrating $Q$-functions $Q^{\pi}(s,a)$ with value functions $V^{\pi}(s)$ in hybrid settings boosts agents' adaptability in distributed systems, enhancing multi-agent coordination and learning. This synergy highlights the complementary roles of these functions in advancing MARL research, particularly for large-scale systems with uncertainty.
\subsubsection{Decentralized Partially Observable Markov Decision Process}
In actual application scenarios, agents typically cannot obtain complete observations. This has driven the emergence of Dec-POMDP, which extends the standard MDP aforementioned to multi-agent settings, where uncertainty and partial observability are key factors \cite{[915]}. Unlike the standard MDP, which assumes full observability for a single agent, Dec-POMDP accommodates multiple collaborative agents working with limited and often noisy information \cite{[914]} and is particularly suited for applications where centralized control is impractical \cite{[906]}. By explicitly modeling uncertainty and enabling optimal coordination, Dec-POMDP provides a strong foundation for addressing multi-agent decision-making challenges in complex dynamic environments.

Specifically, Dec-POMDP can be represented by the tuple $< \mathcal{N}, \mathcal{S}, \mathcal{O}, \mathcal{A}, \mathcal{R}, \mathcal{P}, b_0, h > $, where $\mathcal{N}$ and $\mathcal{S}$ denote the set of collaborative agents and the finite global state space, respectively \cite{[550]}. The joint action space $\mathcal{A} = \mathcal{A}_1 \times \mathcal{A}_2 \times \dots \times \mathcal{A}_n$ encompasses all possible combinations of individual agent actions, and the environment's probabilistic dynamics are governed by the transition function $\mathcal{P}(\cdot|s_t, a_t)$. Furthermore, each agent has
a \emph{local observation} set $\mathcal{O}_i$, while the joint observation space $\mathcal{O} = \mathcal{O}_1 \times \mathcal{O}_2 \times \dots \times \mathcal{O}_n$ accounts for all combinations of individual observations. The observation function $\mathcal{O}(o_t|a_t, s_{t+1})$ defines the probability of observation $o_t$ given joint actions $a_t$ and subsequent states $s_{t+1}$, capturing the uncertainty and partial observability inherent in distributed networks. The reward function $\mathcal{R}(s_t,a_t, s_{t+1})$ assigns a shared reward to state-action pairs $(s_t, a_t)$, motivating agents to collaborate toward a common goal. Unlike single-agent MDPs, agents in Dec-POMDPs independently determine actions based on local observation histories, without knowledge of other agents' actions or observations. Each agent follows a local policy $\pi_i$ mapping observation histories to actions, with no communication or information sharing. The objective is for agents to collectively maximize the cumulative rewards, starting from an initial belief distribution that represents the probability distribution over all possible initial states.

\begin{figure*}[t]
\centering
    \includegraphics[scale=0.45]{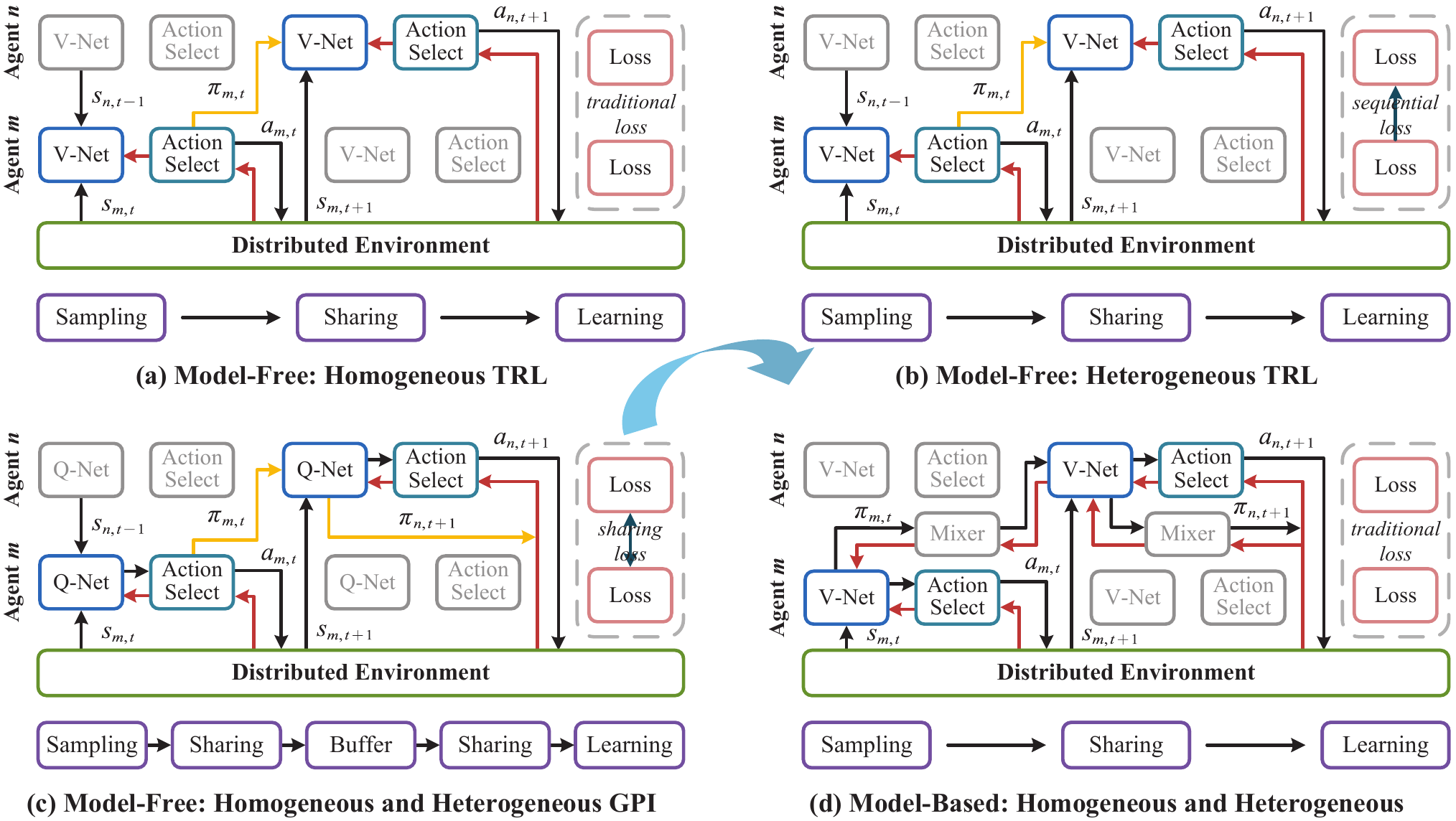}
    \caption{Two typical network frameworks of MARL, including model-free and model-based MARL. Note that ``V-Net" and ``Q-Net" are defined as ``Value Network" and ``Q (state-action value) Network".
    \label{fig1}}
\end{figure*}
However, the excessively high complexity limits the scalability of Dec-POMDPs, especially in large-scale environments. To address these limitations, the following part explores networked MDPs, a related framework that introduces structural constraints to improve scalability \cite{[652]}. By leveraging the inherent networked relationships among agents, networked MDPs strike a balance between computational efficiency and the ability to model decentralized decision-making, offering an attractive alternative for large-scale multi-agent settings.
\subsubsection{Networked Markov Decision Process}
On the other hand, in actual wireless distributed networks, agents and their interactions are often modeled as nodes and edges within a graph, respectively \cite{[301],[302]}, where network structures are dynamic, with varying connectivity and communication levels that significantly impact the strategies employed by agents. While collaborative MDPs and Dec-POMDPs are effective for modeling homogeneous MARL-assisted distributed networks with shared rewards and consistent dynamics, they struggle to capture the complexities of actual applications, particularly in environments where dynamics and interactions evolve over time \cite{[910]}. Furthermore, in scalable distributed networks, the reliance on shared rewards becomes impractical, as computing a collective state-action function requires global information from all agents, which introduces excessive communication overhead, underscoring the need for scalable and decentralized approaches in networked MARL frameworks \cite{[402]}.

To better capture the graph-structured dynamic characteristics of heterogeneous multi-agent settings in wireless distributed networks, researchers have extended MDPs and Dec-POMDPs to the framework of networked MDPs, which exhibit the Markovian property of state transitions \cite{[406]}. In this framework, the distributed network is modeled as $\mathcal{G} = < \mathcal{V}, \mathcal{E} >$, where $\mathcal{G}$ represents the entire graph, $\mathcal{V}$ denotes the set of nodes corresponding to the set of agents $\mathcal{N} = \{1, \dots, N\}$, and $\mathcal{E}$ represents the set of edges capturing interactions and communication between agents \cite{[43]}. Based on this graph structure, the corresponding networked MDP is described by the tuple $<\mathcal{G}, {\mathcal{S}_i, \mathcal{A}_i}, \mathcal{R}, \mathcal{P}, \gamma |{i \in \mathcal{N}}>$, where each agent $i \in \mathcal{N}$ possesses a local state $s_{i,t} \in \mathcal{S}_i$ and selects actions $a_{i,t} \in \mathcal{A}_i$ at any given $t$ time slot. The, the global state can be defined as $s_t = (s_{1,t}, s_{2,t}, \dots, s_{N,t}) \in \mathcal{S}$, while the global action is represented as $a_t = (a_{1,t}, a_{2,t}, \dots, a_{N,t}) \in \mathcal{A}$ The state transitions adhere to the Markov property, with the transition function represented by $\mathcal{P}(\cdot|s_t, a_t)$, respectively.

Building upon this foundation, the networked MDP framework introduces localized policies to better manage the computational and structural challenges inherent in wireless distributed networks. Each agent $i$ adopts a localized policy $\pi_i^{\theta_i}$, parameterized by $\theta_i \in \Theta_i$, which depends only on the states of the agent itself and its immediate neighbors. This design not only enhances scalability but also significantly reduces computational complexity. Using local and neighboring information, each agent computes a localized reward $\mathcal{R}_i$, which serves as the basis for optimizing network performance. The primary objective in this framework is to derive an optimal joint policy that maximizes the global reward, typically measured as the average reward across all agents, $\mathcal{R} = \frac{1}{n} \sum{i=1}^{n} \mathcal{R}_i$, with the corresponding value function defined as follows:
\begin{align}
  V^{\pi}(s) &= \mathbb{E} \bigg[\sum_{t=0}^{\infty} \gamma_t \mathcal{R}(s_t,a_t,s_{t+1})\bigg] = \frac{1}{n}\sum_{i=1}^{n}V^{\pi}_{i}(s).
\end{align}
\subsection{Model-Based MARL Algorithms}
\subsubsection{Introduction to Model-Based MARL}
In large-scale multi-agent settings, sequential decision-making is commonly modeled as a stochastic or Markov game (MG). For $N$ agents, this game is described by the tuple $ <\mathcal{S}, \mathcal{A}_i, \mathcal{R}_i, \mathcal{P}, \gamma|{i \in \mathcal{N}}>$ \cite{[801]}. Here, $\mathcal{S}$ specifies the state space, $\mathcal{A}_i$ represents the action space for agent $i$, and the joint action space is expressed as $\mathcal{A} = \mathcal{A}_1 \times \cdots \times \mathcal{A}_N$. In addition, the reward function $ \mathcal{R}_i: \mathcal{S} \times \mathcal{A}$ determines the reward received, while the transition dynamics $ \mathcal{P}: \mathcal{S} \times \mathcal{A} \to \Delta(\mathcal{S}) $ defines the probability distribution of the next states. Correspondingly, at each time slot $t$, an action $a_{i,t}$ is independently chosen by the agent $i$ based on the current state $s_t$, where the collective actions of all agents except $i$ are represented as $a_{-i,t} = \{a_{j,t}\}_{j \neq i}$. Subsequently, the state transitions to $ s_{t+1} $ and the agent $i$ receives a reward $ r_{i,t} $.  From the perspective of agent $ i $, the combined strategy of the remaining agents is expressed as $ \pi_{-i}(a_{-i,t} | s_t) = \prod_{j \in \{-i\}} \pi^j(a_{j,t} | s_t) $, where $\pi_j:\mathcal{S} \to \Delta(\mathcal{A}_j)$ represents the policy of agent $j$. Then, the objective for each agent is to determine an optimal policy that maximizes the expected cumulative reward \cite{[802]}, which can be defined as:
\begin{align}
    \pi_i^*=\operatorname{argmax} \mathbb{E}_{\tau \sim (\mathcal{P},\pi_i,\pi_{-i})}
    \bigg[\sum_{t=0}^\infty \gamma^{t}R_i(s_{t}, a_{i,t}, a_{-i,t}) \bigg],
\end{align}
where $\tau \in \{(s_{0}, a_{i,0}, a_{-i,0}), (s_{1}, a_{i,1}, a_{-i,1}), \ldots \}$ denotes the sampled trajectory.

Model-based approaches aim to reduce the amount of data needed to reach a specific performance level in distributed networks \cite{[803]}. Agent $i$'s best policy depends on both its actions and the actions of other agents. In stochastic games, an agent's sample complexity splits into two types: i) \emph{dynamics sample complexity}, concerning state samples from the environment \cite{[804]}, and ii) \emph{opponent sample complexity}, involving total action samples from other agents, acquired through direct interaction or decentralized communication \cite{[805]}.
\subsubsection{Efficiency of Model-Based MARL}
In single-agent domains, model-based methods are praised for their theoretical and practical efficiency, mainly due to reduced sample complexity, as illustrated in Fig. 7 \cite{[806]}. They reduce sample demands, enhance stability, and promote exploration. Similarly, in large-scale multi-agent settings, these approaches provide theoretical advantages, where using learned environment models reduces sample complexity, enhancing learning efficiency and better agent coordination.

A typical model-based MARL approach is structured into such two phases: \emph{learning phase}, where an empirical model is constructed using collected data, and \emph{planning phase}, where optimal policies are derived from learned models. When utilizing a generative model, policy updates are made based solely on the generated data. The sample complexity for determining an $ \epsilon $-Nash equilibrium policy in a reward-agnostic setting is $ \mathcal{O}(|\mathcal{S}| |\mathcal{A}^1| |\mathcal{A}^2| (1-\gamma)^{-3} \epsilon^{-2}) $, and in the reward-aware setting, it is reduced to $ \mathcal{O}(|\mathcal{S}| (|\mathcal{A}^1| + |\mathcal{A}^2|) (1-\gamma)^{-3} \epsilon^{-2}) $, where $ \epsilon $ represents the allowed deviation from the Nash equilibrium.
Moreover, the Markov perfect equilibrium MPE of the approximated game serves as an $ \alpha $-approximate equilibrium for the original game, with $ \alpha $ representing the discrepancy between the $ Q $-functions of the two games. The sample complexity for computing an $ \alpha $-approximate equilibrium is $ \mathcal{O}(|\mathcal{S}| |\mathcal{A}| (1-\gamma)^{-2} \alpha^{-2}) $, demonstrating the enhanced sample efficiency.
Robustness analysis establishes a clear boundary for $\alpha$, accounting for approximation errors in the dynamics model and the reward function. While theory points to model-based MARL methods' sample efficiency under certain conditions, there is limited research in broader contexts. In addition, many potential benefits of model-based MARL remain unexplored. These results highlight the sample efficiency of model-based in large-scale multi-agent settings, especially compared to model-free, as detailed in the following subsections.
\begin{figure*}[t]
\centering
    \includegraphics[scale=0.6]{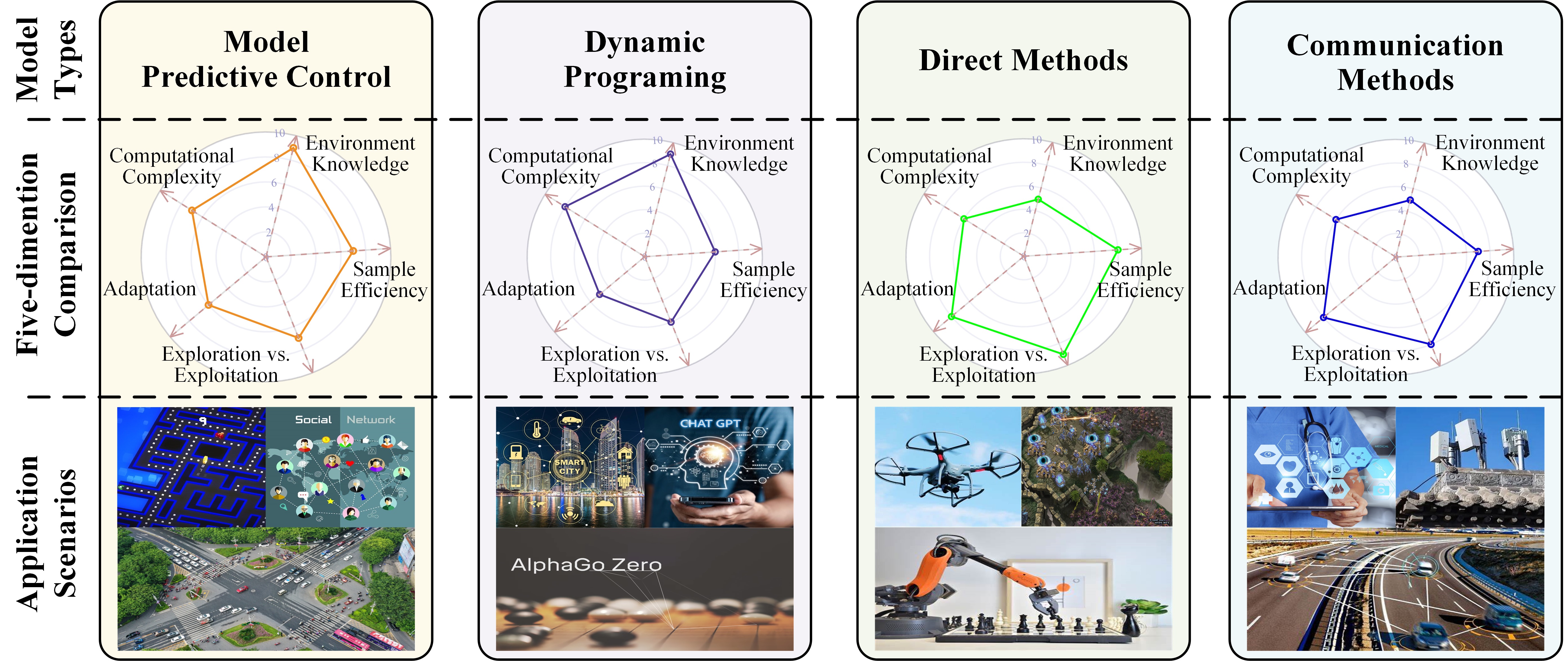}
    \caption{Four types of application scenarios for model-based and model-free MARL. Please refer to \cite{[808]}, \cite{[820]}, \cite{[813]}, and \cite{[828]} for more details.
    \label{fig1}}
\end{figure*}
\subsubsection{Applications of Model-Based MARL}
In the context of model-based MARL, we can categorize their applications into two typical types: model predictive control (MPC) and dynamic programming (DP), as shown in Fig. 8.

\emph{a) Model Predictive Control:}
Model-based MARL provides a robust strategy for approaching intricate decision-making challenges by utilizing models that predict how environments will evolve, similar to MPC \cite{[808]}. In this setup, each agent is tasked with understanding a probabilistic model of its environment's workings, which allows the agent to anticipate future conditions and refine its actions over a set period. This method improves the efficiency of sample usage as it allows ``virtual" interactions within these learned models, thereby minimizing the need for actual real-world experiments \cite{[809]}.

Recurrent world models represent a progressive step in model-based multi-agent reinforcement learning by integrating recurrent neural networks (RNNs). These RNNs adeptly capture temporal dependencies, thus enhancing planning capabilities in environments that are partially observable \cite{[810]}. In parallel, SimPLe emerges as a robust framework for learning world models that are both scalable and adaptable, particularly excelling in handling high-dimensional and continuous action spaces \cite{[811]}. The technique of decentralized policy optimization \cite{[812]} pushes the field further by allowing agents to independently refine their policies through local observations. This approach is particularly suitable for large-scale multi-agent scenarios where communication is limited.

By employing advanced models to predict future actions, model-based MARL can achieve improved sample efficiency along with superior performance in intricate settings. This makes them an attractive method for addressing decision-making challenges in real-world applications, where complexity and unpredictability often present significant hurdles.

\emph{b) Dynamic Programming:}
DP has historically been integral to reinforcement learning, offering robust methods for addressing sequential decision issues. Within MARL, DP is crucial for policy evaluation and optimal strategy determination in intricate, multi-agent settings. A seminal use of DP is its application in deep Q-networks (DQN) for Atari games \cite{[813]}. Here, DP's value iteration principle is instrumental in predicting expected action returns in vast game environments. This value-centric method efficiently approximates optimal action-value functions, enabling agents to pursue decisions that optimize long-term rewards in extensive state-action spaces, a standard challenge in reinforcement learning.

Building on DQN, AlphaGo combines DP with Monte Carlo tree search (MCTS) \cite{[814]}. This novel integration of value iteration with a robust tree search was crucial for AlphaGo's success in Go by evaluating countless scenarios to surpass human skill. Similarly, DP is utilized in zero-sum Markov games \cite{[807]}, where rival agents aim to maximize their gains while hindering opponents. Employing DP, these agents formulate optimal strategies by analyzing their tactics against others, navigating complex strategic interdependence expertly.

MAMBA exemplifies a model-based framework aimed at multi-agent decision-making \cite{[815]}. DP uses it to evaluate actions in both cooperative and competitive contexts. Additionally, the integration of large language models (LLMs) into MARL \cite{[816]} introduces new prospects for utilizing DP in complex, partially observable environments. These LLMs enable agents to ascertain optimal strategies via value iteration and policy evaluation, thus improving decision-making in tasks like natural language processing. The incorporation of DP in LLM-augmented MARL enhances predictive and decision accuracy, thus strengthening agents' ability to navigate intricate environments with limited information.
\subsection{Model-Free MARL Algorithms}
\subsubsection{Introduction to Model-Free MARL}
The primary distinction between model-based and model-free MARL lies in their strategies for learning and decision-making. In model-based methods, a model is created to capture the dynamics of the environment, including transition probabilities and reward functions \cite{[808]}. This model is then utilized for planning and decision-making. In contrast, model-free approaches directly learn a policy or value function from experience, bypassing the need for an explicit environmental model, as shown in Fig. 7. These approaches rely exclusively on interactions with the environment to optimize the policy, without requiring detailed knowledge of the network's internal mechanics \cite{[2]}.

Despite their differences, model-based and model-free should not be regarded as entirely distinct. Rather, they exist on a spectrum with numerous algorithms that combine aspects of both approaches. Although planning efficiency can be improved through the use of learned models in model-based methods, model-free methods are often favored for their simplicity and flexibility. These characteristics make them especially valuable in complex or uncertain environments, where accurately modeling the network's dynamics is challenging or impractical. The following part explores two key strategies within model-free MARL: ``\emph{generalized policy iteration} (GPI)" \cite{[817]} and ``\emph{trust-region learning} (TRL)" \cite{[818]}, as follows:

\emph{a) Generalised Policy Iteration:}
The simplicity of the GPI framework is a key advantage. Although the policy influences both rewards and state visits, the GPI ensures that greedily responding to the value function alone is enough to guide the learning process. $ \forall s\in\mathcal{S}$, we have
\begin{align}
    \label{eq:gpi}
    \pi_{\text{new}}(\cdot|s) = \operatorname{argmax}_{\bar{\pi}(\cdot|s)\in\mathcal{P}(\mathcal{A})} \mathbb{E}_{\textnormal{a}\sim\bar{\pi}}\big[ Q_{\text{old}}(s, \textnormal{a})\big],
\end{align}
which guarantees that the new policy yields higher expected returns for every state, i.e., $V_{\pi_{\text{new}}}(s) \geq V_{\pi_{\text{old}}}(s), \ \forall s \in \mathcal{S}$. Furthermore, this procedure converges to the set of optimal policies, which can be intuitively demonstrated by substituting a fixed-point policy into the equation (\ref{eq:gpi}).
Specifically, in environments with small and discrete spaces, the policy $\pi$ does not need to be explicitly stored for the GPI approximation. Instead, a greedy response to the state-action value function is used. This results in \textit{value-based learning}, in which a Q-function is learned through the Bellman-max update:
\begin{align}
\!\! \! Q_{\text{new}}(s, a) \! =\! \mathcal{R}(s_t, a_t,s_{t+1})\!  +\! \mathbb{E} \bigg[ \gamma\max_{a_{t+1}} Q_{\text{old}}(\textnormal{s}_{t+1}, a_{t+1}) \bigg],
\end{align}
which is known to converge to $Q^{*}$ and has inspired the development of several methods.

As an alternative, an approximate implementation of GPI is through policy gradient (PG) algorithms. These methods optimize the policy $\pi_\theta$ by parameterizing it with $\theta$ and updating the parameters in the direction of the gradient of the expected return, which is given by
\begin{equation}
\nabla_{\theta} \eta(\pi_\theta)= \mathbb{E}_{s \sim \rho_{\pi_{\theta_{\text{old}}}}} \bigg[ \nabla_{\theta} \mathbb{E}_{a \sim \pi_{\theta}} \Big[ Q_{\text{old}}(s_t, a_t) \Big] \bigg].
\end{equation}

The gradient of the GPI optimization goal from equation (8), scaled by $\rho_{\pi_{\theta_{\text{old}}}}$, has a parallel in continuous deterministic policies. Thus, PG-based algorithms estimate the GPI step, resulting in a policy near $\pi_{\theta_{\text{old}}}$ if the update step size $\alpha > 0$ in $\theta_{\text{new}} = \theta + \zeta \nabla_{\theta} \eta(\pi_{\theta}) \big|_{\theta = \theta_{\text{old}}}$ is small enough. PG methods are pivotal in reinforcement learning applications, particularly with continuous action spaces, where value-based methods such as DQN are not feasible.

\emph{b) Trust-Region Learning:}
However, in actual applications, PG-based algorithms often experience high variance in PG estimates and instability during training. To address these challenges, the TRL framework was designed. The core of TRL includes the following policy update:
\begin{align}
    \label{eq:trust-region}
    \pi_{\text{new}}= \operatorname{argmax}\mathbb{E}_{s\sim\rho_{\pi_{\text{old}}}}\big[ A_{\pi_{\text{old}}}(s, a) \big] - C\text{D}_{\text{KL}}^{\text{max}}(\pi_{\text{old}}, \bar{\pi}),
\end{align}
where $C = 4\gamma \max_{s, a}|A_{\pi_{\text{old}}}(s, a)|/(1-\gamma)^2$.It was shown by \cite{[818]} that this update ensures a monotonic improvement in the return, i.e., $\eta(\pi_{\text{new}}) \geq \eta(\pi_{\text{old}})$. Furthermore, the \emph{Kullback-Leibler} (KL)-penalty in the objective function keeps the new policy within the trust region, which is defined as the neighborhood of $\pi_{\text{old}}$. This is crucial for mitigating the instability problems typically observed in PG-based algorithms.
\begin{table}[t!]
  \centering
\fontsize{8.5}{12}\selectfont
\caption{Model-Free MARL vs. Model-Based MARL.}
\label{tab:model_comparison}
\begin{tabular}[b]{ccc}
\toprule
& \bf MF MARL & \bf MB MARL  \\
\midrule
Asymptotic Rewards & \cellcolor{green1!25}$+$ & \cellcolor{yellow!25}$+/-$ \\
Environment Knowledge &\cellcolor{red1!25}$-$& \cellcolor{green1!25}$+$ \\
Exploration vs. Exploitation &\cellcolor{red1!25}$-$& \cellcolor{green1!25}$+$ \\
Computational Complexity & \cellcolor{green1!25}$+$ &\cellcolor{yellow!25}$+/-$\\
Sample Efficiency &\cellcolor{red1!25}$-$ & \cellcolor{green1!25}$+$ \\
Stability and Convergence & \cellcolor{green1!25}$+$  &\cellcolor{red1!25}$-$\\
Scalability &\cellcolor{red1!25}$-$& \cellcolor{green1!25}$+$  \\
Adaption to Changing Rewards &  \cellcolor{red1!25}$-$ & \cellcolor{green1!25}$+$ \\
Adaption to Changing Dynamics &  \cellcolor{red1!25}$-$ & \cellcolor{green1!25}$+$ \\
\bottomrule
\end{tabular}
\end{table}
\subsubsection{Applications of Model-Free MARL}
Similarly, in the context of model-free MARL, their applications can be broadly classified into two typical categories: direct and communication methods, as shown in Fig. 8.

\begin{table*}[tp]
  \centering
  \fontsize{8.5}{12}\selectfont
  \caption{Potential challenges and feasible techniques for MARL.}
  \label{System_ULA}
    \begin{tabular}{
    !{\vrule width1.2pt} m{4.9 cm}<{\centering}
    !{\vrule width1.2pt} m{12.2 cm}<{\centering}
    !{\vrule width1.2pt} }

    \Xhline{1.2pt}
        \rowcolor{gray!30} \bf Potential Challenges  &  \bf Feasible Techniques\cr
    \Xhline{1.2pt}
        Excessive Complexity & \makecell[l]{$\bullet$ Section \uppercase\expandafter{\romannumeral5}-C of ``{information bottleneck-enhanced MARL}", i.e., SubIB} \cr\hline

        Complex Interactive Relationships
        & \makecell[l]{$\bullet$ Section \uppercase\expandafter{\romannumeral5}-A of ``{graph-enhanced MARL}", i.e., relationship modeling} \cr\hline

        Complex Dependence
        & \makecell[l]{$\bullet$ Section \uppercase\expandafter{\romannumeral5}-A of ``{graph-enhanced MARL}", i.e., relationship modeling and hierarchical decoupling} \cr\hline

        Poor Stability
        & \makecell[l]{$\bullet$ Section \uppercase\expandafter{\romannumeral5}-A of ``{graph-enhanced MARL}", i.e., hierarchical decoupling} \cr\hline

        Non-Stationarity
        & \makecell[l]{$\bullet$ Section \uppercase\expandafter{\romannumeral5}-B of ``{collaboration-enhanced MARL}", i.e., collaborative policy and transfer learning} \cr\hline

        Partial Observability
        & \makecell[l]{$\bullet$ Section \uppercase\expandafter{\romannumeral5}-B of ``{collaboration-enhanced MARL}", i.e., collaborative policy and information sharing \\$\bullet$ Section \uppercase\expandafter{\romannumeral5}-D of ``{mirror learning-enhanced MARL}"} \cr\hline

        Poor Scalability
        & \makecell[l]{$\bullet$ Section \uppercase\expandafter{\romannumeral5}-A of ``{graph-enhanced MARL}", i.e., embedded learning \\$\bullet$ Section \uppercase\expandafter{\romannumeral5}-B of ``{collaboration-enhanced MARL}", i.e., collaborative information sharing\\$\bullet$ Section \uppercase\expandafter{\romannumeral5}-D of ``{mirror learning-enhanced MARL}"} \cr\hline

        Poor Sample Efficiency
        & \makecell[l]{$\bullet$ Section \uppercase\expandafter{\romannumeral5}-B of ``{collaboration-enhanced MARL}"\\$\bullet$ Section \uppercase\expandafter{\romannumeral5}-D of ``{mirror learning-enhanced MARL}", i.e., maximum entropy} \cr\hline

        Poor Robustness and Generalization
        & \makecell[l]{$\bullet$ Section \uppercase\expandafter{\romannumeral5}-B of ``{collaboration-enhanced MARL}" \\$\bullet$ Section \uppercase\expandafter{\romannumeral5}-C of ``{information bottleneck-enhanced MARL}"} \cr\hline

        Poor Communication Efficiency
        & \makecell[l]{$\bullet$ Section \uppercase\expandafter{\romannumeral5}-B of ``{collaboration-enhanced MARL}", i.e., collaborative information sharing \\$\bullet$ Section \uppercase\expandafter{\romannumeral5}-D of ``{mirror learning-enhanced MARL}"} \cr\hline

        Slow Convergence Rate
        & \makecell[l]{$\bullet$ Section \uppercase\expandafter{\romannumeral5}-B of ``{collaboration-enhanced MARL}", i.e., collaborative transfer learning} \cr\hline

        Sparse Information Expressiveness
        & \makecell[l]{$\bullet$ Section \uppercase\expandafter{\romannumeral5}-A of ``{graph-enhanced MARL}", i.e., embedded learning \\$\bullet$ Section \uppercase\expandafter{\romannumeral5}-C of ``{information bottleneck-enhanced MARL}"} \cr\hline

        Limited Interpretability
        & \makecell[l]{$\bullet$ Section \uppercase\expandafter{\romannumeral5}-A of ``{graph-enhanced MARL}", i.e., hierarchical decoupling \\$\bullet$ Section \uppercase\expandafter{\romannumeral5}-D of ``{mirror learning-enhanced MARL}"} \cr\hline

        Insufficient Exploration
        & \makecell[l]{$\bullet$ Section \uppercase\expandafter{\romannumeral5}-C of ``{information bottleneck-enhanced MARL}"\\$\bullet$ Section \uppercase\expandafter{\romannumeral5}-D of ``{mirror learning-enhanced MARL}", i.e., transformer} \cr\hline
    \Xhline{1.2pt}
    \end{tabular}
  \vspace{0cm}
\end{table*}
\emph{a) Direct Methods:}
Direct methods in model-free MARL optimize individual policies without requiring explicit communication among agents. These methods, such as centralized critics and decentralized learning frameworks, enable agents to make decisions based on local observations and rewards \cite{[820]}. One of the key advantages is the ability to coordinate in environments where communication is impractical, making them suitable for various real-world applications. A foundational approach in this category is the multi-agent deep deterministic policy gradient (MADDPG), which extends the traditional deep deterministic policy gradient (DDPG) framework to the multi-agent setting \cite{[821]}. By utilizing a shared critic, MADDPG stabilizes the learning process, especially in environments with partial observability.

Value decomposition networks (VDN) and QMIX are value-based strategies that allow decentralized action execution, but capitalize on centralized training \cite{[822],[823]}. VDN separates the global Q-value function into per-agent Q-values, enabling coordination without central control, and QMIX builds on this approach by integrating individual Q-values via a monotonic function, ensuring global optimality. Counterfactual multi-agent (COMA) policy gradients use a counterfactual baseline to enhance credit assignment by more accurately linking rewards to agent actions \cite{[824]}. Similarly, QTRAN maintains a value decomposition that reflects the optimal policy, helping to develop cooperative tactics in complex settings \cite{[825]}.

Further advances include a multi-agent twin delayed deep deterministic policy gradient (MATD3), which builds on MADDPG by incorporating twin Q-functions and target networks to improve stability and convergence \cite{[826]}. This reduces overestimation bias and improves performance in high-dimensional settings. Independent proximal policy optimization (IPPO) represents a decentralized extension of the proximal policy optimization (PPO) algorithm, allowing independent optimization of each agent's policy, making it scalable for multi-agent large-scale settings \cite{[827]}. Overall, direct methods in model-free MARL provide robust frameworks for policy optimization, enabling agents to act optimally and collaborate effectively without explicit communication, making them central to MARL's continued development and application.

\emph{a) Communication Methods:}
Communication methods in model-free MARL focus on enabling agents to exchange information during the learning process, enhancing coordination and decision-making. By facilitating communication, agents can share observations, rewards, or state information, which is crucial for tasks that require cooperation or synchronization \cite{[828]}. Unlike direct methods, where agents operate largely independently, communication-based approaches address the limitations of local information, enabling more effective learning in complex dynamic environments.

One of the first communication-based algorithms is the communication neural net (CommNet), which enables agents to share their hidden states with others, fostering improved coordination and performance in cooperative tasks \cite{[829]}. Expanding on this, approaches such as reinforced inter-agent learning (RIAL) and differentiable inter-agent learning (DIAL) enable agents to exchange states, rewards, and environmental observations while adapting to the behaviors of other agents \cite{[830]}. Similarly, the attentional communication model (ATOC) integrates communication within the actor-critic framework to synchronize actions in long-term decision-making environments, ensuring better coordination in tasks \cite{[831]}.

To address multi-agent environments, advanced techniques, such as multi-agent proximal policy optimization (MAPPO), adapt the PPO framework to allow inter-agent communication during training while keeping execution decentralized \cite{[832]}. Multi-actor-attention-critic (MAAC) further refines coordination by enabling cooperative and competitive interactions \cite{[833]}. The individualized controlled continuous communication model (IC3Net) improves coordinated control by selectively sharing relevant data, enhancing efficiency in complex high-dimensional domains \cite{[834]}. Thus, communication strategies in model-free MARL empower agents to navigate the complexities of decentralized decision-making by enhancing synchronization and information exchange. These strategies are vital for addressing real-world multi-agent scenarios demanding intricate coordination.
\subsection{Analysis and Challenges}
In MARL algorithms, optimizing wireless distributed networks requires both model-based and model-free strategies. Model-based MARL, using pre-existing environmental data, improves sample efficiency and adaptability, effective in non-stationary contexts. Conversely, model-free MARL offers enhanced stability, scalability, and computational efficiency, favoring large, stationary networks where an explicit model may be impractical. Table \ref{tab:model_comparison} shows that the selection between these approaches is based on network needs such as dynamics, resource constraints, and objectives. Hybrid approaches that merge these strategies can effectively address challenges in wireless distributed networks.

Model-based MARL is particularly advantageous for dynamic distributed networks such as UAV swarms or adaptive traffic management systems \cite{[201]} due to its rapid responsiveness and exploration efficiency. In contrast, model-free MARL excels in static distributed networks, such as those that involve wireless communication or cloud RAN \cite{[559],[645]}, where scalability and stability are vital, highlighting the need to tailor MARL tactics to the operational features of the network. Hybrid methodologies are also emerging as effective in achieving a balance between stability, scalability, and adaptability in different contexts. Furthermore, the integration of model-based and model-free MARL in Dec-POMDPs and networked MDPs enhances strategic planning and adaptability, boosting sample efficiency and decision-making in data-limited settings.

\begin{figure*}[t]
\centering
    \includegraphics[scale=0.35]{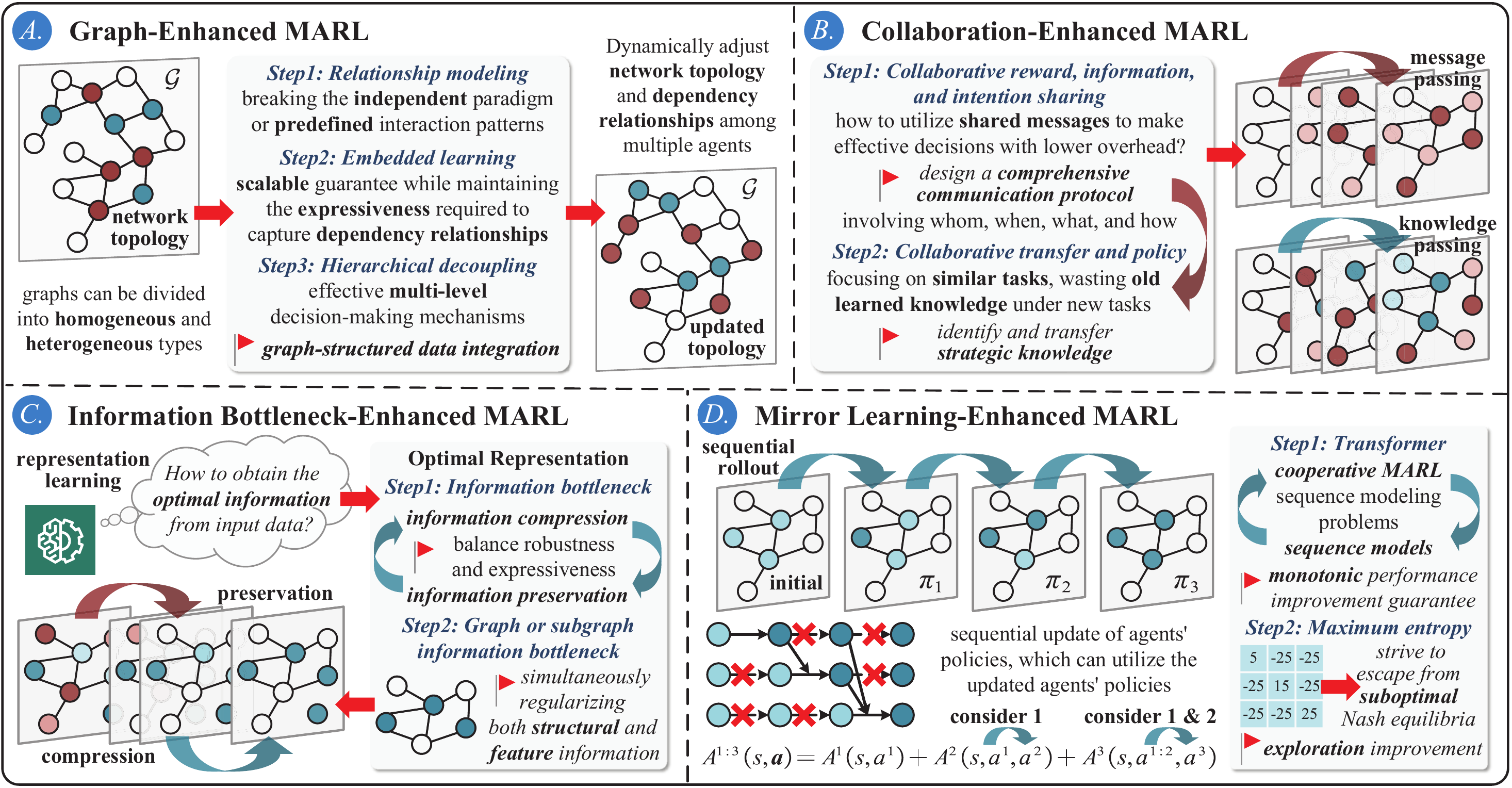}
    \caption{Four emerging techniques for enhancing MARL, including graph-enhanced, collaboration-enhanced, information bottleneck-enhanced, and mirror learning-enhanced. The principles behind each emerging technique are elaborated.
    \label{fig1}}
\end{figure*}
However, this integration poses challenges, particularly the computational overhead involved in managing both models and policies in wireless distributed networks, as shown in Table \uppercase\expandafter{\romannumeral6}. In dynamic settings, model accuracy is vital since errors can harm performance. Additionally, agents must harmonize their strategies with shared goals in multi-agent contexts to balance exploration and exploitation. Overcoming these hurdles is essential to progress the hybrid approaches. Consequently, the remaining challenges are as follows:
\begin{itemize}
\item \emph{Non-Stationarity:}
    The dynamic interplay among agents, where each agent's behavior continuously modifies the environment and affects the behaviors of others \cite{[550]}, poses significant challenges in prediction, often resulting in unstable learning and suboptimal performance.

\item \emph{Partial Observability:}
    Due to the limited information accessible to each agent, partial observability \cite{[559]} prevents agents from observing global information, thereby hindering optimal decision-making and effective coordination.

\item \emph{Poor Scalability:}
    Due to the increasing number of agents and the complexity of interactions in wireless distributed networks, the overhead of coordinating agents grows exponentially \cite{[209]}, making it difficult to effectively scale the network without affecting performance.

\item \emph{Poor Sample Efficiency:}
    It arises from the complexity of coordinating agents in dynamic environments, where high variance in learning outcomes and limited samples per agent, coupled with the need for extensive exploration over large-scale state-action spaces \cite{[407]}, hinder performance improvement within reasonable interactions.

\item \emph{Poor Robustness and Generalization:}
    Traditional MARL often overfit to specific environments or conditions, limiting their ability to generalize in dynamic or unknown settings, especially when faced with new challenges or disturbances \cite{[514],[515]}. This highlights the need for developing more robust and universal models.

\item \emph{Poor Communication Efficiency:}
    Due to constraints such as limited bandwidth or latency, agents find it difficult to share critical information in a timely and reliable manner \cite{[559],[560]}, often resulting in low communication efficiency and hindering agents' decision-making.

\item \emph{Slow Convergence Rate:}
    It significantly hampers responsiveness to rapidly changing environments, leading to high latency and delayed decision-making \cite{[564]}. This requires us to develop algorithms that converge quickly to ensure real-time performance in dynamic environments.

\item \emph{Sparse Information Expressiveness:}
    Due to the fact that agents are usually limited to local observation and difficult to obtain global information, this may lead to the integration of non critical information locally and make suboptimal decisions \cite{[505],[506]}, especially in environments that are prone to interference.

\item \emph{Limited Interpretability:}
    The complexity of multi-agent interactions and model non-linearity makes decision-making processes hard to analyze and explain \cite{[520],[521]}. This opacity poses challenges in understanding and validating, especially in applications where interpretability are essential for reliability and user confidence.

\item \emph{Insufficient Exploration:}
    The dynamic and decentralized nature of the environment, combined with limited coordination among agents, often leads to insufficient exploration \cite{[407]}. This hampers agents' ability to explore diverse states and develop effective strategies within high-dimensional state-action spaces, resulting in suboptimal performance and slower convergence.
\end{itemize}

In summary, traditional MARL algorithms, including both model-based and model-free, encounter significant challenges, including inefficiency in high-dimensional state-action spaces and slow convergence due to limited coordination and communication between agents. These issues become particularly pronounced in decentralized environments with minimal shared information. As environmental complexity increases, traditional methods often struggle to maintain performance. To overcome these limitations, enhanced frameworks have been proposed to better address the demands of complex distributed networks, which are explored in the following section ``\emph{enhanced multi-agent reinforcement learning}".
\section{Enhanced Multi-Agent Reinforcement Learning}
This section explores the limitations of traditional MARL algorithms in scalability, robustness, generalization, stability, and collaboration, which hinder their deployment in actual wireless distributed networks, as shown in Table \uppercase\expandafter{\romannumeral6}. To address these challenges, most recent works have introduced various innovative techniques to enhance MARL, such as ``\emph{graph-enhanced MARL}" in Section \uppercase\expandafter{\romannumeral4}-A, ``\emph{collaboration-enhanced MARL}" in Section \uppercase\expandafter{\romannumeral4}-B, ``\emph{information bottleneck-enhanced MARL}" in Section \uppercase\expandafter{\romannumeral4}-C, and ``\emph{mirror learning-enhanced MARL}" in Section \uppercase\expandafter{\romannumeral4}-D, as shown in Fig. 9. These emerging techniques provide new possibilities for further unleashing the power of MARL in wireless distributed networks.
\subsection{Graph-Enhanced MARL}
What we are familiar with is that the relationships among agents are often complex, such as collaboration or competition, which may leads to the inadequacy of adopting a single MARL to solve the challenges encountered in wireless distributed networks \cite{[574]}. Therefore, with the prosperous development of MARL, recently works are trying to identify connections between MARL and graphs \cite{[560]}, as diverse graph structures have a natural fit with wireless distributed networks, providing strong support for enhancing MARL.

Specifically, agents and their interactions in wireless distributed networks can be represented by the nodes $V$ and edges $E\subset\{(x,y)|x,y\in V\}$ of a graph \cite{[588]}.
By mapping the network topology and sparse communication patterns to graph structures, critical characteristics can be effectively captured.
For instance, graphs enable MARL to identify collaborative relationships among agents, enhancing learning efficiency and policy generalization, as shown in Fig. 9 (a).
This integration provides a powerful tool for analyzing and optimizing wireless distributed networks, which can be specifically divided into three types: relationship modeling \cite{[572]}, embedded learning \cite{[576]}, and hierarchical decoupling \cite{[583]}, each addressing specific challenges, as summarized in Table \uppercase\expandafter{\romannumeral7}.
\subsubsection{Optimizing MARL for Wireless Distributed Networks with Relationship Modeling}
Traditional MARL algorithms typically treat agents as independent entities or predefined interaction patterns, which fail to capture the complexity of wireless distributed networks \cite{[570],[571],[575]}. To address this, relationship modeling-based frameworks utilize graph-structured data $\mathcal{G}=(V,E,\mathcal{A})$ to explicitly model agent relationships \cite{[571],[575]}, where the adjacency set $\mathcal{A}=\{\mathbf{A}_{1},\ldots,\mathbf{A}_{N}\}$ contains adjacency matrices $\mathbf{A}_{n}$ representing communication links among agents. By treating agents as nodes and their interactions as edges, this helps agents better extract structural and time-varying information from dynamic relationship graphs to make informed decisions.
For example, the authors in \cite{[588]} proposed conceptualizing the communication architecture as a learnable graph, which dynamically learns interaction correlations, moving beyond predefined connections to enhance collaboration. This can better capture the spatiotemporal variations inherent in wireless distributed networks. Moreover, the authors in \cite{[560]} successfully applied dynamic relationship graphs to wireless distributed networks, utilizing GNNs to model dependencies such as data flow and information sharing, thereby enhancing coordination. The results show that relationship modeling achieves an EE improvement of 55.44\% higher than traditional collaborative algorithms, especially when there are a large number of UEs.

However, although relationship modeling-based frameworks can effectively capture dependencies and correlations among agents, they still face numerous significant challenges, such as sparse information expressiveness and excessive complexity. These limitations suggest that relationship modeling alone is insufficient for optimizing MARL in wireless distributed networks. By contrast, integrating embedded learning can complement relationship modeling by incorporating global information into the decision-making process, addressing these challenges more effectively, as shown in Table \uppercase\expandafter{\romannumeral6}.
\begin{table*}[t!]
  \centering
  \fontsize{8.5}{12}\selectfont
  \caption{Characteristics for different graph-enhanced MARL algorithms.}
  \label{CE}
   \begin{tabular}{ !{\vrule width1.2pt}  m{1.9 cm}<{\centering}
   !{\vrule width1.2pt}  m{0.65 cm}<{\centering}
   !{\vrule width1.2pt}  m{4  cm}<{\centering}
   !{\vrule width1.2pt}  m{3.5 cm}<{\centering}
   !{\vrule width1.2pt}  m{2.75 cm}<{\centering}
   !{\vrule width1.2pt}  m{2.5 cm}<{\centering} !{\vrule width1.2pt} }
    \Xhline{1.2pt}
        \rowcolor{gray!30} \bf Frameworks  &  \bf Ref. &  \bf Classifications & \bf Algorithm Applications & \bf Pros & \bf Cons \cr
    \Xhline{1.2pt}
        \multirow{6}{*}{\bf \shortstack{Relationship\\ Modeling}}
                    & \cite{[572]}  & Graph Attention (Mean Field)   & Real-World Metropolitan
                    & \multirow{6}{*}{\makecell[l]{$\bullet$ Effectively capture\\ complex dependencies,\\interactive relationships,\\ and correlations    \\$\bullet$ Enhance coordination\\}}
                    & \multirow{6}{*}{\makecell[l]{$\bullet$ Sparse information\\ expressiveness \\$\bullet$ Difficult to model\\ dynamic or unknown\\ relationships}} \\
        \cline{2-4} & \cite{[570]}  & Relational Graph Modeling   & Pommerman  & & \\
        \cline{2-4} & \cite{[571]}  & Communication Graph Modeling  & Predator-Prey \& StarCraft \uppercase\expandafter{\romannumeral2}   & &\\
        \cline{2-4} & \cite{[580]}  & Sequence Graph Modeling   &  StarCraft \uppercase\expandafter{\romannumeral2}  & & \\
        \cline{2-4} & \cite{[573]}  & Knowledge Graph Reasoning   & WN18RR \& NELL995   & & \\
        \cline{2-4} & \cite{[581]}  & Relational Graph Reasoning   & MPE Games  & & \cr\Xhline{1.2pt}

        \multirow{5}{*}{\bf \shortstack{Embedded\\ Learning}}
                    & \cite{[576]}  & Graph Topology Embedded & Virtual Reality (URLLC)
                    & \multirow{5}{*}{\makecell[l]{$\bullet$ Enhance information\\ expressiveness \\$\bullet$ Enhance scalability \\$\bullet$ Enhance adaptability}}
                    & \multirow{5}{*}{\makecell[l]{$\bullet$ Poor robustness \\$\bullet$ Poor stability \\$\bullet$ Difficult to\\ suppress interference}} \\
        \cline{2-4} & \cite{[589]}  & GCN-Based Embedded   &  IEEE-39 Bus Systems  & &\\
        \cline{2-4} & \cite{[577]}  & Joint-Action Embedded   & StarCraft \uppercase\expandafter{\romannumeral2} \& MaMuJoCo  & & \\
        \cline{2-4} & \cite{[579]}  & Satae-Augmentation Embedded  & StarCraft \uppercase\expandafter{\romannumeral2} \& MPE   & &\\
        \cline{2-4} & \cite{[582]}  & Virtual Network Embedded  &  Didactic \& ISP Networks  & &\cr\Xhline{1.2pt}

        \multirow{5}{*}{\bf \shortstack{Hierarchical\\ Decoupling}}
                    & \cite{[583]}  & Subgoal-based Hierarchical  & Mini-Grid \& Trash-Grid
                    & \multirow{5}{*}{\makecell[l]{$\bullet$ Enhance network \\interpretability \\$\bullet$ Enhance stability\\$\bullet$ Reduce computation\\ by decoupling tasks}}
                    & \multirow{5}{*}{\makecell[l]{$\bullet$ Risk of suboptimal\\ coordination across\\multiple levels\\ $\bullet$ Limited flexibility}} \\
        \cline{2-4} & \cite{[585]}  & Attention-based Hierarchical  & Predator-Prey \& Cooperative & & \\
        \cline{2-4} & \cite{[586]}  & Multi-Task Hierarchical   & Parking Navigation & &  \\
        \cline{2-4} & \cite{[584],[587]}  & Policy Hierarchical   & Power-Gas Communication \& Transportation Networks  & &\cr\Xhline{1.2pt}

    \end{tabular}
  \vspace{0cm}
\end{table*}
\subsubsection{Optimizing MARL for Wireless Distributed Networks with Embedded Learning}
Building on the foundation of relationship modeling, embedded learning can effectively address the above challenges by leveraging latent representations to integrate both local and global information \cite{[576]}. Specifically, this framework embeds all agents and their environments into a shared latent space, significantly reducing interaction overhead and achieving better generalization. This makes embedded learning scalable to larger wireless distributed networks while maintaining the expressiveness required to capture complex dependencies. For example, by using contrastive learning \cite{[579]}, it enables agents to learn compact but sufficient representations of their observed information, thereby achieving efficient and adaptive decision-making.

\begin{table*}[t!]
  \centering
  \fontsize{8.5}{12}\selectfont
  \caption{Characteristics for different collaboration-enhanced MARL algorithms.}
  \label{CE}
   \begin{tabular}{ !{\vrule width1.2pt}  m{1.7 cm}<{\centering}
   !{\vrule width1.2pt}  m{0.65 cm}<{\centering}
   !{\vrule width1.2pt}  m{3.875  cm}<{\centering}
   !{\vrule width1.2pt}  m{3.895 cm}<{\centering}
   !{\vrule width1.2pt}  m{2.59 cm}<{\centering}
   !{\vrule width1.2pt}  m{2.59 cm}<{\centering} !{\vrule width1.2pt} }
    \Xhline{1.2pt}
        \rowcolor{gray!30} \bf Frameworks  &  \bf Ref. &  \bf Classifications  & \bf Algorithm Applications & \bf Pros & \bf Cons \cr
    \Xhline{1.2pt}
        \multirow{4}{*}{\bf \shortstack{Collaborative\\ Policy}}
                    & \cite{[546]}  & Unified Multi-Task Policy   & Multi-Task Learning
                    & \multirow{4}{*}{\makecell[l]{$\bullet$ Enhance coordinated\\ decision-making \\$\bullet$ Enhance robustness}}
                    & \multirow{4}{*}{\makecell[l]{$\bullet$ High complexity \\$\bullet$ Poor communication\\ efficiency}} \\
        \cline{2-4} & \cite{[550]}  & Robust Collaborative Policy   & StarCraft \uppercase\expandafter{\romannumeral2} & &  \\
        \cline{2-4} & \cite{[557]}  & Robust Collaborative Policy   & Channel Coding-Modulation  & &  \\
        \cline{2-4} & \cite{[558]}  & Communication-Efficient Policy  & Battle \& Cooperative Spread   & &  \cr\Xhline{1.2pt}

        \multirow{7}{*}{\bf \shortstack{Collaborative\\ Information\\ Sharing}}
                    & \cite{[549]}  & Focus only on When   &  Predator-Prey \& Traffic Junction
                    & \multirow{7}{*}{\makecell[l]{$\bullet$ Effectively promote\\ knowledge sharing\\ among agents \\$\bullet$ High communication\\ efficiency \\$\bullet$ Fast convergence}}
                    & \multirow{7}{*}{\makecell[l]{$\bullet$ Potential for stale or\\ outdated information  \\$\bullet$ Security and privacy\\ risks with shared\\ information\\$\bullet$ Poor transferability}}\\
        \cline{2-4} & \cite{[551]}  & Focus only on How  & Predator-Prey \& Cooperative  & &  \\
        \cline{2-4} & \cite{[555]}  & Both Whom and When  & Predator-Prey \& Traffic Junction  & &  \\
        \cline{2-4} & \cite{[552]}  & Both Whom and What   & Traffic Junction \& Mixed  & &  \\
        \cline{2-4} & \cite{[567]}  & Whom, When, What, and How  & Cooperative \& Mixed  & &  \\
        \cline{2-4} & \cite{[559],[560]}  & Whom, When, What, and How & Wireless Communications  & &   \cr\Xhline{1.2pt}

        \multirow{4}{*}{\bf \shortstack{Collaborative\\ Transfer\\ Learning}}
                    & \cite{[566]}  & Equilibrium Transfer  & Grid-World \& Wall \& Soccer
                    & \multirow{4}{*}{\makecell[l]{$\bullet$ Accelerates learning\\ by transferring learned\\knowledge \\$\bullet$ Enhance robustness}}
                    & \multirow{4}{*}{\makecell[l]{$\bullet$ Overreliance on task\\ similarity\\ $\bullet$ Poor stability\\ $\bullet$ Transfer risk}}\\
        \cline{2-4} & \cite{[563]}  & Policy Transfer   &  Pac-Man \& MPE  & &  \\
        \cline{2-4} & \cite{[564]}  & Lateral Transfer   & MPE  & &   \\
        \cline{2-4} & \cite{[565]}  & Knowledge Transfer   & StarCraft \uppercase\expandafter{\romannumeral2}  & &   \cr\Xhline{1.2pt}

    \end{tabular}
  \vspace{0cm}
\end{table*}
Moreover, embedded learning enhances adaptability in dynamic environments by continuously updating latent representations during training to reflect evolving network states. This is particularly valuable in wireless distributed networks with frequently changing conditions, such as power grid communication networks \cite{[589]} and mobile communication \cite{[582]}. Meanwhile, embedded learning also enables implicit communication among agents, as the shared latent space encodes relevant information without the need for explicit message passing. Similarly, embedded learning-based frameworks also introduce significant challenges, such as noise interference and poor robustness. To overcome these challenges, hierarchical decoupling provides a structured approach for managing multi-level dependencies, thereby enhancing MARL's ability to tackle complex problems in wireless distributed networks.
\subsubsection{Optimizing MARL for Wireless Distributed Networks with Hierarchical Decoupling}
Wireless distributed networks typically contain hierarchical structures, such as local and global decision layers, which support hierarchical decoupling by introducing multi-level decision-making mechanisms to enhance MARL. Specifically, hierarchical decoupling decomposes the optimization problem into multiple subtasks \cite{[583],[587]}, where agents align with higher-level objectives while optimizing local tasks. This simplifies the decision-making process, improving interpretability and cross-level coordination. For instance, the authors in \cite{[584]} proposed a two-layer hierarchical MARL algorithm to address uncertainties and extreme events. In this framework, high-level actions guide switching decisions between power-gas communication and transportation networks, while low-level actions are constructed by MARL to compute the routing and repairing decisions of repair crews.
The obtained results indicate that hierarchical decoupling can effectively capture dynamic characteristics, with a performance improvement of 26.29\% compared to traditional MARL algorithms.

Nevertheless, implementing hierarchical decoupling also presents challenges, such as designing effective hierarchical structures and decomposing rewards across levels, which require domain-specific knowledge that is often difficult to obtain in actual wireless distributed networks. Additionally, the increased complexity of hierarchical models can lead to higher computational demands during the training phase.

\emph{\textbf{Lessons Learned:}} By integrating relationship modeling, embedded learning, and hierarchical decoupling into graph-enhanced MARL, their synergistic fusion enables scalable and globally aligned solutions for wireless distributed networks, facilitating more effective deployment in actual scenarios.
\subsection{Collaboration-Enhanced MARL}
Due to constraints such as limited bandwidth, low latency, and high capacity in wireless distributed networks \cite{[540]}, establishing efficient communication and transfer frameworks is crucial \cite{[542],[543]}, as demonstrated in Fig. 9 (b). These frameworks can be specifically divided into five types: collaborative policy \cite{[557],[558]}, information sharing \cite{[559],[560]}, reward \cite{[561]}, transfer learning \cite{[563],[564],[565]}, and intention sharing \cite{[556]}, as shown in Table \uppercase\expandafter{\romannumeral8}. They enable the design of enhanced-MARL for distributed networks, addressing previous challenges from different perspectives.

However, due to environmental complexity, information asymmetry, and self-interest, agents often focus on maximizing their own rewards, leading to decoupled rather than collaborative actions. To overcome this, agents must learn to balance exploration and exploitation while ensuring their actions complement those of others. Then, we mainly discuss other three frameworks in the following part, including policy, information sharing, and transfer learning.
\subsubsection{Optimizing MARL for Wireless Distributed Networks with Collaborative Policy}
An important idea for optimizing MARL is collaborative policies, where agents not only optimize their individual behavior but also improve collective performance. Unlike traditional MARL, where each agent $i$ learn separate policies $\pi_{i,j}$ for different tasks $j \in \mathcal{J}$, collaborative policies allow agents to share insights and learn a unified policy $\pi_{i}$. This is particularly useful in actual distributed networks with inconsistent or multiple tasks. For instance, the authors in \cite{[546]} introduced a frameworks for distilling multiple single-task policies $\pi_{i,j}$ into a unified policy $\pi_{i}$ that performs well across multiple related tasks, enhancing coordination and decision-making without explicitly distinguishing tasks.

Moreover, collaborative policies can also effectively address the challenge of noisy interference in wireless distributed networks. For example, the authors in \cite{[550],[557]} proposes effective frameworks for joint learning and communication to guide the design of collaboration-enhanced MARL.
Specifically, the first framework mitigates noisy components by controlling message variance during training, ensuring that agents retain only useful information. And the second framework incorporates noisy communication channels into the environment model, encouraging agents to adapt to and effectively communicate in such conditions. Both frameworks emphasize collaborative policies to enhance performance in noisy environments while maintaining robustness, providing valuable insights for improving wireless network performance.

However, implementing collaborative policies in wireless distributed networks presents significant challenges.
As the number of agents grows, the complexity of collaboration across all agents increases exponentially, leading to longer training times and difficulties in convergence. Additionally, excessive global communication can result in interaction delays and bandwidth constraints, negatively affecting communication efficiency. To address these challenges, collaborative information sharing-based frameworks are essential in wireless distributed networks, focusing on the \emph{whom}, \emph{when}, \emph{what}, and \emph{how} of information exchange to optimize communication while maintaining performance \cite{[559],[560]}.
\subsubsection{Optimizing MARL for Wireless Distributed Networks with Collaborative Information Sharing}
Unlike collaborative policy-based frameworks, which focus on aligning agents' behaviors with overall optimization goals, collaborative information sharing \cite{[549],[551],[555],[552],[559],[560]} emphasizes efficient decision-making through strategic sharing of information while minimizing communication overhead.  This encourages us to develop a systematic communication protocol for collaboration-enhanced MARL from multiple perspectives, including \emph{whom}, \emph{when}, \emph{what}, and \emph{how}, as follows:
\begin{itemize}
\item \emph{Whom:} Each agent needs to decide with whom to communicate \cite{[555],[552],[559],[560]}, including 1) \emph{broadcasting} information to all agents, such as public; 2) \emph{multicasting} information to selected agents, such as those with shorter distances or similar features; and 3) \emph{unicasting} information directly to a special agent, such as private.

\item \emph{When:} Each agent needs to decide whether to \emph{transmit} shared information or \emph{skip} communications \cite{[549],[555],[559],[560]}, which can be predetermined before the training phase or obtained through learning.

\item \emph{What:} Each agent needs to decide what information to share \cite{[552],[559],[560]}, including 1) \emph{existing} information, such as their past observations and actions; and 2) \emph{predicted} information, such as their intended states.

\item \emph{How:} Each agent needs to decide how to transmit or receive shared information \cite{[551],[559],[560]}, including 1) utilizing compression techniques to effectively \emph{transmit information}, and 2) utilizing attention mechanisms to determine the priority of \emph{receiving information}.
\end{itemize}

\begin{figure}[t]
\centering
    \includegraphics[scale=0.5]{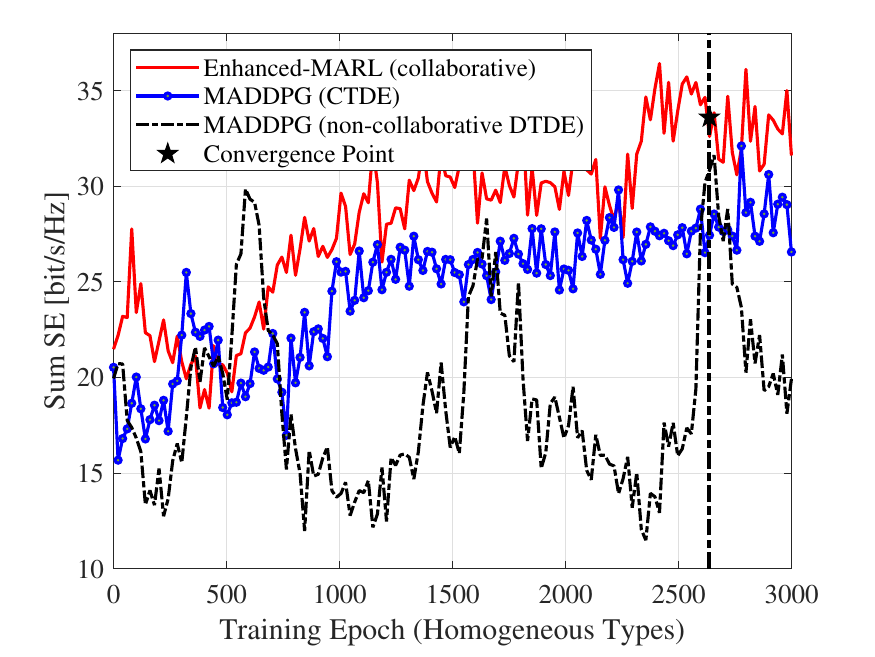}
    \caption{Test reward (sum SE) curves of homogeneous distributed networks under different mechanisms. Please refer to \cite{[560]} for more details.
    \label{fig1}}
\end{figure}
\begin{figure}[t]
\centering
    \includegraphics[scale=0.5]{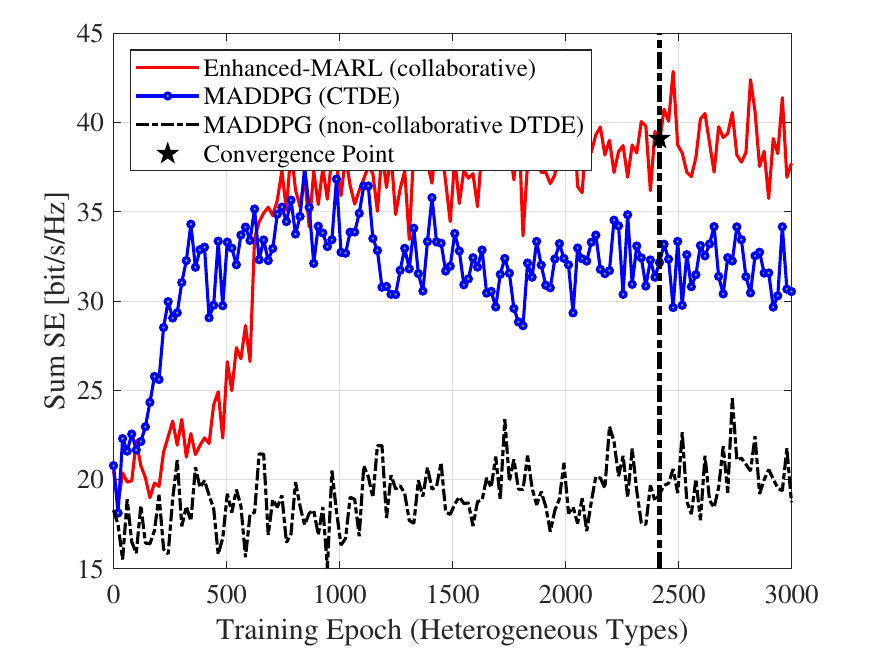}
    \caption{Test reward (sum SE) curves of heterogeneous distributed networks under different mechanisms. Please refer to \cite{[560]} for more details.
    \label{fig1}}
\end{figure}
Specifically, Fig. 10 and Fig. 11 illustrate the test reward curves for homogeneous and heterogeneous distributed networks \cite{[560]}. The obtained results highlight that designing systematic communication protocols to foster collaboration significantly enhance network performance, e.g., achieving SE improvements for 19.39\% and 20.46\% in homogeneous and heterogeneous networks, respectively, compared to traditional CTDE paradigms. This underscores the importance of effective collaborative information sharing.
However, non-stationarity caused by continuous environmental changes limits further performance improvement, which requires emerging techniques to address, such as integrating transfer learning to help agents adapt to policy and environmental shifts.
\subsubsection{Optimizing MARL for Wireless Distributed Networks with Collaborative Transfer Learning}
Collaborative transfer learning-based frameworks focus on identifying and sharing strategic knowledge that is universal across multiple agents, such as environmental structures or collaborative policies \cite{[563],[564],[565]}. These frameworks allow agents to reuse previously learned knowledge in different tasks without having to start learning from scratch. This not only accelerates the learning process but also improves the generalization ability of the learned strategies to address non-stationarity, making them better suited for actual distributed networks.

As shown in Table \uppercase\expandafter{\romannumeral8}, collaborative transfer learning-based frameworks have been applied to various multi-task or cross-task learning \cite{[563],[564],[565]}. For example, the authors in \cite{[564]} introduced a novel multi-task approach with knowledge transfer in collaboration-enhanced MARL, shifting learning from distillation to environmental rewards. The obtained results reveal that this approach contributes to multi-task learning surpassing specific single-task learning, highlighting their generalization. Moreover, the authors in \cite{[565]} integrated an attention module to improve collaborative transfer learning, enabling knowledge transfer among heterogeneous agents and avoiding negative transfer when tasks differ, demonstrating the designed frameworks' flexibility.

\emph{\textbf{Lessons Learned:}} By effectively integrating various collaborative frameworks into distributed networks, a better balance between superior performance and low overhead can be achieved, overcoming the challenges commonly encountered in traditional MARL algorithms, as illustrated in Table \uppercase\expandafter{\romannumeral6}.

\begin{table*}[t!]
  \centering
  \fontsize{8.5}{12}\selectfont
  \caption{Characteristics for different information bottleneck-enhanced MARL algorithms.}
  \label{CE}
   \begin{tabular}{ !{\vrule width1.2pt}  m{2.05 cm}<{\centering}
   !{\vrule width1.2pt}  m{0.65 cm}<{\centering}
   !{\vrule width1.2pt}  m{3.7  cm}<{\centering}
   !{\vrule width1.2pt}  m{3.9 cm}<{\centering}
   !{\vrule width1.2pt}  m{2.5 cm}<{\centering}
   !{\vrule width1.2pt}  m{2.5 cm}<{\centering}   !{\vrule width1.2pt} }

    \Xhline{1.2pt}
        \rowcolor{gray!30} \bf Frameworks  &  \bf Ref. &  \bf Classifications  & \bf Algorithm Applications & \bf Pros & \bf Cons \cr
    \Xhline{1.2pt}
        \multirow{4}{*}{\bf \shortstack{Basic\\ Information \\ Bottleneck}}
                    & \cite{[505]} & Universal IB & Representation Learning
                    & \multirow{4}{*}{\makecell[l]{$\bullet$ Balance robustness\\ and expressiveness \\$\bullet$ Avoid overfitting \\of learning models}}
                    & \multirow{4}{*}{\makecell[l]{$\bullet$ Unable to access\\ graph-structured data \\$\bullet$ Poor applicability\\ and scalability}} \\
        \cline{2-4} & \cite{[506]}  & Variational IB & Representation Learning & &  \\
        \cline{2-4} & \cite{[507]}  & Task-Oriented IB & Classification \& Clustering & &   \\
        \cline{2-4} & \cite{[508]}  & Self-Supervised IB & Deep Multi-View Learning  & &  \cr\Xhline{1.2pt}

        \multirow{8}{*}{\bf \shortstack{Graph\\ Information \\ Bottleneck}}
                    & \cite{[509]}  & Universal GIB  & Classification \& Clustering
                    & \multirow{8}{*}{\makecell[l]{$\bullet$ Effectively integrate \\graph-structured data\\$\bullet$ Enhance robustness \\in complex dynamic \\environments\\$\bullet$  Effectively suppress \\information loss}}
                    & \multirow{8}{*}{\makecell[l]{$\bullet$ Overreliance on \\global graphs, which \\is difficult to obtain\\ in practice \\$\bullet$ Poor interpretability \\of predicted or \\output results}} \\
        \cline{2-4} & \cite{[510]}  & Variational GIB  & Classification \& Denoising & &  \\
        \cline{2-4} & \cite{[517]}  & Cross-Channel GIB & Classification \& Clustering  & &  \\
        \cline{2-4} & \cite{[519]}  & Self-Supervised GIB  & Deep Multi-View Learning  & &  \\
        \cline{2-4} & \cite{[512]}  & Dynamic GIB  & Link Prediction & &  \\
        \cline{2-4} & \cite{[514]}  & Robust GIB  & Classification \& Link Prediction & &  \\
        \cline{2-4} & \cite{[515]}  & Robust Edge-GIB & Signal Processing & &  \\
        \cline{2-4} & \cite{[511]}  & Communication-Efficient GIB & StarCraft \uppercase\expandafter{\romannumeral2} Games & &  \cr\Xhline{1.2pt}

        \multirow{6}{*}{\bf \shortstack{Subgraph\\ Information \\ Bottleneck}}
                    &  \cite{[520]}  & Universal SubIB & Classification \& Denoising
                    & \multirow{6}{*}{\makecell[l]{$\bullet$ Effectively remove \\noise and redundancy \\from global graphs \\$\bullet$ Significantly reduce \\overhead and latency}}
                    & \multirow{6}{*}{\makecell[l]{$\bullet$ Performance and \\convergence are \\ sensitive to the \\ selection of critical\\ subgraphs}} \\
        \cline{2-4} & \cite{[521]}  & Variational SubIB & Classification \& Clustering & &  \\
        \cline{2-4} & \cite{[524]}  & Conditional GIB & Interaction Prediction & &  \\
        \cline{2-4} & \cite{[525]}  & Heterogeneous GIB & Classification \& Clustering & &  \\
        \cline{2-4} & \cite{[526]}  & Self-Supervised SubIB &  Classification \& Clustering & &  \\
        \cline{2-4} & \cite{[523]}  & Adaptive SubIB & Classification \& Denoising & &  \cr\Xhline{1.2pt}
    \end{tabular}
  \vspace{0cm}
\end{table*}
\begin{figure*}[t]
\centering
    \includegraphics[scale=0.35]{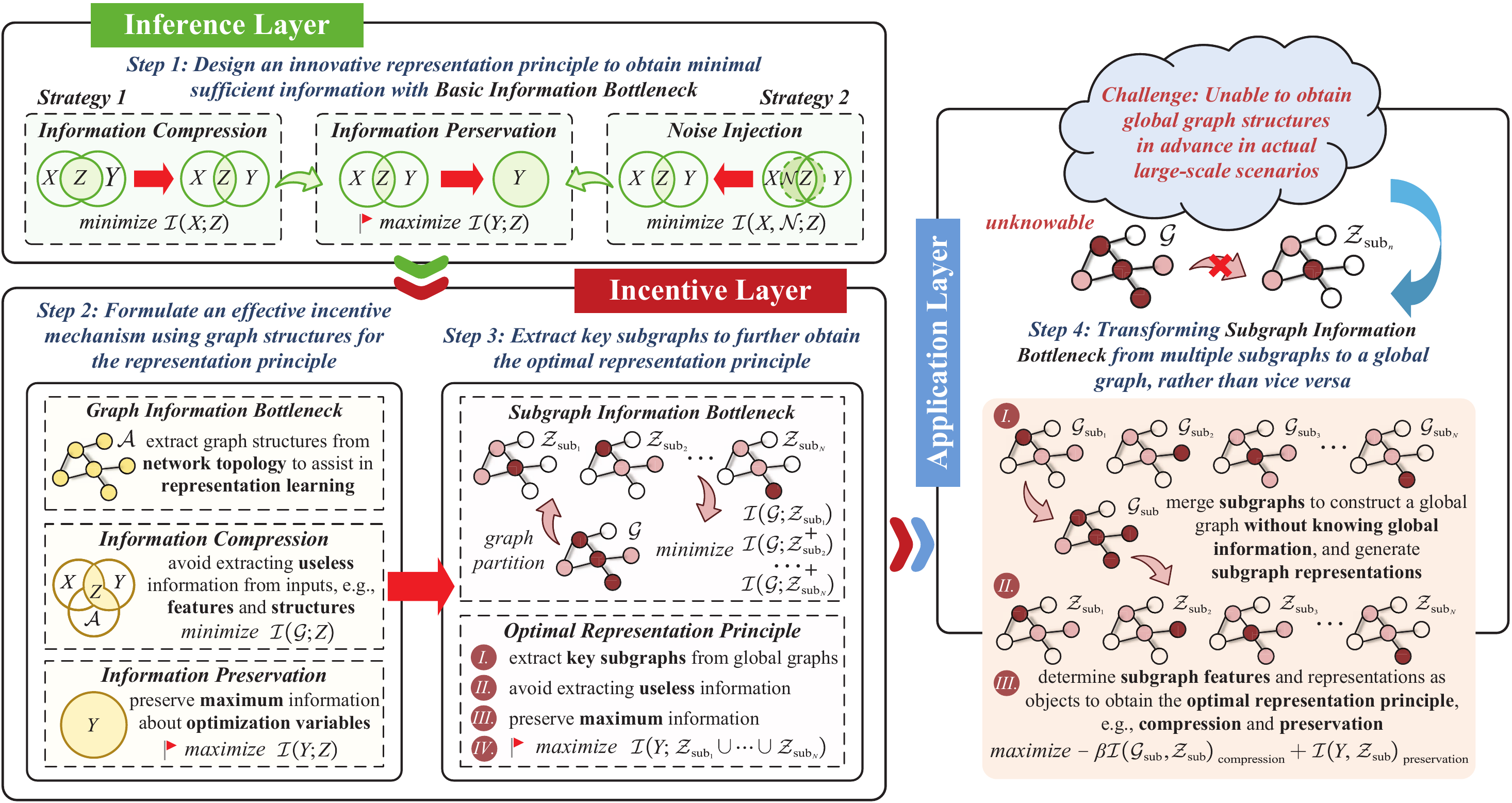}
    \caption{Three frameworks of information bottleneck-enhanced MARL, including basic IB, GIB, and SubIB, all reflect an innovative representation principle of learning minimal sufficient information. Please refer to \cite{[505]}, \cite{[509]}, \cite{[520]}, and \cite{[523]} for more details.
    \label{fig1}}
\end{figure*}
\begin{table}[t]
  \centering
\fontsize{8.5}{11}\selectfont
\caption{Comparison between RD theory and various IB theories.}
  \label{paper}
\begin{tabular}{ccccc}
\toprule
\bf Theories &  \bf Inputs &  \bf Outputs\\
\midrule
RD &  Data $X$ & Representations $Z$ & \\
Basic IB  &  Data $X$ and  $Y$ & Representations $Z$  &\\
GIB   &  Data $\mathcal{G}=(X,\mathcal{A})$ and  $Y$ & Representations $Z$ &\\
SubIB  &  Data $\mathcal{G}=(X,\mathcal{A})$ and  $Y$ & Representations $Z_\text{sub}$ & \\
\bottomrule
\end{tabular}
\end{table}
\subsection{Information Bottleneck-Enhanced MARL}
This subsection mainly utilizes the innovative IB theory to provide an innovative principle for optimizing traditional MARL algorithms, which is to transform the original representation (e.g., hidden or communication representations) into one that contains minimal sufficient information, while greatly avoiding the aggregation of irrelevant information \cite{[501],[502],[503]}, as shown in Fig. 9 (c). Therefore, we first shortly introduce the concept of mutual information, which represents measuring the information contained by one random variable $X$ about another $Y$ and lays an important foundation for the development of IB theory \cite{[531]}, satisfying
\begin{equation}
\setcounter{equation}{10}
\mathcal{I}(X;Y)=\mathbb{E}_{p(x,y)}\left[\text{\rm log}\frac{p(x)p(y)}{p(x,y)}\right],
\end{equation}
where $p(x)$ and $p(y)$ are the marginal distributions of variables $X$ and $Y$, respectively, and $p(x,y)$ denotes the joint probability distribution of variables $X$ and $Y$,

Moreover, the definition of mutual information also can be divided into two categories for different application conditions: discrete and continuous, satisfying
\begin{subequations}
\begin{align}
\mathcal{I}(X;Y)&=\sum_{x,y}p(x,y)\text{\rm log}\frac{p(x)p(y)}{p(x,y)},\\
\mathcal{I}(X;Y)&=\int_{x,y}p(x,y)\text{\rm log}\frac{p(x)p(y)}{p(x,y)}dxdy,
\end{align}
\end{subequations}

Then, we briefly review the rate distortion (RD) theory, as the introduced IB theory belongs to a special case of it \cite{[500]}. Specifically, the RD theory provides a general framework for learning optimal information compression, including lossy compression and distortion assurance, as follows:
\begin{itemize}
\item \emph{Lossy Compression:} The former seeks the optimal compact representation $Z$ of input information $X$, so as not to lose too much information about $X$. An effective method to measure how much information about input value $x \in X$ is contained in its representation $z\in Z$ is to utilize the above mutual information, i.e., $\mathcal{I}(X;Z)$, where the compression can be formally defined by a mapping probability distribution $p(z|x)$, and lower $\mathcal{I}(X;Z)$ means more compact representation $Z$ about $X$.
\item \emph{Distortion Assurance:} The latter indicates that the distortion between the input value $x \in X$ and its compact representation $z\in Z$ satisfies the determined tolerance $D^*$ to ensure that $z$ can better extract critical information from $x$, i.e., the expected distortion $E(D)=\sum_{x,z}p(x)p(z|x)d(x,z)\leqslant D^*$, where the distortion function $d(x,z)$ represents how distorted $x$ and $z$ are.
\end{itemize}

Based on the above analysis, the RD theory can be formulated as follows:
\begin{equation}
\setcounter{equation}{12}
\mathcal{R}(D)=\min_{p(z|x): E(D)\leqslant D^*}\mathcal{I}(X;Z),
\end{equation}
which has been widely applied in various emerging fields such as data compression, feature selection, and pattern classification. However, the difficulty in obtaining the distortion function $d(x,z)$ and the unpredictability of the tolerance $D^*$ without any prior knowledge hinder the application of RD theory in real-world scenarios \cite{[500]}.

By contrast, the introduced IB theory can utilize a special loss function $\mathcal{L}_\text{IB}(\cdot)$ and the Lagrangian formulation of the constrained optimization to effectively overcome the aforementioned challenges. In the following part, we mainly discuss various typical IB frameworks and their improved variants, which can be classified into basic IB \cite{[505],[506],[507],[508]}, GIB \cite{[509],[510],[527]}, and SubIB \cite{[520],[521]}, and the corresponding characteristic comparison is shown in Table \uppercase\expandafter{\romannumeral9} and Table \uppercase\expandafter{\romannumeral10}. Moreover, we introduce how IB theories optimize MARL, focusing on various representations.
\subsubsection{Basic Information Bottleneck}
Unlike RD theory, which only focuses on obtaining a compact representation $Z$ from inputs $X$, basic IB theory introduces an additional auxiliary variable $Y$ to maximally preserve relevant information about it (e.g., optimization variables or labels) when obtaining representations $Z$ \cite{[505],[506]}. Specifically, IB theory utilizes two types of mutual information $\mathcal{I}(X;Z)$ and $\mathcal{I}(Y;Z)$ to establish relationships between $Z$ and related variables $X$ and $Y$, with the goal of ensuring the optimal representation $Z$ to provide maximum information about $Y$ and avoid obtaining irrelevant information from $X$, as shown in Fig. 12 (a) (\emph{step 1}). Then, basic IB theory can be formulated as
\begin{equation}
\setcounter{equation}{13}
\min_{p(z|x)\in \Omega}\mathcal{L}_\text{IB}(X,Y;Z) = -\mathcal{I}(Y;Z) + \beta \mathcal{I}(X;Z),
\end{equation}
where $\Omega$ is the space of the conditional distribution $p(z|x)$ of $z \in Z$ given $x \in X$, and $\beta$ is the trade-off between preservation and compression. This innovative principle enables the learned model to naturally avoid overfitting and effectively balance robustness and expressiveness, as follows:

\begin{itemize}
\item \emph{Robustness:} $\mathcal{I}(X;Z)$ denotes the compression from inputs $X$ to representations $Z$, and a lower $\mathcal{I}(X;Z)$ indicates less dependence of $Z$ on $X$, which reduces the impact of inaccurate $X$ in interference environments.
\item \emph{Expressiveness:} $\mathcal{I}(Y;Z)$ denotes how much information about variables $Y$ is preserved by representations $Z$, and a higher $\mathcal{I}(Y;Z)$ indicates that $Z$ can better express relevant information about $Y$, which is beneficial for solving various optimization problems.
\end{itemize}

As shown in Table \uppercase\expandafter{\romannumeral10}, basic IB theory has been applied to various networks to optimize representations \cite{[505],[506],[507],[508]}. Typical works include \cite{[506]}, where a variational approximation of traditional IB was developed, enabling networks to parameterize IB models and using the reparameterization trick for efficient training. The obtained results demonstrate superior generalization and robustness of this variational IB framework. Additionally, the authors in \cite{[507]} applied this variational IB framework to graph classification, examining the rate-distortion trade-off under both static and dynamic channel conditions. An important characteristic of IB frameworks is their focus on learning optimal representations $Z$ using only inputs, which captures the minimal sufficient information.

However, extending basic IB theory to graph-structured data presents two major challenges: the interdependence and discreteness of graph-structured data \cite{[509]}. These issues complicate model training under the IB theory and hinder its application to wireless distributed networks, which are inherently graph-structured. Therefore, extracting the minimal sufficient information from such data is a key challenge when applying IB theory to wireless distributed networks.
\subsubsection{Graph Information Bottleneck}
Unlike basic IB theory, which handles continuous data with independent and identically distributed (i.i.d.) characteristics, GIB theory is designed to extract relevant information from graph-structured data $\mathcal{G}=(X,\mathcal{A})$ with local dependencies. It aims to learn underlying data patterns by exploring correlations among graph structures $\mathcal{A}$, as shown in Fig. 12 (b) (\emph{step 2}). Specifically, GIB theory retains the core principle of minimal sufficient information from basic IB \cite{[509],[510]} and extends it to graph-structured data. Correspondingly, GIB theory can be formulated as:
\begin{equation}
\setcounter{equation}{14}
\min_{p(z|x,\mathcal{A})\in \Omega_\text{G}}\mathcal{L}_\text{GIB}(\mathcal{G},Y;Z) = -\mathcal{I}(Y;Z) + \beta \mathcal{I}(\mathcal{G};Z),
\end{equation}
where $\Omega_\text{G}$ is the space of the conditional distribution $p(z|x,\mathcal{A})$ of $z \in Z$ given $x \in X$ and additional graph structures $\mathcal{A}$.

Recent years have witnessed the successful integration of GIB theory, rather than traditional limited IB theory, in various networks \cite{[509],[510],[512],[517],[518],[519],[513]}, as summarized in Table \uppercase\expandafter{\romannumeral10}.
For example, the authors in \cite{[517],[518],[519]} improved existing GIB frameworks to better suit applications such as graph classification, clustering, and deep multi-view learning. While learning mechanisms vary from supervised to self-supervised, these frameworks adhere to the unified principle of minimal sufficient information, ensuring that only relevant information is retained and preventing the aggregation of irrelevant information. Moreover, considering that previous works have focused on static graphs, which are contrary to actual environments carrying complex spatiotemporal features. This prompts the authors in \cite{[512]} to combine dynamic graphs with GIB theory to address this challenge, thereby enhancing robustness in complex dynamic environments.

\begin{figure*}[t]
\centering
    \hspace{-0.28cm}
    \includegraphics[scale=0.57]{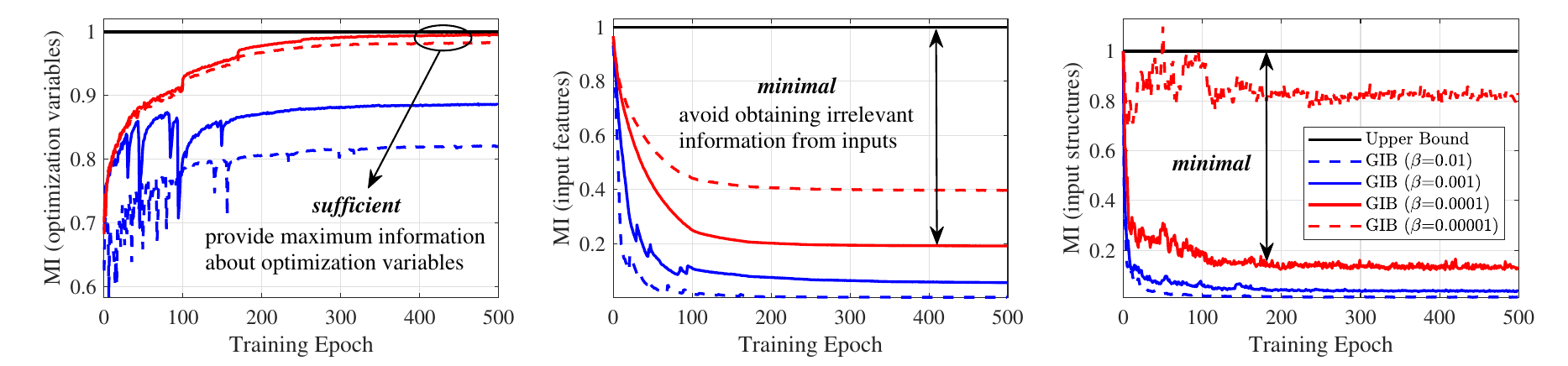}
    \caption{Mutual information curves of MARL-assisted wireless distributed networks under the typical GIB theory, including $\mathcal{I}(Y;Z)$ between representations and optimization variables, $\mathcal{I}(Y;X)$ between representation and input features, and $\mathcal{I}(Y;\mathcal{A})$ between representation and input structures, where $\mathcal{I}(Y;X)$ and $\mathcal{I}(Y;\mathcal{A})$ are decomposed from $\mathcal{I}(Y;\mathcal{G})$. Please refer to \cite{[515]} for more details.
    \label{fig1}}
\end{figure*}
However, the aforementioned works primarily focus on selecting single images or vertices as update objects, leading to significant information loss and weak expressiveness due to dimensional compression. This limitation has motivated the exploration of updating edges in graph structures, which has proven effective in downstream tasks like classification \cite{[514]} and signal processing \cite{[515]}. Specifically, the authors in \cite{[514]} transformed traditional objects into edges, while the authors in \cite{[515]} expanded this concept, further transforming objects from single-source edges in homogeneous graphs (i.e., two-dimensional) to multi-source hyper-edges in heterogeneous graphs (i.e., multidimensional). The obtained results indicate that updating representations through edges or hyper-edges can effectively suppress information loss. As shown in Fig. 13, applying GIB theory helps avoid the aggregation of irrelevant information (\emph{robustness}) while achieving performance close to the centralized upper bound (\emph{expressiveness}), thereby ensuring their applicability in disturbed environments. For example, the proposed robust Edge-GIB \cite{[515]} approach the performance upper bound with an error tolerance of nearly 72.23\% in $\mathcal{I}(Y;X)$ and 87.29\% in $\mathcal{I}(Y;\mathcal{A})$ for data transmission to better cope with random interference noise.

Moreover, note that as the dimensionality of updated objects increases (e.g., two-dimensional, three-dimensional, and multidimensional), the expressiveness of hidden representations also improves, which can better optimize downstream tasks.
\subsubsection{Subgraph Information Bottleneck}
Both basic IB and GIB theories focus on learning global hidden representations $Z$, while noise and redundancy in graph-structured data $\mathcal{G}$ hinder effective representation learning \cite{[520],[521]}. On the other hand, they also lack interpretability of predicted or output results $Y$. Therefore, identifying predictable but compressed subgraphs to remove redundancy and obtain interpretable parts of the global graph is an important solution. Unlike traditional IB and GIB theories, emerging SubIB frameworks focus on learning critical subgraph hidden representations $Z_{\text{sub}}$, which effectively removes noise and redundancy from the original graphs $\mathcal{G}$, as shown in Fig. 12 (c) (\emph{step 3}). This innovative SubIB theory can be formally defined as
\begin{equation}
\setcounter{equation}{15}
\min_{p(z|x,\mathcal{A})\in \Omega_\text{G}}\!\!\!\!\!\!\mathcal{L}_\text{SubIB}(\mathcal{G},Y;Z) = -\mathcal{I}(Y;Z_\text{Sub}) + \beta \mathcal{I}(\mathcal{G};Z_\text{Sub}),
\end{equation}

Moreover, in actual dynamic wireless distributed networks, obtaining global graph structures in advance can be extremely challenging.
This prompts us to transform traditional SubIB theory to start from multiple subgraphs rather than relying on the global graph, thereby eliminating the dependency on global graph structures, as shown in Fig. 12 (d) (\emph{step 4}). Then, we can define the improved SubIB theory as follows:
\begin{equation}
\setcounter{equation}{16}
\min_{p(z|x,\mathcal{A})\in \Omega_\text{G}}\!\!\!\!\!\!\mathcal{L}_\text{SubIB}(\mathcal{G},Y;Z) = -\mathcal{I}(Y;Z_\text{Sub}) + \beta \mathcal{I}(\mathcal{G}_\text{Sub};Z_\text{Sub}),
\end{equation}

\begin{table*}[t!]
  \centering
  \fontsize{8.5}{12}\selectfont
  \caption{Characteristics for different mirror learning-enhanced MARL algorithms.}
  \label{mirror}
   \begin{tabular}{
   !{\vrule width1.2pt}  m{2 cm}<{\centering}
   !{\vrule width1.2pt}  m{0.65 cm}<{\centering}
   !{\vrule width1.2pt}  m{3.35  cm}<{\centering}
   !{\vrule width1.2pt}  m{3.7 cm}<{\centering}
   !{\vrule width1.2pt}  m{2.8 cm}<{\centering}
   !{\vrule width1.2pt}  m{2.8 cm}<{\centering}   !{\vrule width1.2pt} }
    \Xhline{1.2pt}
        \rowcolor{gray!30} \bf Frameworks  &  \bf Ref. &  \bf Classifications &  \bf Algorithm Applications &  \bf Pros  & \bf Cons \cr
    \Xhline{1.2pt}
        \multirow{5}{*}{\bf \shortstack{Maximum\\ Entropy}}
                    & \cite{[407]}  & Sample Efficiency   & Starcraft \uppercase\expandafter{\romannumeral2} \& MPE \& Soccer
                    & \multirow{5}{*}{\makecell[l]{$\bullet$ Effectively decompose\\ soft policy update \\$\bullet$ Significantly enhance\\ sample efficiency}}
                    & \multirow{5}{*}{\makecell[l]{$\bullet$ Difficult to represent\\ multi-modal complex\\ dynamics\\$\bullet$ Poor long-term action\\ prediction}} \\
        \cline{2-4} & \cite{[427]}  & General Parameterization   &  CartPole \& Acrobot & &  \\
        \cline{2-4} & \cite{[428]}  & Linear Convergence   & Channel Coding-Modulation  & &  \\
        \cline{2-4} & \cite{[429]}  & Sublinear Convergence & MaMuJoCo & &  \\
        \cline{2-4} & \cite{[411]}  & Sample Efficiency & Starcraft \uppercase\expandafter{\romannumeral2} \& MaMuJoCo  & &  \cr\Xhline{1.2pt}

        \multirow{6}{*}{\bf \shortstack{Transformer}}
                    & \cite{[431]}  & Exploration Enhanced  & Starcraft \uppercase\expandafter{\romannumeral2} \& Flatland
                    & \multirow{6}{*}{\makecell[l]{$\bullet$ Effectively achieve\\ exploration-exploitation\\ trade-off \\$\bullet$ Effectively achieve\\ accurate and consistent\\ long-term forecasting}}
                    & \multirow{6}{*}{\makecell[l]{$\bullet$ Ignore Dec-POMDPs'\\ distinct features \\$\bullet$ Excessive complexity \\$\bullet$ Excessive resource\\ demands}} \\
        \cline{2-4} & \cite{[432]}  & Exploration Enhanced  & LLMs & &  \\
        \cline{2-4} & \cite{[433]}  & State Uncertainty   & Two-Player \& MPE & &  \\
        \cline{2-4} & \cite{[434]}  & Expansion Enhanced  & Combinatorial Optimization  & &  \\
        \cline{2-4} & \cite{[435]}  & Centralized Aggregation  & Starcraft \uppercase\expandafter{\romannumeral2} & &  \\
        \cline{2-4} & \cite{[436]}  & Expansion Enhanced  & Starcraft \uppercase\expandafter{\romannumeral2} \& MaMuJoCo & &  \cr\Xhline{1.2pt}

    \end{tabular}
  \vspace{0cm}
\end{table*}
Note that learning subgraph hidden representations $Z_{\text{sub}}$ rather than global hidden representations $Z$ can further reduce overhead and latency, making this SubIB theory more conducive to expanding to large-scale downstream tasks.
\subsubsection{Optimizing MARL for Wireless Distributed Networks with Information Bottleneck}
Thanks to the powerful information compression (\emph{robustness}) and preservation (\emph{expressiveness}) of various IB theories, they have been successfully applied to solve various challenges in MARL-assisted wireless distributed networks, as shown in Table \uppercase\expandafter{\romannumeral6}.
Correspondingly, we can adopt applicable frameworks to address the corresponding challenges, and their specific tutorials are as follows:
\begin{itemize}
\item \emph{Challenging Problem 1:} Poor robustness and generalization often result from learning representations that overfit to known environments, making it challenging to generalize to new and unknown scenarios. To address this, we can apply critical IB theories to improve representation learning, ensuring that optimal representations preserve relevant data about optimization variables and prevent the aggregation of irrelevant data, thereby enhancing robustness and generalization. Moreover, optimizing variable $Y$ is often not obtainable in advance, which prompts us to update equations (4), (5), and (6) as follows:
    \begin{equation}
    \setcounter{equation}{17}
    \min_{z}\mathcal{L}_{\text{MARL}}(\mathcal{G};Z) = -\mathcal{L}_{\text{MARL}}(Z) + \beta \mathcal{I}(\mathcal{G};Z),
    \end{equation}
    where $\mathcal{L}_{\text{MARL}}(Z)$ denotes the loss function of MARL. Note that input data $\mathcal{G}$ can be transformed to $X$ under basic IB theory, and hidden representations $Z$ can be transformed to $Z_{\text{sub}}$ under SubIB theory.
\item \emph{Challenging Problem 2:} For poor communication efficiency, we can adopt a similar principle to learn optimal collaborative representations $M$, which clearly defines the communication strategy for each agent, mainly focusing the ``\emph{what}" in collaboration-enhanced MARL, thereby achieving efficient communication even with limited resources. Then, equation (7) can be updated to
    \begin{equation}
    \setcounter{equation}{18}
    \begin{aligned}
    \min_{m,z}\mathcal{L}_{\text{MARL}}(\mathcal{G}&;M,Z)= -\mathcal{L}_{\text{MARL}}(Z|M) \\
    &+ \beta_1 \mathcal{I}(\mathcal{G};M) + \beta_2 \mathcal{I}(\mathcal{G};Z|M),
    \end{aligned}
    \end{equation}
    where the Markovian process is transformed from the original $<\mathcal{G} \rightarrow Z \rightarrow Y>$ to $<\mathcal{G} \rightarrow M \rightarrow Z \rightarrow Y>$.
\item \emph{Challenging Problem 3:} For insufficient exploration in sparse-reward environments, it is crucial to provide agents with useful signals to guide their policy exploration. Rather than relying on aimless exploration, a target bottleneck can be employed to facilitate more effective policy transfer, which can be formulated as follows:
    \begin{equation}
    \setcounter{equation}{19}
    \min_{z}\mathcal{L}_{\text{MARL}}(\mathcal{G},\mathcal{T};Z) = -\mathcal{L}_{\text{MARL}}(Z) + \beta \mathcal{I}(\mathcal{T};Z|\mathcal{G}),
    \end{equation}
    where $\mathcal{T}$ provides information about the environmental reward structure for each agent. This enables more efficient exploration policies, especially when tasks are ambiguous or ill-defined.
\item \emph{Challenging Problem 4:} The excessive complexity in large-scale environments results in significant communication latency and network overhead, limiting the actual deployment of MARL in wireless distributed networks. By contrast, low-complexity SubIB theory can be leveraged to learn subgraph representations $Z_{\text{sub}}$. By using interpretable subgraphs that remove redundancy to train the original MARL networks, latency and overhead in wireless distributed networks can be significantly reduced. Similarly, we can formalize it as
    \begin{equation}
    \setcounter{equation}{20}
    \min_{z}\mathcal{L}_{\text{MARL}}(\mathcal{G};Z_{\text{sub}})=-\mathcal{L}_{\text{MARL}}(Z_{\text{sub}}) + \beta \mathcal{I}(\mathcal{G};Z_{\text{sub}}).
    \end{equation}

    Note that only SubIB theory is applicable for reducing complexity, while IB and GIB theories are not applicable due to learning global representations $Z$ rather than $Z_{\text{sub}}$.
\end{itemize}

\begin{figure*}[t]
\centering
    \includegraphics[scale=0.525]{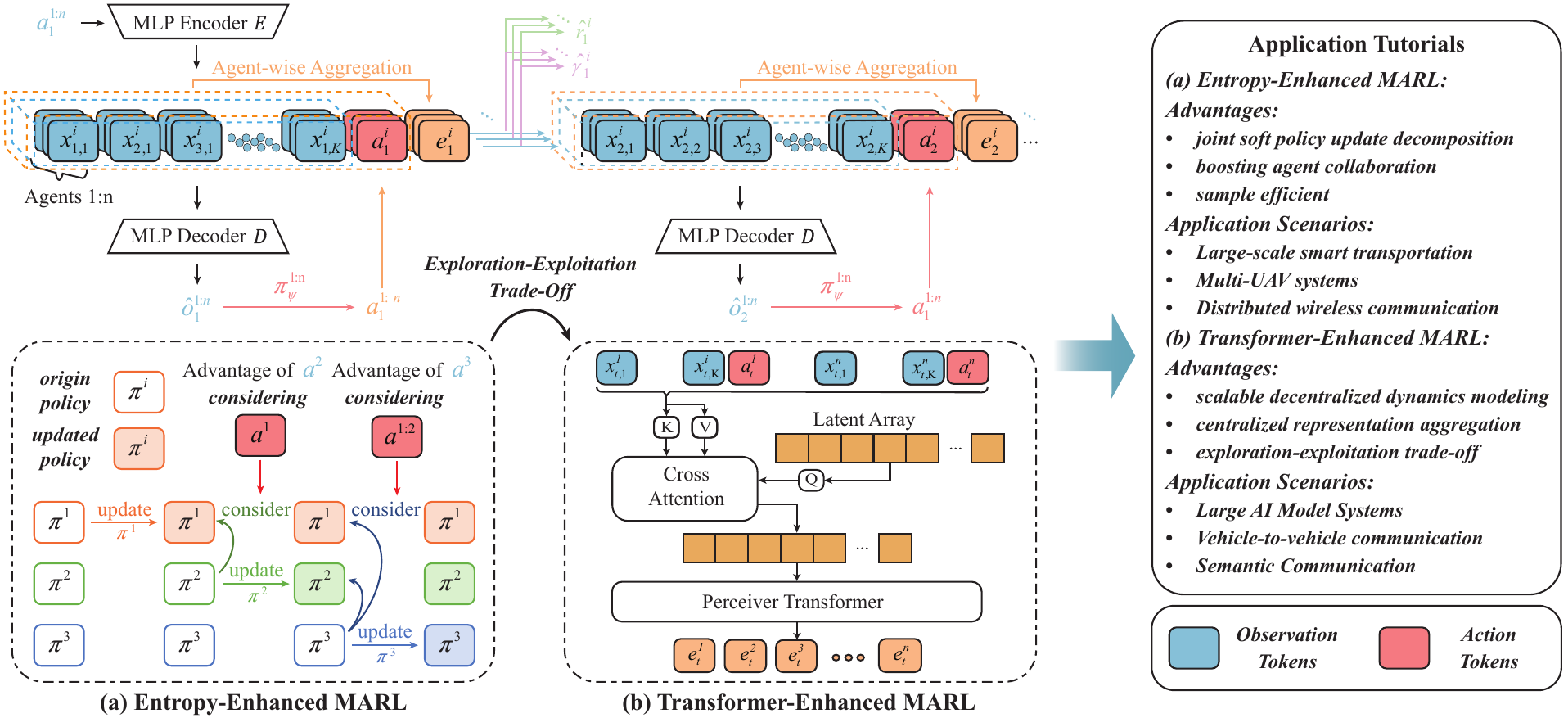}
    \caption{Two frameworks of mirror learning-enhanced MARL, including entropy and Transformer. Please refer to \cite{[407]} and \cite{[431]} for more details.
    \label{fig1}}
\end{figure*}
\emph{\textbf{Lessons Learned:}} By effectively integrating various IB theories into MARL-assisted wireless distributed networks, it helps to achieve superior robustness and expressiveness of learned representations, overcoming multiple challenges encountered in traditional MARL algorithms.
\subsection{Mirror Learning-Enhanced MARL}
Traditional frameworks such as GPI \cite{[817]} and TRL \cite{[818]} face scaling challenges in MARL algorithms, often relying on regularization or heuristics for practical success \cite{[401]}. However, these results are empirically driven and lack robust theoretical support, typically originating from approximations rather than genuine theoretical foundations \cite{[402]}. To address this issue, \emph{mirror learning} was proposed in \cite{[403]}. This framework provides theoretical assurances applicable to numerous MARL algorithms and deviates from the limiting assumptions in GPI and TRL, as illustrated in Fig. 9 (d). Mirror learning optimizes an objective through a drift functional to track policy divergence, ensuring consistent return improvements and convergence to optimality \cite{[404]}. Thus, this framework bridges the gap between algorithm creation and theoretical integrity, solving the scalability and reliability challenges encountered by traditional MARL approaches \cite{[405],[406]}.

Mirror learning-enhanced MARL algorithms, though theoretically robust, face high sample complexity and poor exploration \cite{[407]}. On-policy methods such as heterogeneous-agent proximal policy optimisation (HAPPO) and heterogeneous-agent trust region policy optimisation (HATRPO) \cite{[406]} demand new data for each gradient update, resulting in high costs with increasing task complexity or agent numbers \cite{[408]}. Although off-policy methods ease sample demands, they often become unstable and overly sensitive to hyperparameters \cite{[409]}. Thus, strategic development is needed for effective sample use, stable learning, and a balanced exploration-exploitation between agents. Existing strategies are mainly categorized into maximum entropy \cite{[406],[411],[427],[428],[429]} and Transformer approaches \cite{[431],[432],[433],[434],[435],[436]}, detailed in Table \uppercase\expandafter{\romannumeral11}. The following sections explore these frameworks, focusing on improved exploration strategies and off-policy learning techniques.
\subsubsection{Optimizing MARL for Wireless Distributed Networks with Maximum Entropy}
To formalize the concept of learning stochastic policies $ \boldsymbol{\pi}(\cdot | s_t) $ and address the issue of insufficient exploration, the challenge of collaborative MARL is framed within the probabilistic graphical model framework \cite{[410]}. An optimality variable $ \mathcal{O}_t $ is introduced based on the methodology outlined in \cite{[412]}. This variable, which assumes binary values, represents whether the joint actions executed by all agents are optimal. Specifically, $ \mathcal{O}_t = 1 $ signifies that the joint action at $t$ time slot is optimal. The probability of this event is modeled as $ \mathcal{P}(\mathcal{O}_t = 1 | s_t, a_t) \propto \exp(\mathcal{R}({s}_t, a_t)) $, where $ \mathcal{R}({s}_t, a_t) $ denotes the reward for performing actions $ a_t $ in states $ {s}_t $.

Structured variational inference is employed to approximate the posterior distribution $\mathcal{P}(\tau | \mathcal{O}_{1:T} = 1) $, given by
\begin{align}
\bigg[\mathcal{P}(s_1) \prod_{t=1}^T\mathcal{P}(s_{t+1} | s_t, a_t) \bigg] \exp\bigg( \sum_{t=1}^T \mathcal{R}(s_t, a_t) \bigg),
\end{align}
over the trajectory $ \tau $. The variational distribution is defined as
\begin{align}
Q(\tau) = Q(s_1) \prod_{t=1}^T Q(s_{t+1}|s_t, a_t) Q(a_t|s_t).
\end{align}

In this formulation, the environment dynamics are fixed by setting $ Q(s_1) =\mathcal{P}(s_1) $ and $ Q(s_{t+1}|s_t, a_t) =\mathcal{P}(s_{t+1}|s_t, a_t) $. This constraint ensures the avoidance of risk-seeking behaviors, as outlined in \cite{[412]}.
\begin{figure}[t]
\centering
    \includegraphics[scale=0.5]{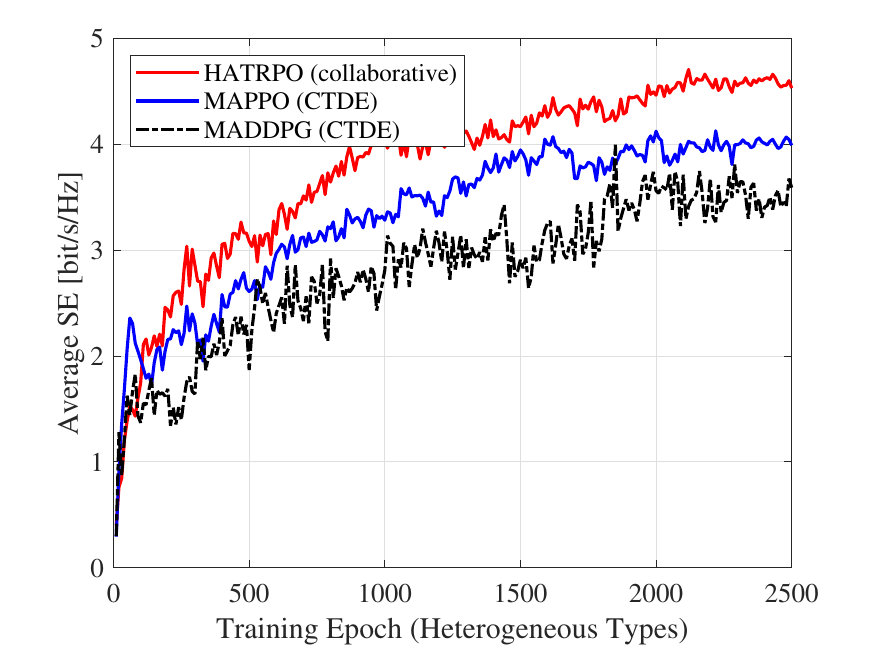}
    \caption{Test reward (average SE) curves of heterogeneous distributed networks under different models. Please refer to \cite{[626]} for more details.
    \label{fig1}}
\end{figure}
This inference procedure ultimately leads to the \emph{maximum entropy} objective for mirror learning-enhanced MARL, as shown in Fig. 14 (a), which aims to optimize the joint policy while ensuring sufficient exploration of the state-action space:
\begin{align}
\label{eq3}
\!\!\!\!J(\boldsymbol{\pi})\!=\!\mathbb{E}\bigg[ \sum_{t=1}^{T} \Big(\mathcal{R}\left(s_t, a_t,s_{t+1}\right) +\alpha\sum_{i=1}^{n}{\mathcal{H}\left(\pi_i\left(\cdot | s_t\right)\right)}\Big)\bigg],
\end{align}
\normalsize
where $\alpha$ represents the temperature constant, which governs the balance between maximizing rewards and incorporating entropy regularization, and $\mathcal{H}$ is the entropy function.

A key proposition presented by \cite{[411]} demonstrates that the joint soft policy update can be decomposed into a sequence of local policy updates, which is significant as it shows that a mirror learning-enhanced MARL problem can be reformulated as the aggregation of $n$ independent mirror learning-enhanced RL problems. This decomposition enables agents to efficiently improve a shared soft policy in multi-agent learning by optimizing their own KL-divergence, aiding global optimization. As depicted in Fig. 15, the simulations confirm that mirror learning in MARL surpasses traditional methods concerning GPI and TRL, demonstrating its superiority in boosting agent collaboration. This underscores mirror learning's reliability in tackling issues in traditional MARL, reinforcing its theoretical and practical benefits in wireless networks.
\subsubsection{Optimizing MARL for Wireless Distributed Networks with Transformer}
To address the challenges, especially exploration-exploitation trade-off, that inherent in multi-agent environments, a robust world model was proposed in \cite{[426]}, combining scalable decentralized dynamics modeling with centralized representation aggregation, adhering to the CTDE architecture. Decentralized dynamics learning was formulated as a sequence modeling problem over discrete tokens to ensure accurate and consistent long-term predictions despite non-stationary local dynamics. The approach centers on the highly expressive Transformer architecture, as shown in Fig. 14 (b). Significantly, this work introduces the first Transformer-based world model specifically tailored for multi-agent settings. Within this framework, a trajectory $\tau_{i}$ for agent $i$ is defined as comprising $T$ local observations and actions, as
\begin{align}
    \tau_{i} = (o_{i,1}, a_{i,1}, \ldots, o_{i,t}, a_{i,t}, \ldots, o_{i,T}, a_{i,T}).
\end{align}

To efficiently utilize the Transformer architecture, representing trajectories as sequences of tokens is crucial. A simplistic discretization method often divides continuous observations into $ m $ equal bins per dimension \cite{[425]}. However, this ignores Dec-POMDPs' distinct features, where agents experience shared and individual observations. Instead, compressing observations into tokens offers a more compact representation and enables token reuse across multiple agents' observations.

\emph{\textbf{Lessons Learned:}} By effectively integrating multiple mirror learning-based theories into MARL-assisted wireless distributed networks, it helps to achieve superior exploration efficiency and low overhead, overcoming several challenges encountered in traditional MARL algorithms.

To achieve this, a vector quantized variational autoencoder (VQ-VAE) is utilized, serving a role similar to that of a tokenizer in natural language processing (NLP) \cite{[423],[422]}. It comprises three primary components: an encoder, a decoder, and a codebook, which is a discrete set of code vectors sized $N$, with fixed dimensionality $n_z$. The encoder generates latent variables from the input observations, each as a vector of $n_z$ sized, transformed into $K$ tokens. A nearest-neighbor search in the codebook maps these tokens to the closest vectors by minimizing distance. The decoder then reconstructs the observation from these tokens, translating each codebook index back into the original approximation. The codebook efficiently compresses local observations into succinct token sequences, improving sequence modeling by streamlining inputs.

\begin{figure*}[t]
\centering
    \includegraphics[scale=0.3975]{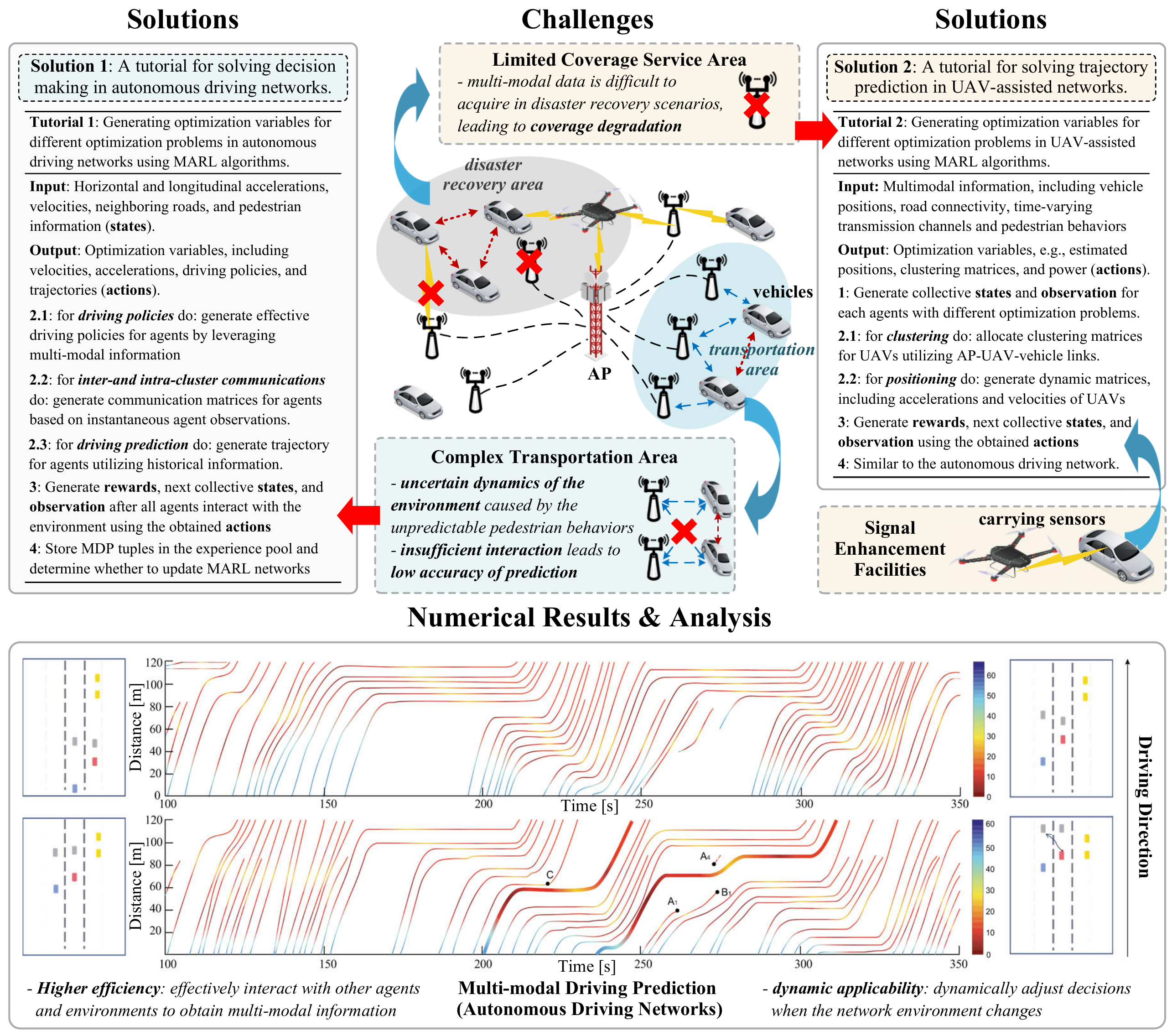}
    \caption{Key challenges faced and feasible techniques in autonomous driving for 6G. Challenge 1: Complex transportation areas caused by uncertain dynamics of the environment and insufficient interaction, which can be addressed utilizing inter-and intra-cluster communications and driving prediction. Challenge 2: Limited coverage service area caused by natural disasters, which can be addressed utilizing effective UAVs. Please refer to \cite{[659]} for more details.
    \label{fig1}}
\end{figure*}
When working on model prediction, we explore how attention maps function by employing the cross-attention mechanism derived from both shared Transformer and Perceiver models. In predicting local dynamics, we can pinpoint two unique attention strategies. One strategy follows a Markovian pattern, basing predictions mostly on the latest transitions. Conversely, the second strategy displays a more structured approach, directing focus towards specific markers over several previous transitions. While aggregating agent data, we notice two primary approaches: focusing on individual entities and identifying commonalities among agents. These varied attention strategies, observed in Transformer and Perceiver models, are crucial for forming precise and reliable representations of intricate local dynamics.

\emph{\textbf{Lessons Learned of Section \uppercase\expandafter{\romannumeral5}:}}
Analyzing recent studies reveals the essential integration of MARL and emerging techniques in addressing various challenges of wireless distributed networks. Specifically, emerging techniques enhance MARL through their unique advantages, such as effective communication protocols in collaboration-enhanced MARL and innovative representation principles in IB-enhanced MARL, which are crucial for achieving robust, scalable, and efficient communication in applications like autonomous driving and wireless communications. This synergy provides new possibilities for further unleashing the full power of MARL in wireless distributed networks for 6G.
\section{Application Scenarios}
In this section, we explore several application scenarios where MARL can be effectively utilized to enhance performance and unleash the full power of wireless distributed networks, including homogeneous and heterogeneous paradigms. These scenarios demonstrate the versatility and potential of MARL in tackling various challenges inherent in wireless distributed networks for 6G.
\subsection{MARL for UAV-Assisted Communication Networks}
UAV-assisted communication networks \cite{[438],[439]} play an important role in applications such as disaster rescue and monitoring \cite{[637]}, especially when traditional infrastructure is damaged, as they can quickly deploy a large number of UAV-assisted APs to establish temporary communication links and enhance resource-carrying capacity \cite{[639],[209]}. However, how to achieve efficient resource allocation and secure collaboration among UAVs in complex dynamic environments is a crucial aspect. MARL \cite{[640]} offers significant potential for facilitating communication, leveraging its decentralized nature, especially in unpredictable settings.

For example, the authors in \cite{[640]} conducted a comprehensive survey and in-depth analysis of autonomous multi-UAV wireless networks that support MARL, and summarized multiple cutting-edge applications of MARL-enabled multi-UAV wireless networks. This further demonstrates the effectiveness of MARL in enhancing UAV-assisted communication networks by optimizing key factors in 6G. Then, based on the strict demands of actual distributed networks, we introduce two typical applications, such as resource allocation \cite{[645],[646]} and secure communications \cite{[643],[647]}, and their optimization tutorials with MARL are as follows:
\begin{itemize}
\item \emph{{Resource Allocation:}} Given the limited spectrum resources and onboard hardware constraints of miniaturized UAVs, designing efficient resource allocation schemes is essential to ensure the effectiveness of UAV-assisted communication networks \cite{[640]}. In this context, the authors in \cite{[645]} proposed a MARL-based collaborative scheme for resource allocation in UAV-assisted networks, where each UAV is modeled as an agent. Through collaboration, UAVs dynamically select deployment positions, transmission power, and sub-channel allocations. This decentralized scheme significantly mitigates issues like communication latency, compared to centralized paradigms.

    In contrast, the authors in \cite{[646]} introduced a MARL-based dynamic scheme for resource allocation that operates without information sharing. Each agent communicates solely with associated ground UEs and learns the optimal policy based on local observations, without utilizing the information observed by neighboring agents. The obtained results indicate that this scheme achieves a better balance between performance and overhead.

\item \emph{{Secure Communications:}} Due to the high altitude of UAVs, aerial-to-ground communication links are primarily characterized by line-of-sight (LoS) propagation, which significantly increases the risk of eavesdropping \cite{[643]}. This vulnerability underscores the urgent need for robust countermeasures to safeguard the confidentiality and integrity of communication. One promising approach to address this challenge is adversarial MARL \cite{[647]}, which models the interaction between legitimate and eavesdropping UAVs as a zero-sum game. Each UAV alternates between acting as an agent to optimize its respective adversarial objectives. Furthermore, the authors in \cite{[645]} introduced a federated learning framework that enables UAVs to share non-sensitive data while ensuring confidentiality is preserved.
\end{itemize}
\begin{figure*}[t]
\centering
    \includegraphics[scale=1.85]{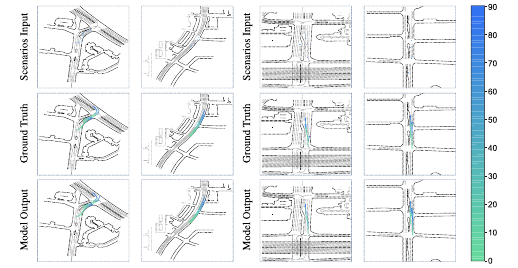}
    \caption{Description of time-space diagram of a 3-lane arterial illustrating the effect of bus stops for 6G, including scenarios input, ground truth, and model output. The left part of the figure highlights lane-changing movements, demonstrating MARL's proficiency in forecasting varied and precise trajectories amidst dynamic driving situations, while the right part showcases its adeptness in navigating complex intersection contexts, underscoring its robustness and precision in distributed conditions. Please refer to \cite{[662]} for more details.
    \label{fig1}}
\end{figure*}

In the above works, the agents belong to the same type, fulfilling the consistency condition typical of homogeneous distributed networks.
However, when both ground and aerial APs coexist in a UAV-assisted network \cite{[642]}, the network evolves into a heterogeneous air-ground integrated paradigm. In such a scenario, the ground APS are stationary, and coordinating agents with varying capabilities and limitations becomes a key challenge.
Moreover, this concept can be further extended to a heterogeneous space-air-ground integrated paradigm \cite{[653]}, where factors like satellite-based communication links and coordination across multiple layers are critical.

\emph{\textbf{Lessons Learned:}} The integration of MARL into applications such as resource allocation and secure communications holds significant promise for enhancing UAV-assisted communication networks. MARL offers a powerful approach to address challenges like high network overhead, communication latency, and information leakage. In summary, the insights from this subsection highlight MARL's ability to dynamically adapt strategies in unpredictable environments, making it a key enabler for improving collaboration among UAVs.
\begin{figure*}[t]
\centering
    \includegraphics[scale=0.3975]{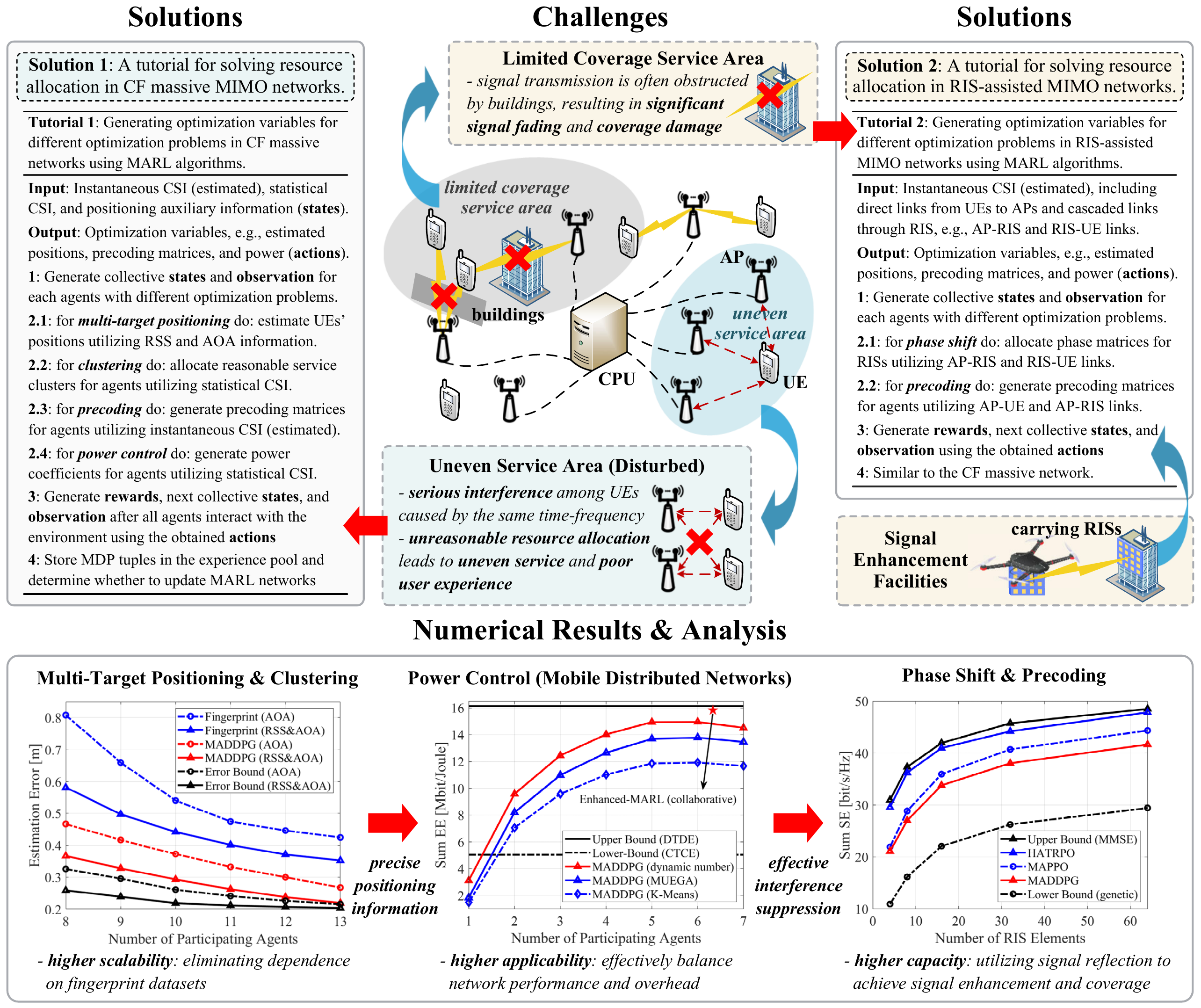}
    \caption{Key challenges faced and feasible techniques in CF massive MIMO networks for 6G. Challenge 1: Uneven service areas caused by severe interference and unreasonable resource allocation, which can be addressed utilizing multi-target positioning, clustering, precoding, and power control. Challenge 2: Limited coverage service area caused by building obstruction, which can be addressed utilizing effective RISs. Please refer to \cite{[619]}, \cite{[623]} and \cite{[626]} for more details.
    \label{fig1}}
\end{figure*}
\subsection{MARL for Autonomous Driving Networks}
Autonomous driving are increasingly deployed to enhance traffic safety, reliability, and efficiency in modern distributed networks \cite{[201],[651]}, due to the flexibility and real-time decision-making capabilities of autonomous vehicles in dynamic environments. However, the inherent limitations of sensors may lead to erroneous decisions, potentially causing severe accidents \cite{[652]}. By contrast, integrating MARL with autonomous vehicles \cite{[650]} offers a powerful approach, compensating for sensor deficiencies and enhancing safety through its decentralization and collaboration mechanisms. Then, we review recent research on MARL-assisted autonomous driving networks, focusing on advanced decision-making strategies and communication mechanisms \cite{[648],[649],[650]}.

For example, the authors in \cite{[650]} reviewed key studies of MARL-related autonomous driving networks and exploring their critical components such as environment modeling and algorithm design. This provides valuable insights into applying MARL to autonomous driving networks. One fundamental area where MARL shows promise is motion planning \cite{[650]}, and the optimization tutorial with MARL is as follows:
\begin{itemize}
\item \emph{{Motion Planning:}} As an important aspect of autonomous driving, motion planning mainly involves real-time planning and adjustment based on perceived environments and traffic conditions \cite{[654],[656]}. This is essential for ensuring safety and stability in unpredictable environments, enabling vehicles to effectively avoid obstacles, pedestrians, and other traffic participants \cite{[437]}. In this context, the authors in \cite{[648],[649],[650]} adopted MARL from numerous emerging techniques to address motion planning problems, which can help vehicles collaborate with multiple neighboring vehicles and better adapt to complex dynamic conditions in wireless distributed networks.
\end{itemize}

Fig. 16 displays driving paths designed for real urban landscapes, utilizing features like maps and initial paths, yielding outputs that detail spatiotemporal interactions, speed, and acceleration. The progression of time is represented by varying the intensities of the color \cite{[659]}. The left segment illustrates lane-change dynamics, while the right segment demonstrates the model's stability in intersections. This highlights that trajectory predictions can reach an average displacement error of 2.045 and a minimum average displacement error of 1.472, emphasizing the criticality of coordinated decision-making in autonomous driving networks. Furthermore, Fig. 17 specifically depicts the agent behavior on a three-lane road with two bus stops, demonstrating path creation through a suitable exploration policy. The results clearly indicate that agents can flexibly adjust their strategies to accommodate changes in environmental conditions, effectively showcasing the adaptability of the model in dynamic scenarios \cite{[662]}.

However, sharing only states or actions (e.g., position, velocity, and acceleration) does not provide sufficient context for vehicles to fully understand their environment or neighboring vehicles' intentions, limiting effective collaboration in actual distributed networks. To overcome this, sharing visual information from onboard cameras and LIDAR can enhance situational awareness \cite{[659]}, allowing vehicles to detect obstacles and predict the behavior of neighboring vehicles more accurately, thereby fostering more efficient collaboration.

\emph{\textbf{Lessons Learned:}} While traditional sharing schemes for autonomous driving networks are useful as they typically involve sharing specific data, they fall short in accurately identifying obstacles or predicting the behavior of other traffic participants. This limitation increases the risk of collisions or accidents. Thus, expanding information sharing to include environmental perception data, such as visual information from onboard cameras and LIDAR, is crucial. By incorporating this richer context, vehicles can gain a better understanding of the surrounding traffic dynamics, enabling more effective collision avoidance. In summary, capturing and sharing key environmental data through MARL is vital for enhancing collaboration and safety in autonomous driving networks.
\subsection{MARL for CF Massive MIMO Networks}
Among emerging techniques, CF massive MIMO \cite{[610],[611],[612]} is considered an evolutionary paradigm of massive MIMO technique in meeting the complex demands of 6G use cases, which embraces an extraordinarily large number of geographically distributed APs rather than cellular cells, thereby promoting effective communication and overcoming performance constraints, such as spectrum efficiency (SE) and energy efficiency (EE) \cite{[617]}.
Recently, various MARL algorithms have been extensively applied to address various optimization problems in CF massive MIMO networks, including multi-target positioning \cite{[619],[620]}, clustering \cite{[621],[622],[623],[624],[628]}, precoding \cite{[626],[627]}, and power control \cite{[629],[630],[631]}. For different optimization problems, we can categorize common CF massive MIMO networks into uplink and downlink, where UEs in the uplink and APs in the downlink can both be defined as heterogeneous types of agents. Moreover, Fig. 18 illustrates the tutorials of MARL for different optimization problems, and their specific details are discussed as follows:
\begin{itemize}
\item \emph{{Multi-Target Positioning (downlink):}} Multi-target positioning involves accurately determining the positions of targets through the dense deployment of APs \cite{[660],[661]}, which helps to achieve efficient resource allocation. However, traditional schemes rely heavily on high-overhead fingerprint databases, which limit scalability in actual distributed networks. This is where MARL shines in CF massive MIMO networks \cite{[619],[620]}, as it enables adaptive multi-target positioning due to its decentralized nature. Specifically, the authors in \cite{[619],[620]} introduced a joint positioning and correction framework, which achieved a positioning accuracy improvement of over 51.58\% using MARL and eliminated the need for fingerprint databases, thus addressing the challenges of high-dimensional signal processing.

\item \emph{{Clustering (coexistence of uplink and downlink):}} Clustering in CF massive MIMO networks mainly involves AP, UE, or antenna clusters \cite{[621],[622],[623],[624],[628]}, all of which aim to achieve efficient communication. Specifically, MARL allows agents to autonomously learn the optimal policy and select suitable neighboring agents for intra-cluster collaboration, which is beneficial for reducing overhead, e.g., achieving a 66.38\% reduction \cite{[623]}. Compared to traditional global schemes, they focus on achieving a balance between performance and overhead, rather than blindly pursuing performance. Thus, optimization objectives under clustering are often to maximize EE rather than SE \cite{[622],[623]}.

\item \emph{{Precoding (downlink):}} Due to the dense user distribution and numerous antennas, interference management has become a serious challenge in CF massive MIMO networks, as signal overlap can degrade performance \cite{[610],[611]}. By contrast, precoding offers a viable solution by using known CSI to adjust transmitted signals, directing them to the intended UEs while reducing interference. However, imperfect CSI caused by unpredictable environments complicates accurate beamforming decisions. This prompts the authors \cite{[626],[627]} to attempt utilizing MARL to address this, where each agent autonomously adjusts its transmission direction based on local observations, thus reducing reliance on perfect CSI. Moreover, the robust Edge-GIB framework proposed in \cite{[515]} can further assist in extracting key information from imperfect CSI to improve beamforming decisions, especially in environments prone to interference.

\item \emph{{Power Control (coexistence of uplink and downlink):}} On the other hand, power control is equally crucial for interference management in CF massive MIMO networks, typically relying on statistical CSI instead of instantaneous CSI to optimize power allocation \cite{[610],[611]}. However, imperfect CSI also complicates effective power control, highlighting the need for MARL to achieve accurate power coefficients. In this context, the authors in \cite{[629],[630],[631]} proposed MARL-based intelligent power control schemes, with the former \cite{[629],[630]} focusing on studying power allocation among UEs, while the latter \cite{[631]} extending this to antennas to further suppress interference. This decentralized approach eliminates reliance on centralized control and perfect CSI, achieving a better balance between performance and overhead.
\end{itemize}

It is evident that MARL algorithms have demonstrated strong potential in wireless distributed networks, particularly in balancing performance and overhead, thanks to their decentralization and collaboration mechanisms. However, implementing MARL-assisted wirelesss distributed networks in actual 6G settings presents opportunities but also significant challenges, such as complexity. Unlike static settings, adaptation to rapid changes such as user dynamics and channel variations requires swift decision-making. This requires complex algorithms to manage high-dimensional data with minimal delay. Additionally, deploying MARL in 6G presents obstacles such as the need for distributed computing, agent communication, and scalable designs. Synchronizing agents becomes difficult in large systems with limited connectivity.

\emph{\textbf{Lessons Learned:}} This subsection summarizes four typical optimization problems addressed by MARL in CF massive MIMO networks. Unlike traditional schemes, integrating MARL eliminates reliance on centralized control and perfect CSI, enabling more accurate decision-making. Additionally, various IB theories can be applied to extract essential data, overcoming the challenges posed by imperfect CSI in actual distributed networks, especially in environments prone to interference.
\begin{figure*}[t]
\centering
    \includegraphics[scale=0.1375]{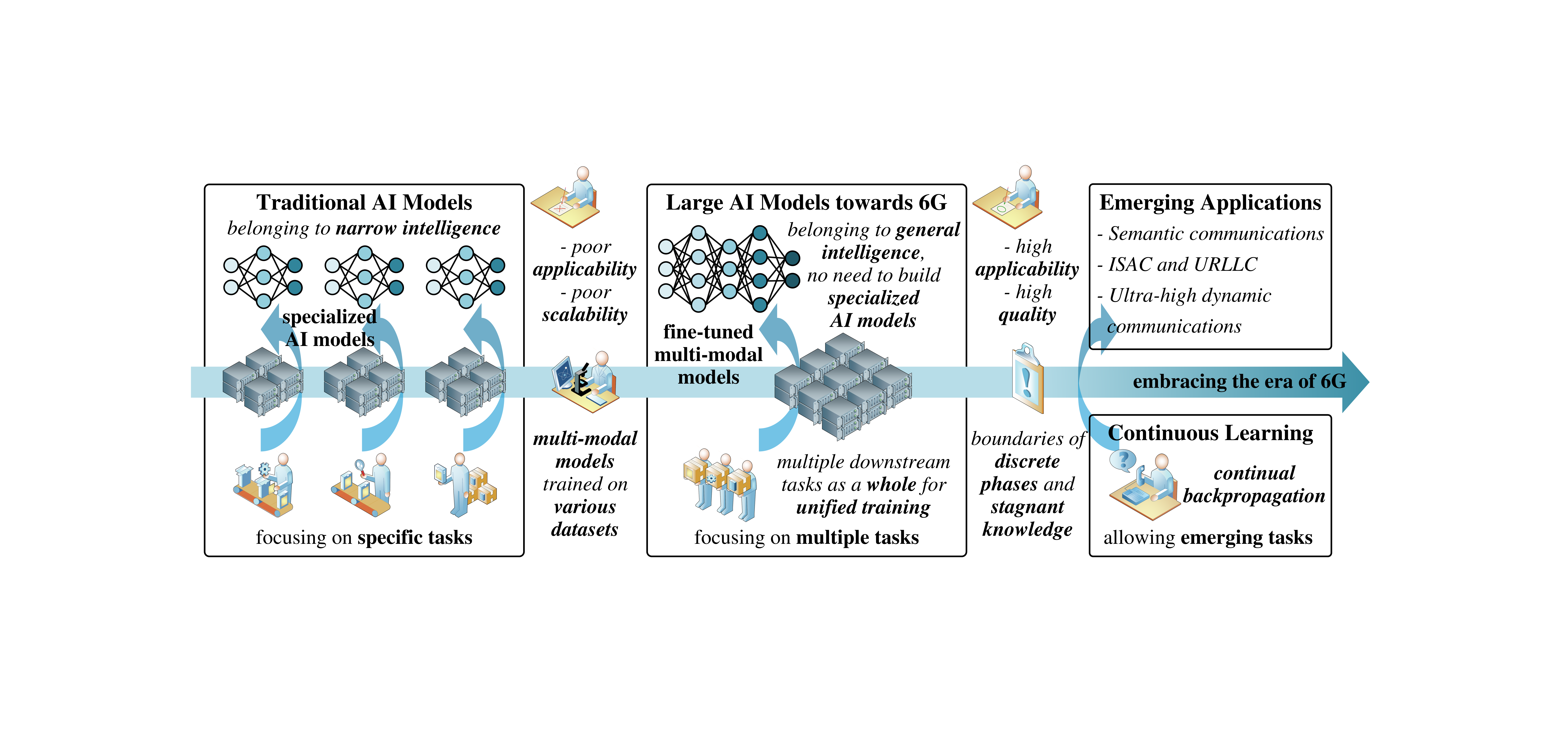}
    \caption{The evolution of future directions for MARL-assisted wireless distributed networks, including from traditional AI models to large AI models towards 6G, from transient learning to continuous learning, and from traditional communications to emerging communications.
    \label{fig1}}
\end{figure*}
\subsection{MARL for RIS-Assisted MIMO Networks}
RIS \cite{[632],[633]} has also gained significant attention as an emerging technique for wireless distributed networks, offering the ability to alter radio waves at the electromagnetic level without requiring complex processors or power amplifiers \cite{[634]}. Similarly, many studies have proposed MARL algorithms for RIS-assisted MIMO networks to unleash their potential, with typical applications including phase shift \cite{[635],[636]} and precoding \cite{[626],[627]} (similar to precoding discussed in Section \uppercase\expandafter{\romannumeral6}-C), and the specific details of the former are discussed as follows:
\begin{itemize}
\item \emph{Phase Shift (downlink):} Phase shift involves adjusting the phase of each element in RISs to optimize signal propagation, enhancing signal synthesis (i.e., constructive interference) or reducing signal attenuation (i.e., destructive interference) \cite{[615],[634]}. However, when multiple APs and RISs are involved, designing collaborative phase shifts becomes challenging \cite{[615]}, especially with imperfect CSI, which can increase mutual interference and degrade performance. Taking this into account, the authors in \cite{[635],[636]} addressed this challenge by utilizing MARL, where each RIS autonomously adjusts its phase shift based on local feedback, reducing the need for centralized control and mitigating interference, even under imperfect CSI conditions. Moreover, various IB theories can also be utilized to suppress the impact of imperfect CSI \cite{[515]}.
\end{itemize}

Correspondingly, Fig. 16 also illustrates the tutorials of MARL algorithms for different optimization problems in CF RIS-assisted MIMO networks, and the obtained results further reveal the potential of MARL in wireless distributed networks.

\emph{\textbf{Lessons Learned:}} Emerging CF RIS-assisted massive MIMO networks require advanced communication techniques to manage an immense number of devices, especially in environments with multiple APs and RISs.  MARL can better fulfill this responsibility with powerful decentralized nature, as it defines each RIS as an agent that autonomously adjusts its phase shift, reducing mutual interference and improving performance. Even under imperfect CSI, a common challenge in actual scenarios, MARL has been proven to be effective. In summary, this subsection highlights MARL's crucial role in efficient resource allocation across actual distributed networks.
\section{Future Directions}
In this section, we discuss several future research directions for MARL-assisted wireless distributed networks, which can be mainly divided into three categories, such as structural expansion, dimensional expansion, and application expansion.
\subsection{Dimensional Expansion: AI Models to Large AI Models}
As we move towards the deployment of 6G networks, which aim to provide higher reliability, stability, and service quality than existing techniques, thereby further enhancing user experience. This advancement has driven the rise of large-scale AI models (LAMs) and fundamentally transformed the role of AI \cite{[709]}. Specifically, compared to traditional AI models \cite{[707],[443]} that focus on specific tasks, LAMs \cite{[708],[710],[442]} demonstrate the potential for a more general and extensive form of intelligence, capable of handling the enormous complexity and scalability demands of 6G, as shown in Fig. 19. Moreover, LAMs are also expected to usher in a new era of wireless distributed networks \cite{[708],[710]}, which can fine-tuned multi-mode models trained on various datasets to handle multiple downstream tasks, thereby eliminating the need to build and train specialized AI models for each specific task. Note that the key advantage of LAMs lies in their ability to handle large amounts of data, learn from different sources, and support real-time decision-making at an unprecedented scale.

Correspondingly, the evolution of AI models in MARL-assisted wireless distributed networks towards LAMs can be understood through the following important transformations:
\begin{itemize}
\item \emph{{Breaking the Restriction of Static Knowledge Transfer:}} AI models in MARL-assisted wireless distributed networks typically focus on specific tasks and static knowledge transfer mechanisms, which makes it difficult for them to adapt to rapidly changing environments or expand into new tasks. By contrast, LAMs utilize advanced \emph{continuous learning} techniques and large-scale knowledge representation to achieve dynamic incremental learning while maintaining plasticity. This enables agents to seamlessly transfer knowledge between numerous tasks and contexts, enhancing their generalization in different environments.

\item \emph{{Breaking the Barrier of Model Scalability:}} AI models in MARL-assisted wireless distributed networks typically operate on relatively small network scales, limiting their ability to process and analyze typically large and complex datasets in 6G. By contrast, LAMs aim to effectively enhance scalability, allowing agents to train powerful models to handle massive datasets with high-dimensional inputs. This scalability unlocks the ability to process vast amounts of data in real-time, empowering empowering to make more informed decisions and seamlessly adapt to the growing demands of 6G networks.
\end{itemize}

This indicates that under the catalysis of LAMs, this paradigm shift in AI can enable MARL-assisted wireless distributed networks to autonomously adapt, learn, and optimize in various complex dynamic environments. Therefore, integrating LAMs into MARL-assisted wireless distributed networks represents a transformative leap, enabling the next generation of intelligent and scalable networks.
\subsection{Structural Expansion: Transient to Continuous Learning}
On the other hand, embracing the era of 6G networks, traditional transient learning focuses on training models on limited and discontinuous datasets, leading to the necessity of exploring more advanced techniques that can adapt to the dynamic and continuous characteristics of complex networks.
This has driven the evolution from traditional transient learning to continuous learning \cite{[701],[702]}, as shown in Fig. 19. Specifically, the evolution towards continuous learning is not only a technological enhancement, but also a fundamental shift in how complex networks learn, adapt, and optimize over time.

Correspondingly, for MARL-assisted wireless distributed networks, the key transformations brought about by continuous learning compared to transient learning are as follows:
\begin{itemize}
\item \emph{{Breaking the Boundaries of Discrete Training Phases:}} Transient learning typically involves episodic training where agents are trained over a fixed duration and environment. This makes it difficult to cope with continuous changes in network demands. By contrast, continuous learning enables agents to continuously update and improve their strategies \cite{[705]}, thereby achieving lifelong learning. This adaptability ensures that the network can respond in real-time to new conditions and unforeseen challenges without the need for resetting or restarting.

\item \emph{{Breaking the Barrier of Stagnant Knowledge:}} Transient learning typically forgets previously learned knowledge when adapting to new tasks, which may lead to agents focusing on specific short-term tasks. By contrast, continuous learning allows networks to continuously update their knowledge while retaining old knowledge as new datasets flow in, thereby achieving continuous real-time adaptation \cite{[703]}. This effectively addresses the challenges brought by ``\emph{catastrophic forgetting}".

\end{itemize}

Continuous learning requires algorithms that introduce ongoing diversity into networks to maintain infinite plasticity, such as continual backpropagation \cite{[703]}, which is a variant of backpropagation where a small portion of less-used units are continuously and randomly reinitialized. Thus, this transformation enhances MARL's effectiveness in complex distributed networks, where sustained adaptability is crucial for success.
\subsection{Application Expansion: Emerging Communications}
In addition to the aforementioned extensions, it is imperative to explore more advanced emerging communication paradigms that can enhance the intelligence and effectiveness of future 6G networks,
such as semantic communications, ISAC, URLLC, and ultra-high dynamic communications, as shown in Fig. 19. Their specific details are as follows:
\begin{itemize}
\item \emph{{Semantic Communications:}}
    Semantic communications focus on conveying meaning rather than raw data. Unlike traditional networks that prioritize high data rates and low errors, semantic communication emphasizes the relevance and context of exchanged information. This is crucial in applications like video streaming, remote healthcare, and autonomous networks, where context drives decision-making \cite{[835]}. Traditional approaches often transmit redundant information inefficiently, while MARL allows agents to adapt to communication semantics, optimizing data interpretation and transmission. Collaborative learning enhances data compression, signal representation, and the transmission of only relevant information, making communication networks more efficient.

\item \emph{{Integrated Sensing and Communications:}}
    ISAC integrates communication and sensing into a unified framework, enhancing spectrum utilization and network efficiency by enabling simultaneous data transmission and precise environmental awareness \cite{[216]}. This is critical in dynamic environments like autonomous driving and industrial automation, where real-time sensing drives decision-making. Traditional networks struggle to balance data transmission rates with sensing capabilities. In contrast, MARL approaches can significantly improve network functionality by enabling agents to collaborate effectively in resource distribution and decision-making.

\item \emph{{Ultra-Reliable Low-Latency Communications:}}
    URLLC is vital for networks like the tactile Internet and autonomous driving, which require high reliability and low latency. Integrating URLLC into MARL-driven distributed systems, such as CF massive MIMO, enhances reliability and scalability by decentralizing communication \cite{[714]}. MARL's multi-modal perception ensures ultra-reliable links, essential for intelligent transportation systems that rely on synchronized visual and tactile inputs \cite{[711]}. Effective URLLC scheduling minimizes latency for critical data, crucial in applications like autonomous vehicles and UAVs, where real-time resource allocation is needed.

\item \emph{{Ultra-High Dynamic Communications:}}
    Unlike traditional dynamic communications, which often focuses on low-speed nodes such as UAV swarms or autonomous vehicles, ultra-high dynamic communications focus on high-speed rail scenarios, characterized by rapidly changing network topology and fluctuating demands. This requires effective response to Doppler frequency shift caused by high-speed mobility while achieving real-time decision-making to cope with time-varying channels \cite{[916]}. In this context, traditional designed algorithms and communication protocols often struggle to maintain performance. By contrast, collaboration-enhanced MARL provides the possibility of maintaining service quality in ultra-high dynamic settings through efficient communication mechanisms, enabling agents to continuously learn and adapt to changing network conditions.
\end{itemize}

The integration of various emerging communications with MARL-assisted wireless distributed networks is an area of ongoing exploration. When integrated with MARL, these emerging communication paradigms hold the potential to further optimize intelligent policies, enabling agents to learn and adapt collaboratively in dynamic environments, paving the way for more scalable, adaptable, and context-sensitive distributed networks in future 6G.
\section{Conclusions}
In this tutorial paper, we have presented a comprehensive review of MARL-assisted wireless distributed networks, which are an integral part of 6G. Firstly, we have introduced the evolution and distinctive features of wireless distributed networks and MARL, particularly highlighting their connection. Subsequently, various preliminaries and basic components for wireless distributed networks were proposed, and a comprehensive investigation of different structures was conducted, including homogeneous and heterogeneous. Also, we have discussed the basic concepts of MARL and presented two typical categories, i.e., model-based and model-free, providing essential guidance and insights for integration into wireless distributed networks. Furthermore, we have summarized the integration of emerging techniques with MARL-assisted wireless distributed networks and offered valuable insights into their implementation. Last, we have presented many MARL-empowered application scenarios and potential future directions in wireless distributed networks. This tutorial paper can serve as a guideline for primary research works on MARL-assisted wireless distributed networks for 6G from the perspective of potential connections, network analysis, algorithm design, the integration of emerging techniques, MARL-empowered application scenarios, and promising future directions.
\bibliographystyle{IEEEtran}
\bibliography{IEEEabrv,Ref}
\end{document}